\documentclass[a4paper,fleqn,usenatbib,useAMS]{mnras}

\usepackage{newtxtext,newtxmath}
\usepackage[T1]{fontenc}
\usepackage[applemac]{inputenc}
\usepackage{ae,aecompl}

\usepackage{graphicx}	% Including figure files
\usepackage{amsmath}	% Advanced maths commands
\usepackage{amssymb}	% Extra maths symbols

\title[S0 galaxies are faded spirals]{S0 galaxies are faded spirals: clues from their angular momentum content}

\author[F. Rizzo et al.]{
Francesca Rizzo,$^{1,2}$\thanks{E-mail: frizzo@MPA-Garching.MPG.DE}
Filippo Fraternali,$^{3,1}$
Giuliano Iorio$^{1}$
\\
$^{1}$Dipartimento di Fisica e Astronomia, Università di Bologna, Viale Berti Pichat 6/2, I-40127, Bologna, Italy\\
$^{2}$Max Planck Institute for Astrophysics, Karl-Schwarzschild-Strasse 1, 85740 Garching, Germany\\
$^{3}$Kapteyn Astronomical Institute, University of Groningen, Landleven 12, 9747 AD Groningen, The Netherlands
}

\date{7th February 2018}

\pubyear{2017}

\begin{document}
\label{firstpage}
\pagerange{\pageref{firstpage}--\pageref{lastpage}}
\maketitle

\begin{abstract}
The distribution of galaxies in the stellar specific angular momentum versus stellar mass plane ($j_{\star}$-$M_{\star}$) provides key insights into their formation mechanisms. In this paper, we determine the location in this plane of a sample of ten field/group unbarred lenticular (S0) galaxies from the CALIFA survey. We performed a bulge-disc decomposition both photometrically and kinematically to study the stellar specific angular momentum of the disc components alone and understand the evolutionary links between S0s and other Hubble types. We found that eight of our S0 discs have a distribution in the $j_{\star}$-$M_{\star}$ plane that is fully compatible with that of spiral discs, while only two have values of $j_{\star}$ lower than the spirals. These two outliers show signs of recent merging. Our results suggest that merger and interaction processes are not the dominant mechanisms in S0 formation in low-density environments. Instead, S0s appear to be the result of secular processes and the fading of spiral galaxies after the shutdown of star formation.
\end{abstract}

\defcitealias{rf12}{RF12}

\section{Introduction}
One of the greatest challenges for modern extragalactic astrophysics is understanding the formation and evolution of the different types of galaxies. A number of physical processes are involved in galaxy shaping and, despite decades of research, a number of questions remain unanswered concerning the mechanisms involved in the evolutionary history of the different morphological types and their relative importance in different environments. Observations indicate that the morphology of galaxies is strongly correlated with their angular momentum content \citep[][\citetalias{rf12} hereafter]{fall83, fall13, cortese16, rf12}. An early analysis of galaxy angular momenta \citep{fall83}, confirmed by more sophisticated later studies \citep[][\citetalias{rf12}]{fall13, cortese16}  indicated that all morphological types lie along different but nearly parallel sequences in the plane of specific angular momentum versus stellar mass, $j_{\star}$-$M_{\star}$, with an offset of a factor $\sim$\,5 \citep{fall13} between discs and ellipticals. These results were confirmed by \citet{cortese16} who analysed the distribution of galaxies in the $j_{\star}$-$M_{\star}$ plane for the largest sample studied until now using IFU observations from the SAMI survey \citep{sami}. They showed that the scatter in the $j_{\star}$-$M_{\star}$ distribution is related to the stellar light distribution and the morphology of galaxies. To explain the $j_{\star}$-$M_{\star}$ relations of the different morphological types, galaxy formation and evolution models have to specify how various physical processes set the values of the specific angular momentum and mass of a galaxy.\\
What characterizes the $j_{\star}$-$M_{\star}$ diagram is that the initial conditions in it are well known \citep[][\citetalias{rf12}]{fall83}. According to the current model of structure formation, indeed, the dark matter halo and its gas component should have comparable angular momenta because baryons must have experienced the same tidal torques as dark matter before virialization, radiative dissipation and other non linear effects \citep[\citetalias{rf12};][]{genel}. The angular momentum of dark matter halos is well known in the $\mathrm{\Lambda CDM}$ framework \citep{mo10}. The halos are not completely symmetric and exert tidal torques on each other, inducing a net angular momentum, which is often described in terms of a dimensionless spin parameter 
\begin{equation}\label{eq:lj}
	\lambda=\frac{J |E|^{1/2}}{GM^{5/2}}
\end{equation}
where $J$ is the angular momentum, $E$ is the total energy (kinetic plus potential) and $M$ is the total mass \citep{peeble}. Cosmological N-body simulations \citep[e.g.][]{bullock, maccio} showed that the spin parameter distribution of collapsed dark matter halos is well approximated by a lognormal distribution. What characterizes this distribution is that the values of the median $\overline \lambda(\sim 0.035)$ and of $\sigma_{\mathrm{ln\lambda}}(\sim 0.5)$ depend only weakly on cosmological parameters, redshift and halo mass \citep{mo10}. Cosmological hydrodynamical simulations \citep{bryan} suggested that these statistical properties of dark matter haloes are not significantly influenced by baryonic processes associated with galaxy formation.\\ 
The specific angular momentum, $j_{\mathrm{DM}}=J/M$, of dark matter halos can be expressed in terms of $\lambda$ and M using the expression
\begin{equation}
	j_{\mathrm{DM}}\propto\lambda\,M^{2/3},
	\label{jdm}
\end{equation} 
whose derivation is described in Appendix \ref{app:lambda}. For the reasons explained above, the same proportionality as that in equation~(\ref{jdm}) is expected between angular momentum and mass of the baryonic component near the time of virialization. In addition, in the standard theory of disc galaxy formation, it is assumed that baryons retain their specific angular momentum as they collapse to halo centres \citep{fall80, mo98}. However, this assumption could be considered simplistic because there are numerous phenomena that can modify the initial distribution of $j$ and can cause a loss of angular momentum, for example, by transferring it from the baryonic to dark matter component \citepalias{rf12}. Under the assumption that this loss of angular momentum is not accompanied by a concomitant change in mass \citepalias{rf12}, we can parametrize it introducing a retention factor $f_{\mathrm{j}}$, that is the ratio between the stellar and dark matter specific angular momentum. In addition, considering that only a fraction $f_{\star}$ of the cosmological baryon fraction $f_{\mathrm{b}}$ is converted into stars, we can translate equation~(\ref{jdm}), valid for dark matter, to its stellar counterpart:
\begin{equation}
	j_{\star}\propto f_{\mathrm{j}}\lambda \left(\frac{M_{\star}}{f_{\mathrm{b}}\,f_{\star}}\right)^{2/3}.
	\label{jstar}
\end{equation}
Observations showed that $j_{\star}\propto\mathrm{M_{\star}^{2/3}}$ for spiral discs and elliptical galaxies \citep[][\citetalias{rf12}]{fall83}, as predicted by this simple model. Moreover, under reasonable assumptions for $f_{\star}$ and assuming the median value, $\overline{\lambda}$, the zero-points of the obsevational relations tell us that $f_{\mathrm{j}}\sim0.8$ and $f_{\mathrm{j}}\sim0.1$ for spirals and ellipticals respectively \citep{fall13}. The difference in $f_{\mathrm{j}}$ between these two morphological types is generally ascribed to merging processes, considered responsible for the formation of ellipticals from progenitor spirals, that change the $j_{\star}/M_{\star}^{2/3}$ ratio in a roughly mass-independent way \citep{fall13, fall79, genel}. However, recent studies \citep[e.g.][]{lagos, zavala} showed that the distribution of ellipticals in the $j_{\star}$-$M_{\star}$ plane can be explained by an early star-formation quenching that led these galaxies to form the bulk of their stars at the turnaround of their dark matter halos. If this is the case the $j_{\star}$ is strongly related to the evolution of the inner (within $10\%$ of the virial radius) $j_{\mathrm{DM}}$, so that the stars and the dark matter subclumps in which the stars are locked transfer $\sim80\%$ of their angular momentum to the outer halo as a consequence of mergers between dark matter subclumps. On the contrary, the $j_{\star}$ of spirals  is dominated by the contribution of stars formed after turnaround by accreted gas that has high $j_{\star}$ \citep{zavala}. \citet{posti} and \citet{shi} found that the observed distribution of galaxies in the $j_{\star}$-$M_{\star}$ plane can be explained by a biased collapse scenario, in which the angular momentum retention factor $f_{\mathrm{j}}$ is linked to the star formation efficiency, inflowing gas fraction and formation redshift.\\
Both the distribution of different morphological types in the $j_{\star}$-$M_{\star}$ diagram and the interpretation in terms of $f_{\mathrm{j}}$ were confirmed by cosmological hydrodynamical simulations  \citep[e.g.][]{genel, teklu}. However, \citet{genel} showed that the value of $f_{\mathrm{j}}\sim1$ for late types can also be explained taking into account feedback processes (galactic winds caused by stellar or AGN feedback), instead of considering a full retention factor. In particular, in their simulations, galaxies, in the presence of powerful galactic fountains, are subjected to a combination of loss and gain of angular momentum that leads to a final angular momentum consistent with a full retention of the initial one. On the other hand, the offset of ellipticals from disc-dominated galaxies is always well explained by taking into account that mergers tend, on average, to decrease $j_{\star}$ and increase $M_{\star}$. Another mechanism that could change the initial position of a galaxy in the $j_{\star}$-$M_{\star}$ plane is gas accretion. In the so-called inside-out formation scenario \citep{larson} the specific angular momentum of disc galaxies increases with time because the outer parts, with higher specific angular momenta, should form later than the inner ones thanks to accretion of cold gas \citep[e.g.][]{pez}. However, as shown by \citet{pezzulli15} the radial and mass growth of the stellar discs are such that the disc $j_{\star}$-$M_{\star}$ relation does not significantly evolve with time. This building of the disc through cold gas accretion is considered, for example, by \citet{gram15} the main formation mechanism of S0 galaxies, as opposed to mergers, which may be insufficient to produce the number of observed S0 galaxies. To sum up, the $j_{\star}$-$M_{\star}$ diagram can be used as a tool to understand which mechanism is the most important for the origin of a certain morphological type. The position of a galaxy in this parameter space, indeed, is not arbitrary but it varies during its life as the different evolutionary processes cause a change of its initial $j_{\star}$ and $M_{\star}$ values.\\
In this paper we take advantage of the offset between the $j_{\star}$-$M_{\star}$ sequences of different morphological galaxy types to obtain clues on the evolutionary connections between them. In particular we compare the distribution of a sample of S0 discs with respect to spiral discs and ellipticals in the $j_{\star}$-$M_{\star}$ plane. S0 galaxies are, indeed, of particular interest: the presence of both a bulge and a disc, as well as the absence of spiral arms, place them in an intermediate position between ellipticals and spirals in Hubble’s tuning fork diagram \citep{hubble}. This implies that S0s have always been considered  as galaxies with intermediate properties between these two classes. However, within the framework of the current galaxy formation end evolutionary models, there is an open debate \citep[e.g][]{capps0, van76} on whether S0s are formed in galaxy mergers \citep[e.g][]{bekki98, falcon15, que}, in a similar manner as the elliptical galaxies, or they originated from secular evolution of spiral galaxies after the removal of their cold gas reservoir \citep[e.g][]{laur10, vandenbergh09, will}. The distribution of S0 galaxies in the $j_{\star}$-$M_{\star}$ plane can give us clues to distinguish which of the two scenarios is the most plausible to explain their formation, especially in the more controversial low density environments. A merger scenario should result in a distribution of S0s similar to that of ellipticals, while a passive scenario (consumption of cold gas) should give a distribution similar to that of spirals.\\
The estimation of the stellar angular momentum, one of the main purposes of this paper, can be obtained from observations by measuring two quantities: the surface brightness and the rotation velocity. The derivation of these two quantities from the data will be described in Sections~\ref{section1} and \ref{sec:kinematic} respectively. In particular, in Section~\ref{section1}  we present the method used to perform a photometric bulge-disc decomposition of the galaxies of our sample that allows us to obtain the surface brightness of these two components. In Section~\ref{sec:kinematic} we describe the kinematic bulge-disc decomposition, obtained thanks to the development of new software. In Section~\ref{rcurve} we describe the derivation of disc rotation curves. In Section~\ref{sec:jandm} we derive the distribution of our S0 discs in the $j_{\star}$-$M_{\star}$ plane and we discuss its implication (Section~\ref{sec:discussion}). Finally, in Section~\ref{sec:conclusion} we summarize our main results.

\section{Sample and data}
\subsection{Sample selection}
The sample of galaxies studied in this work is drawn from the CALIFA (Calar Alto Legacy Integral Field Area) Survey \citep{sanchez}. To our knowledge, CALIFA data have not yet been used to study the $j_{\star}$-$M_{\star}$ relation, while SAMI data have been analysed by \citet{cortese16} but without performing the separation of the disc component from the whole galaxy. The morphological classification of these galaxies was carried out through visual inspection of \textit{r}-band SDSS images by members of the CALIFA collaboration. Thanks to this classification, all galaxies were assigned a value for each one of the following categories \citep{walcher2014}:
\begin{enumerate}
\item E for elliptical, S for spiral, I for irregular
\item 0-7 for ellipticals; 0, 0a, a, ab, b, bc, c, cd, d, m for spirals or r for irregulars
\item  B for barred, A for unbarred, AB if unsure
\item Presence of merger features or not
\end{enumerate}
We select all the galaxies from the CALIFA Second Public Data Release \citep{garcia15} that fulfill the following criteria: unbarred (A), S0 or S0a galaxies without clear signs of merger features. The selection of S0a galaxies\footnote{The distinction between S0 and S0a galaxies is based on the visual prevalence of the disc.} should not be considered as a bias for the purpose of this work: in S0a galaxies the disc appears visually more prominent with respect to S0s but they are characterized by the absence of spiral arms, which is the main feature of the S0 morphology. It is now widely accepted that S0s form a parallel sequence to spirals galaxies that span a large range of B/T ratios \citep[e.g][]{van76, k12, capps0}, while they differ by their absence of spiral arms.
We select only unbarred galaxies because both the photometric and spectroscopic decomposition are easier if only two components, bulge and disc, are included in the analysis. However, since the barred galaxies studied in \citetalias{rf12} do not show any systematic difference in the $j_{\star}$-$M_{\star}$ plane from the unbarred ones, we do not expect that an analysis of barred S0s results in significant deviations from the results found in this work.\\
In Table \ref{nome}, for each galaxy we report the luminosity distance (columns 3), calculated considering the redshift (column 4) taken from NED, corrected for the most secure \citep{cap11} contribution for the infall into Virgo attractor and assuming the cosmological parameters: $H_{\mathrm{0}}=70$ km\,s$^{-1}$\,Mpc$^{-1}$, $\Omega_{\mathrm{M}}=0.3$, $\Omega_{\mathrm{\Lambda}}=0.7$ and the indication of the galaxy environment (column 6), taken from NED.

\begin{table*}
	\centering
	\caption{S0 galaxies studied in this paper. Column 1: names of the galaxies. Column 2: Morphological types taken from visual classification by the CALIFA group \citep{walcher2014}. Column 3: luminosity distances. Column 4: redshifts taken from NED, corrected for infall into the Virgo attractor. The uncertainties reported are estimated as explained in the main text. Column 5: scale. Column 6: Environment indication taken from NED.}
	\label{nome}
	\begin{tabular}{l l l c l l}
	\hline
	Galaxy & Type &$D_{\mathrm{L}}$(Mpc) & z & scale(kpc/arcsec) & Environment\\
 	\hline
	NGC\,7671 & S0   & 61.0$\pm$1.6 & 0.0141$\pm$0.0004 & 0.287$\pm$0.007 & Pair\\
	NGC\,7683 & S0   & 55.1$\pm$1.4 & 0.0127$\pm$0.0003 & 0.261$\pm$0.006 & Isolated \\
	NGC\,5784 & S0   & 82.3$\pm$3.5 & 0.0189$\pm$0.0008 & 0.38$\pm$0.01 & Group\\
	IC\,1652 & S0a     & 76.7$\pm$3.4 & 0.0177$\pm$0.0008 & 0.36$\pm$0.01 & Group\\
	NGC\,7025 & S0a & 74.7$\pm$0.9 & 0.0172$\pm$0.0001 & 0.350$\pm$0.001 & Isolated\\
	NGC\,6081 & S0a & 78.2$\pm$6.0 & 0.0180$\pm$0.0014 & 0.37$\pm$0.03 & Field\\
	NGC\,0528 & S0   & 71.4$\pm$3.3 & 0.0165$\pm$0.0008 & 0.33$\pm$0.01 & Group\\
	UGC\,08234 & S0  & 122.6$\pm$1.1 & 0.0280$\pm$0.0002 &0.562$\pm$0.005 & Field\\
	UGC\,10905 & S0a &118.4$\pm$3.1& 0.0271$\pm$0.0007 & 0.54$\pm$0.01 & Field\\
	NGC\,0774 & S0    & 67.0$\pm$2.4 & 0.0154$\pm$0.0005 & 0.31$\pm$0.01 & Field\\
	\hline
    \end{tabular}
\end{table*}

\subsection{Stellar kinematics}
We derive the stellar kinematics of our galaxies using the spectral datacubes of the Second Public Data Release from the CALIFA Survey \citep{garcia15} \footnote{We use the Second Public Data Release from the CALIFA Survey (DR2), even though the DR3 \citep{dr3} is now available. The main difference between DR2 and DR3 is that a correction was applied to DR3 data cubes to match their spectrophotometry to that of the SDSS DR7. However, as showed in \citet{dr3} the shape of the spectra is not influenced by this correction and the flux of the integrated spectrum changed by less than a few percent. Since we use normalised spectra this correction does not influence the results of our analysis.}. The data used in this work are the result of observations made using the integral-field spectroscopic instrument PMAS in PPAK mode, mounted on the 3.5-m telescope at the Calar Alto Observatory. The PPAK Integral Field Unit has a FoV of 74$\times$64 arcsec$^2$ sampled by 331 fibers of 2.7 arcsec diameter each, concentrated in a single hexagon bundle. To obtain well resolved stellar kinematics we use the CALIFA spectral setup with the best spectral resolution (V1200). This covers a nominal wavelength range of 3650-4840\,\AA\,with a spectral resolution R$\sim$1650 at $\lambda\sim$4500\AA\,(FWHM$\sim$2.7\AA, i.e. $\sigma\sim\,80$\,km\,s$^{-1}$). The datacubes used in this work have been calibrated by the CALIFA team using version 1.5 of the reduction pipeline \citep{sanchez, garcia15}  and consist of spectra with a sampling of 1$\times$1 arcsec$^2$ per spatial pixel.\\
In our sample the CALIFA FoV includes typically a coverage of the galaxies out to $\sim 2 R_{\mathrm{d}}$. In Section \ref{sec:jind} we describe the impact of this extension on the estimate of $j_{\star}$. Prior to proceeding with the extraction of the stellar kinematics we spatially binned the datacube using the Voronoi 2D binning algorithm of \citet{cap-bin} to obtain a signal-to-noise ratio of $\sim20$ \citep{sanchez} to ensure reliable stellar kinematics. 

\section{Photometric bulge-disc decomposition}\label{section1}
\subsection{Two-dimensional decomposition}
We assume that S0 galaxies are primarily two-component systems formed by a stellar disc and a bulge. Under this assumption the surface-brightness distribution of each galaxy of the sample can be modelled using a Sérsic function for the bulge and an exponential function for the disc. Beside the parameters that define these two components, this decomposition method provides an estimate for the bulge-to-disc luminosity ratio, B/D, in each spatial pixel. As we show in Section~\ref{bsutsec}, we use this ratio as a constraint to perform a two component kinematic fit to the spectra and derive the rotation velocities and velocity dispersions of the disc and bulge components. \\
The galaxy decomposition in bulge and disc components is performed using the 2D fitting routine GALFIT (\textit{version 3.0.5}, \citealt{peng2010}). The parameters that specify the axisymmetric profiles selected to model the two components are:
\begin{enumerate}
\item \textbf{Bulge:} the effective radius ($R_{\mathrm{e}}$), the surface brightness  at $R_{\mathrm{e}}$ ($I_{\mathrm{e}}$) and the index $n$ of the Sérsic profile
\begin{equation}
 	I_{\mathrm{b}}(R)=I_{\mathrm{e}}\exp\left\{-\kappa(n) \left[\left(\frac{R}{R_{\mathrm{e}}}\right)^{1/n}-1\right]\right\};
	\label{eq:bulge1}
\end{equation}
The parameter $\kappa$ in equation (\ref{eq:bulge1}) is coupled to $n$ so that half of the total luminosity is within $R_{\mathrm{e}}$ \citep{binandmer}.
\item \textbf{Disc:} the scale length ($R_{\mathrm{d}}$) and the central surface brightness ($I_{\mathrm{0}}$) of the exponential profile
  \begin{equation}
 	I_{\mathrm{d}}(R)=I_{0}\exp(-R/R_{\mathrm{d}}).
 	\label{eq:disco1}
 \end{equation}
 \end{enumerate} 
The model profiles defined in equations~(\ref{eq:bulge1}) and  (\ref{eq:disco1}) are 1D and they are used to generate 2D images. To take into account the sky projection of a model component, GALFIT uses generalised elliptical isophotes, where the radial coordinate is described by:
\begin{equation}
	R(x,y)=\left[(x-\xi_{0})^{2}+\left(\frac{y-\eta_{0}}{q}\right)^{2}\right]^{\frac{1}{2}}
\end{equation}
The origin ($\mathrm{\xi_{0}, \eta_{0}}$) is at the galaxy centre, taken to be the same for the two components. The $x$-axis is along the apparent major-axis, whose orientation is defined by the position angle, $PA$, while the parameter $q$ is the ratio of the minor to major axis of an ellipse. These two parameters can be different for the two components. \\
To estimate the geometrical and photometric parameters of the exponential and Sérsic profiles, GALFIT minimises the $\mathrm{\chi^{2}}$ residuals between the galaxy image and the 2D models using the non-linear Levenberg-Marquardt algorithm.  

\subsection{Pre-processing of galaxy images} \label{sec:maskphoto}
Galaxy decomposition is carried out on SDSS Data Release 7\footnote{http://www.sdss.org/dr7} \textit{r}-band images. In particular the data for each galaxy of the sample are extracted from fpC files, i.e. SDSS corrected images that have been bias-subtracted, flat-fielded, and purged of bright stars \citep{stough}.  Prior to fitting we subtracted from each image the ‘softbias’ level of 1000 counts added to all pixels by the SDSS pipeline. Although GALFIT allows the sky to be included as a free fitting component, we estimate the sky background level by averaging values obtained in regions not influenced by the galaxy flux or other sources. This sky level is held fixed during the fitting process. This procedure allows us to minimise the number of parameters to fit. We verified for a couple of galaxies (NGC\,7671, NGC\,7025) that using DR13, characterized by an improved sky subtraction, would not affect our results as the fitted parameters are compatible within 1$\sigma$ between the two releases. Postage-stamp images centered on the target source are used for the fitting. An SDSS pipeline\footnote{https://code.google.com/archive/p/sdsspy} is used to reconstruct the PSF at the position of the galaxy\footnote{The PSF reconstruction is based on an algorithm (Lupton et al.\,2001) which performs a PSF estimation using a combination of point sources from the SDSS image and expanding them in terms of Karhunen-Love functions.}, with typical values of the FWHM $\sim$1.3 arcsec. The PSF modeling is not critical for the estimate of the bulge effective radii, because they are all larger than 1.3 arcsec. Indeed, GALFIT convolves the model image with this input PSF to take into account the effect of the seeing, before comparing it with the data. To convert from galaxy model counts to surface brightness the zero-point is calculated from the SDSS zero-point, extinction and airmass terms associated with each image. Additionally, for most sources of the sample, prior to proceeding with the fitting process we mask the bright stars and/or dust lanes that contaminate the image of the galaxy. This process is done using PYBLOT, a task of the GIPSY software package \citep{van}, through which one can select by hand the regions of an image to be blanked (see the lower left panels of Fig.~\ref{n08234} and \ref{n7683}-\ref{n0774}).

\subsection{Fitting process}\label{sec:fitting process}
\begin{figure*}
	\includegraphics[width=16cm]{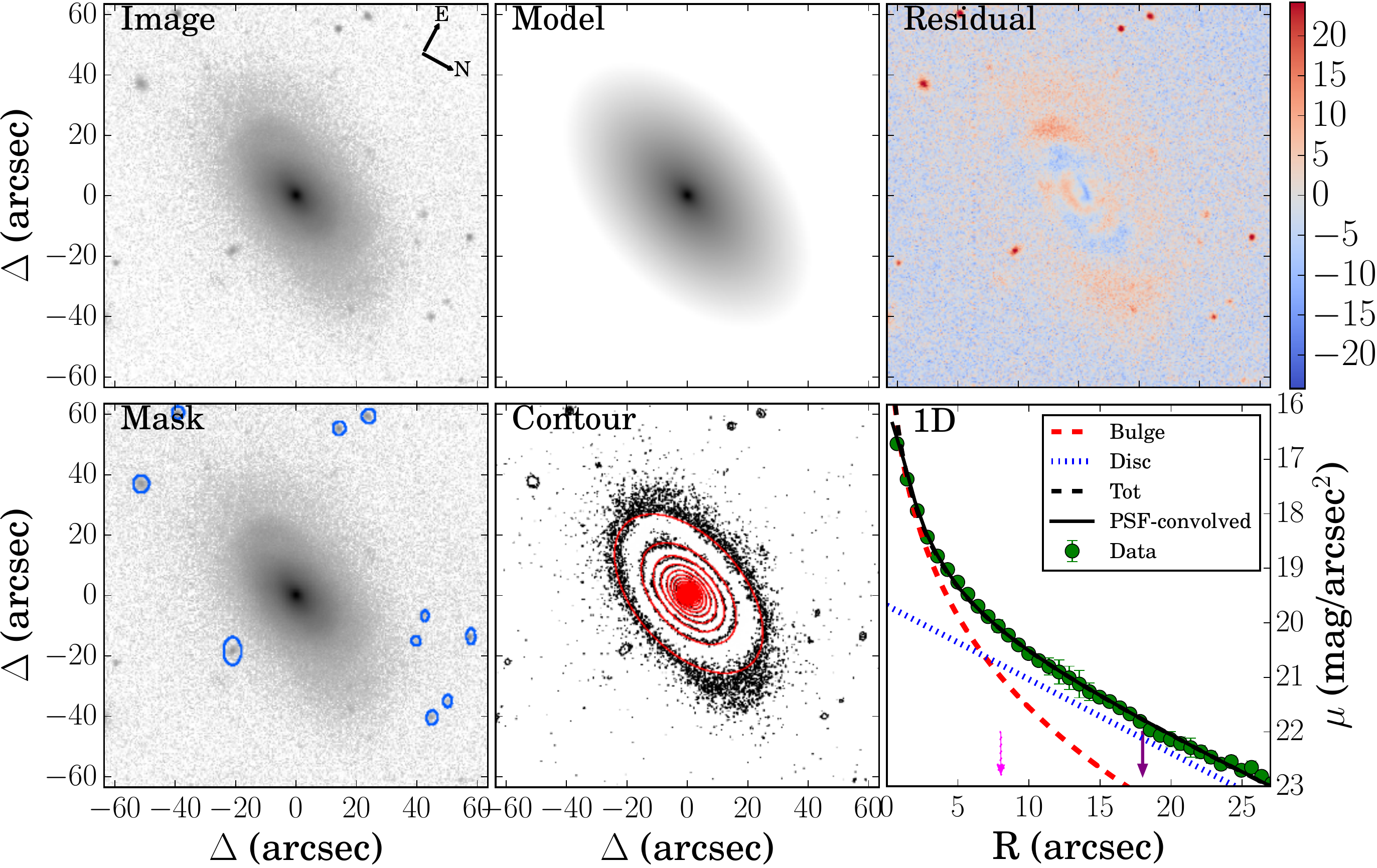}
    \caption{\textbf{Upper panels}: (\textit{Left}) SDSS \textit{r}-band data image of UGC\,08234. The north direction is at -99.6$^\circ$ as indicated by the black arrows. (\textit{Middle}) Two-component GALFIT model. (\textit{Right}) Relative residuals, obtained considering the quantity (data$-$model)/data*100 for each pixel. \textbf{Lower panels}: (\textit{Left}) SDSS image of the galaxy with the mask overlaid-contours. The regions inside the blue contours are those excluded during the fitting process (see Section~\ref{sec:maskphoto}). (\textit{Middle}) Map contour levels: the lowest level is at 3$\mathrm{\sigma_{sky}}$ and the next levels are incremented by a factor of 1.2 each. (\textit{Right}) 1-D surface brightness profile. Green circles: data extracted in a wedge of 10$^\circ$ along the major axis of the disc. Dashed and dotted lines are seeing-free model profiles for individual components and their respective sum, as indicated in the legend. The black solid line represents the PSF-convolved total profile which fits the observed data. Magenta dotted arrow: disc scale length; purple arrow: maximum radius for which we have kinematic data from CALIFA for this galaxy.}
 	\label{n08234}
\end{figure*}

To find the best parameters that describe the bulge and disc components, we excute successive runs. First we leave free all the parameters during the fitting process, with the exception of the sky background (see Section~\ref{sec:maskphoto}). The $\xi_{0}$ and $\eta_{0}$ positions that specify the centre are then kept fixed at the values found in this first run. In the successive runs the simultaneous fits therefore uses a total number of nine free parameters: five for bulge (surface brightness at the effective radius, effective radius, Sérsic index, axis ratio and position angle for its major axis) and four for disc (central surface brightness, disc scale-length, axis ratio and position angle for its semi-major axis) (Table~\ref{tab:photo}). For some galaxies, however, in the final runs the position angles of bulge and disc are forced to be equal; these cases are discussed individually in Appendix \ref{app:a}. For one galaxy (NGC\,0528) the Sérsic index is kept fixed at a standard value of 4, because letting it be free does not produce physically meaningful fits (see discussion in Appendix~\ref{sec:n0528}). In a few cases the photometric decomposition turned out to be more complicated because of the presence of other central substructures like an embedded bar. However, we decided to consider only the two main components that represent the disc and the bulge contributions, disregarding the presence of residuals that may indicate the presence of other central substructures. These extra components are never dominant, as is clear from inspection of the relative residuals (upper right panels of Fig.~\ref{n08234} and Fig.~\ref{n7683}-\ref{n10905}), which are always of the order of 5\%, except for the galaxies with prominent dust lanes. The only exception to this two-component decomposition is the galaxy NGC\,0774 (see discussion in Appendix~\ref{ph:n0774} and Fig.~\ref{n0774}). This is the only galaxy of our sample for which the addition of a second disc was necessary. The presence of two discs, with the second one that represents an excess of light above the inner exponential profile, is relatively common among disc galaxies, especially early types \citep[e.g.][]{erwin08}. The profiles that show this upturn from inner exponential, the so-called ‘type III’ profiles \citep{erwin05}, are considered the result of tidal interactions or of minor \citep{laurik2001, pen} and major mergers \citep{borlaff}.\\
The parameters derived through these bulge-disc decomposition are used to calculate the global bulge-to-total luminosity ratio, B/T  (Table~\ref{tab:bsut}). 
\begin{table*}
	\caption{Parameters of the two-component best-fit model for the S0 galaxies in our sample. The $PA$ is measured from North to East. The measurement uncertainties are calculated as explained in Section~\ref{incertezze}. The parameters that are kept fixed during the fitting process are those without errors. For NGC\,0774 it was necessary to add a second disc component, indicated as D2; the first is indicated with D1, while the one with suffix D shows the parameters for the luminosity weighted disc (see the Appendix \ref{ph:n0774} for further details).}
 \label{tab:photo}
	\begin{tabular}{l l l l l l l l l l}
	\hline
Galaxy &$\mu_{\mathrm{e,b}}$ & $R_{\mathrm{e,b}}$ & $n$ & $q_{\mathrm{b}}$ & $PA_{\mathrm{b}}$
 & $\mu_{\mathrm{0,d}}$ & $R_{\mathrm{d}}$ & $q_{\mathrm{d}}$ & $PA_{\mathrm{d}}$ \\
&$\mathrm{mag/arcsec^{2}}$ &  arcsec &      &     & degrees & $\mathrm{mag/arcsec^{2}}$ &  arcsec &           & degrees\\
\hline
NGC\,7671 &17.94$\pm$0.11 & 1.92$\pm$0.13 & 2.34$\pm$0.14 & 0.91 $\pm$0.01 & 134.78&
19.35$\pm$0.10& 10.59$\pm$0.65  &0.57$\pm$0.01& 134.78  \\
 \\
NGC\,7683 & 19.29$\pm$0.02 & 5.06$\pm$0.12 & 2.26$\pm$0.08 & 0.70$\pm$0.01& 137.25 
& 20.24$\pm$0.08 & 17.79$\pm$0.92  &0.49$\pm$0.04 & 137.25  \\
\\
NGC\,5784 & 18.71$\pm$0.03 & 3.16$\pm$0.08 & 2.08$\pm$0.02 & 0.74$\pm$0.01 & 81.18$\pm$0.26
& 19.33$\pm$0.04 & 8.53$\pm$0.22 &  0.85$\pm$0.01 & 57.55\\
\\
IC\,1652 & 20.26$\pm$0.27 & 3.40$\pm$0.56 & 4.59$\pm$0.42 & 0.75$\pm$0.02 & -9.77 
& 18.99$\pm$0.02 & 10.02$\pm$0.11  &0.19$\pm$0.01 & -9.77  \\
\\
NGC\,7025 & 19.36$\pm$0.23 & 5.18$\pm$0.79 & 2.35$\pm$0.19 & 0.84 $\pm$0.01& 44.78 
 & 20.35$\pm$0.15 & 20.88$\pm$1.25  &0.62$\pm$0.01 & 44.78  \\
\\
NGC\,6081 & 18.91$\pm$0.01 & 2.98 $\pm$0.04 & 1.31$\pm$0.01 & 0.76$\pm$0.01& 141.49$\pm$0.83 
& 19.71$\pm$0.06 & 13.78$\pm$0.52  &0.35$\pm$0.01 & 128.45$\pm$0.01 \\
\\
NGC\,0528 & 19.93$\pm$0.07 & 4.73$\pm$0.33 & 4.00  & 0.76$\pm$0.01& 68.31$\pm$1.2
 & 19.48$\pm$0.08 & 10.30$\pm$0.29  &0.40$\pm$0.01 & 56.83$\pm$0.08  \\
\\
UGC\,08234 & 19.18$\pm$0.18 & 3.63$\pm$0.43 & 4.63$\pm$0.53 & 0.65$\pm$0.01& 140.57 
& 19.68$\pm$0.13 & 8.02$\pm$0.17  &0.56$\pm$0.01 & 140.57   \\
\\
UGC\,10905 & 19.73$\pm$0.22 & 4.46$\pm$0.62 & 3.98$\pm$0.23 & 0.64$\pm$0.01& 174.75 
 & 20.45$\pm$0.18 & 12.53$\pm$0.92  &0.48$\pm$0.01 & 174.75 \\
\\
NGC\,0774 D & 20.54$\pm$0.30 & 2.67$\pm$0.67 & 4.65$\pm$0.39 & 0.90$\pm$0.01& 164.56 
 & 18.76$\pm$0.09 & 6.33$\pm$0.79  & 0.67$\pm$0.01 & 164.56\\
NGC\,0774 D1 & 20.54$\pm$0.30 & 2.67$\pm$0.67 & 4.65$\pm$0.39 & 0.90$\pm$0.01& 164.56 
 & 18.56$\pm$0.04 & 4.42$\pm$0.15  &0.65$\pm$0.01 & 164.56 \\
NGC\,0774 D2 & 20.54$\pm$0.30 & 2.67$\pm$0.67 & 4.65$\pm$0.39 & 0.90$\pm$0.01& 164.56 
 & 21.09$\pm$0.20 & 15.15$\pm$1.32 &0.76$\pm$0.01 & 164.56 \\
\hline
 	\end{tabular}
\end{table*}

\begin{table}
	\centering
	\caption{Column 2: bulge-to-total $r$-band luminosity ratio. The uncertainties are the result of the standard formula for error propagation. Column 3: bulge-to-disc ratio conversion factor ($L_{\mathrm{d}}/L_{\mathrm{b}}$ in equation \ref{k}) between the $r$-band and the spectral range $\left[3899, 4476 \right] \AA$.}
	\label{tab:bsut}
	\begin{tabular}{l l l}
	\hline
	Galaxy & B/T & $L_{\mathrm{d}}/L_{\mathrm{b}}$\\
	\hline
	NGC\,7671 & 0.35$\pm$0.05 & 0.77\\
	NGC\,7683 & 0.43$\pm$0.03 & 0.88\\
	NGC\,5784 & 0.36$\pm$0.02 & 0.79\\
	IC\,1652     & 0.35$\pm$0.09 & 0.87\\
	NGC\,7025 & 0.37$\pm$0.09 &  0.88\\
	NGC\,6081 & 0.34$\pm$0.02 &  0.84\\
	NGC\,0528 & 0.52$\pm$0.06 &  0.91\\
	UGC\,08234 & 0.64$\pm$0.16 & 0.92\\
	UGC\,10905 & 0.54$\pm$0.17 & 0.87\\
	NGC\,0774 & 0.09$\pm$0.04 &  0.81\\
	\hline
	\end{tabular}
\end{table}

\subsection{Estimating the uncertainties}
\label{incertezze}
GALFIT computes the parameter uncertainties from the covariance matrix produced during the least-square minimisation. These uncertainties would be correct if the dominant contribution were only Poisson noise. However, in reality these uncertainties are strong underestimates of the true errors as these are often dominated by systematics, such as the assumptions of the number and the types of components used in the fitting, non-uniformness of the sky, errors in the determination of the point spread function and of the shape of the mask. The choice of the model profiles is one of the dominant factor in the systematic uncertainties. Some galaxies of our sample, for example, when modelled with two components, return residuals with a symmetric pattern which may indicate the presence of extra components. In these galaxies, the residuals are not due to Poisson noise only, but mostly to structures which cannot be subtracted away perfectly. Another factor that could have a dominant contribution to the parameter uncertainties is the sky estimation \citep{gram}. Different authors \citep{vika12,haubler} found that a variation of the mean sky value within 1$\sigma$ leads to uncertainties of 12\% and 14\% for the bulge effective radius and Sérsic index respectively. On the other hand, the disc parameters seem more robust and less affected by the various systematic uncertainties \citep{vika14}. All the above is confirmed by our decomposition: the addition of further components or the variation of the mean sky has a bigger effect on the bulge parameters rather than on those of the disc. In conclusion as the mean sky value is one of the dominant contributors to the errors and it is known a priori (Section \ref{sec:maskphoto}), we estimate the fitting parameter errors considering the sky uncertainty. This is done considering the variation of the parameters caused by changing the mean sky level by $\pm2\sigma$. The choice of $2\sigma$ should be considered a conservative choice to take into account other uncertainties, which we cannot quantify, such as for example the shape of the masks.

\section{Kinematic bulge-disc decomposition: double-Gaussian fit}\label{sec:kinematic}
Since one of the main aims of this work is to derive the kinematics of bulge and disc components of our galaxy sample separately, we devise a method that consists of the following steps:
\begin{enumerate}
\item We assume that the observed line-of-sight velocity distribution (LOSVD) is produced by the contribution of bulge and disc components that are, in principle, characterized by different kinematics. 
\item In order to describe these different kinematics, we model the LOSVD at each point of the galaxy with two Gaussians that have different amplitudes, means ($V_{\mathrm{b}}$, $V_{\mathrm{d}}$)  and dispersions ($\sigma_{\mathrm{b}}$, $\sigma_{\mathrm{d}}$) .
\item Since the spectrum at a certain location across the galaxy can be considered as a luminosity-weighted sum of bulge and disc spectra, we fix the amplitude of the two Gaussians using the information on the bulge-to-disc luminosity ratio (Section \ref{bsutsec}) given by our photometric decomposition (Section~\ref{section1}).
\item We fit the kinematic parameters ($V_{\mathrm{b}}$, $V_{\mathrm{d}}$, $\sigma_{\mathrm{b}}$, $\sigma_{\mathrm{d}}$) using a Markov Chain Monte Carlo (MCMC) routine.
\end{enumerate}
Under the above mentioned assumptions, we model the galaxy spectrum at a certain location (Voronoi bin), identified by the coordinates $\mathrm{(\xi,\eta)}$ on the sky, see Section \ref{sec:coordinates}, using a model spectrum $S_{\mathrm{mod}}$ such that
\begin{equation}
 	S_{\mathrm{mod}}(u;\xi,\eta)=S_{\mathrm{mod,b}}(u;\xi,\eta)+S_{\mathrm{mod,d}}(u;\xi,\eta)
  	\label{eq:2gaussiane}
\end{equation}
with 
\begin{equation}
	S_{\mathrm{mod,b}}(u;\xi,\eta)=[i_{\mathrm{b}}(\xi,\eta)F_{\mathrm{b}} \ast T_{\mathrm{b}}](u)
  	\label{eq:gaub}
 \end{equation}
 and
\begin{equation}
 	S_{\mathrm{mod,d}}(u;\xi,\eta)=[i_{\mathrm{d}}(\xi,\eta)F_{\mathrm{d}} \ast T_{\mathrm{d}}](u)
 	\label{eq:gaud}
\end{equation}

where: 
 \begin{enumerate} 
 \item$*$ denotes the convolution;
 \item $u=c\,\ln\lambda$ is the spectral velocity \citep{binandmer} (c is the velocity of light and $\lambda$ is the wavelength);
 \item $i_{\mathrm{b}}(\xi,\eta)$ and $i_{\mathrm{d}}(\xi,\eta)$ are the dimensionless fluxes of bulge and disc respectively, so $i_{\mathrm{b}}(\xi,\eta)/i_{\mathrm{d}}(\xi,\eta)$ represents the bulge-to-disc flux ratio at the position ($\xi$, $\eta$);
 \item $T_{\mathrm{b}}$ and $T_{\mathrm{d}}$ are the spectrum templates for bulge and disc respectively;
 \item $F_{\mathrm{b}}$ and $F_{\mathrm{d}}$ are their LOSVDs, described by mean velocities, $V_{\mathrm{b}}$ and $V_{\mathrm{d}}$ and dispersions, $\sigma_{\mathrm{b}}$ and $\sigma_{\mathrm{d}}$,
\begin{equation}
	F_{\mathrm{b}}(v_{\mathrm{los}})=\frac{1}{\sqrt{2\pi}\,\sigma_{\mathrm{b}}}\exp \left[\frac{-(v_{\mathrm{los}}-V_{\mathrm{b}})^{2}}{2 \sigma^{2}_{b}}\right]
 	\label{eq:bulge}
\end{equation}
\begin{equation}
	F_{\mathrm{d}}(v_{\mathrm{los}})=\frac{1}{\sqrt{2\pi}\,\sigma_{\mathrm{d}}}\exp\left[\frac{-(v_{\mathrm{los}}-V_{\mathrm{d}})^{2}}{2 \sigma^{2}_{d}}\right].
 	\label{eq:disco}
 \end{equation}
 \end{enumerate}
In the following subsections we describe how we determine the quantity $i_{\mathrm{b}}(\xi,\eta)/i_{\mathrm{d}}(\xi,\eta)$, equations~(\ref{eq:gaub}) and (\ref{eq:gaud}), and choose the templates. All these quantities are used in the MCMC routine that we developed to determine the best parameters of  $F_{\mathrm{b}}$ ($V_{\mathrm{b}}$ and $\sigma_{\mathrm{b}}$) and  $F_{\mathrm{d}}$ ($V_{\mathrm{d}}$ and $\sigma_{\mathrm{d}}$) that allow us to reproduce the observed galaxy spectrum in each spatial Voronoi bin, $\mathrm{S_{obs}(u;\xi,\eta)}$. 

\subsection{Description of the coordinates}\label{sec:coordinates}
We refer to the coordinates on the plane of the sky as ($\xi$,$\eta$). The $\xi$-axis is parallel to the direction of the right ascension and points westward and the $\eta$-axis is parallel to the direction of the declination and points northward, with the origin fixed at the galaxy centre found in Section~\ref{section1}. The quantities $\xi$ and $\eta$ in equation~(\ref{eq:2gaussiane}) are the coordinates of the bin centroids, as defined by the Voronoi 2D binning technique. In the next section we will estimate the bulge-to-disc luminosity ratio at each point ($\xi$,$\eta$), using the information on the structural parameters of these two components, given by our previous photometric decomposition (see Section~\ref{section1}). However, the radius $R$ in the expressions of Sérsic and exponential profiles (equations~(\ref{eq:bulge1}) and (\ref{eq:disco1})) is not the same quantity at fixed ($\xi$,$\eta$) because bulge and disc isophotes have different ellipticities and in some cases different position angles. The radius $R$ that enters in the equations~(\ref{eq:bulge1}) and (\ref{eq:disco1}) for calculation of bulge and disc surface brightness is given by the following expressions respectively \citep{mendez08}:
\begin{multline}
	R_{\mathrm{Bulge}}(\xi,\eta)=[(-\xi\sin{PA_{\mathrm{b}}}+\eta\cos{PA_{\mathrm{b}}})^{2}+\\
	-(\xi\cos{PA_{\mathrm{b}}}+\eta\sin{PA_{\mathrm{b}}})^{2}/q_{\mathrm{b}}^{2}]^{1/2}
	\label{rbulge}
\end{multline}
and
\begin{multline}	
	R_{\mathrm{Disc}}(\xi,\eta)=[(-\xi\sin{PA_{\mathrm{d}}}+\eta\cos{PA_{\mathrm{d}}})^{2}+\\
	-(\xi\cos{PA_{\mathrm{d}}}+\eta\sin{PA_{\mathrm{d}}})^{2}/q_{\mathrm{d}}^{2}]^{1/2}
	\label{rdisc}
\end{multline}
where $PA_{\mathrm{b}}$, $PA_{\mathrm{d}}$, $q_{\mathrm{d}}$, $q_{\mathrm{b}}$ are those obtained in Section~\ref{section1}.

\subsection{Template spectra}\label{template}
As templates, we use the stellar population models of \citet{vazdekis}, based on the Jones stellar library\footnote{These stellar population models were retrieved directly from http://www.iac.es/proyecto/miles} which covers the spectral range 3855-4476\,\AA\,with a FWHM$\sim$1.8\,\AA. We choose models with a Kroupa IMF \citep{kroupa}, with 53 ages spanning t=0.03-14 Gyr and 12 metallicities [M/H] from -2.27 to 0.40.\\
We do not employe all the template spectra during the fitting process, but we determine for each galaxy of our sample the best combination of them that is able to reproduce the observed spectrum. In order to find this combination we use the penalised pixel-fitting routine pPXF \citep{cap-ppxf}. This program minimizes the $\chi^{2}$ between an observed galaxy spectrum and a model galaxy spectrum, $S_{\mathrm{mod}}$, described by the following expression:
\begin{equation}
	S_{\mathrm{mod}}(u)=\sum_{\mathrm{i}=1}^{N}a_{\mathrm{i}}[F \ast T_{\mathrm{i}}](u),
	 \label{eq:gmod}
\end{equation}
where $u$ spans the spectral velocities corresponding to the wavelength range: [3899, 4476] \AA, in which there are relevant absorption lines (see Section~\ref{sec:fit} for further details). The quantity $S_{\mathrm{mod}}$ in equation~(\ref{eq:gmod}) is defined not only by the parameters of the LOSVD, F,  but also by coefficients $a_{\mathrm{i}}$ assigned to each template spectrum $T_{\mathrm{i}}$. \\
We used the best-fit coefficients $a_{\mathrm{i}}$, found by pPXF, to construct optimal template spectra for the bulge and the disc of each galaxy of our sample. To find the template spectra for the bulge and disc components we apply the above method to two different galaxy spectra, representative of the bulge and the disc. To determine the representative bulge spectrum, we use the spectrum obtained as the average of spectra in the region within (at most) the bulge effective radius (from our photometry, see Section~\ref{section1}). This assures a negligible contamination by the disc component. To have a representative disc spectrum we average spectra in the most external regions (the width of them is typically $\sim$5-7 arcsec) along the disc major axis where the contribution of the bulge is negligible. The application of pPXF to these representative bulge and disc spectra allows us to determine the coefficients $a_{\mathrm{i}}$ of equation~(\ref{eq:gmod}). \\
In the end, for each of the ten galaxies we have a template spectrum for the bulge and another for the disc, which can be expressed as
\begin{equation}
	T_{\mathrm{b}}(u)=\sum_{i=1}^{M}a_{\mathrm{i,b}}T_{\mathrm{b,i}}(u)
 	\label{eq:templateb}
\end{equation}
\begin{equation}
	T_{\mathrm{d}}(u)=\sum_{i=1}^{N}a_{\mathrm{i,d}}T_{\mathrm{d,i}}(u)
	\label{eq:templated}
\end{equation}
where $a_\mathrm{{i, b/d}}$ are the coefficients found by pPXF, equation~(\ref{eq:gmod}). In practice, we apply a constraint on the selection of optimal bulge template $T_\mathrm{{b,i}}$, such that only those with age t$\geqslant$10 Gyr are selected during the pPXF fitting process. This assumption reflects the typical ages of S0 bulges found by different studies \citep{kim, falcon02}. Furthermore it is in line with the negative age gradients usually detected in galaxies and the expectation of the common inside-out scenario for galaxy formation \citep{gonzalez15}.  However, there is not a general agreement about the ages of the bulges in S0 galaxies, e.g. \citet{john14} studied a sample of 21 S0s from the Virgo Cluster and they found that the bulges contain younger and more metal rich stellar populations than the discs. We note that while the choice of optimal templates is important for a reliable estimate of the velocity dispersion, it has no significant effect on the line of sight velocity field, which is the critical parameter for estimating the specific angular momentum \citep{cortese16}. We verify that after releasing the constraint on the ages of bulge templates, those chosen by pPXF in the full range available (t=0.03-14 Gyr) differed negligibly from those found under our assumption. A detailed description of the physical properties of the templates and of the values of the coefficients $a_{\mathrm{i}}$ found for each galaxy of our sample is given in Appendix \ref{app:b}. 

\subsection{Bulge-to-disc ratio}\label{bsutsec}
The amplitudes $i_{\mathrm{b}}(\xi,\eta)$ and $i_{\mathrm{d}}(\xi,\eta)$ of the two gaussians in equations~(\ref{eq:gaub}) and (\ref{eq:gaud}) represent the relative luminosity-weighted contributions of the bulge and disc components. We determine this contribution using our photometric bulge-disc decomposition (Section~\ref{section1}). This decomposition was obtained in the SDSS \textit{r}-band, while the fitted galaxy spectrum covers the wavelength range 3899-4476\,\AA. Thus, we have to convert our bulge-to-disc ratio, $I_{\mathrm{b}}/I_{\mathrm{d}}$, from one band to the other. \\
To perform this conversion we use the stellar population templates that reproduce the bulge and disc spectra for each galaxy of our sample (see Section~\ref{template}) in the 3899-4476\,\AA\,range and we estimate the contribution of these templates in the \textit{r}-band. We assume that the relative amplitude of bulge and disc gaussians, $i_{\mathrm{b}}(\xi,\eta)/i_{\mathrm{d}}(\xi,\eta)$ (equations~(\ref{eq:gaub})-(\ref{eq:gaud})) has the same dependence on the coordinates in the interval 3899-4476\,\AA\,as in \textit{r}-band. This is equivalent to assuming the absence of strong colour gradients in the bulge and disc component within $2R_{\mathrm{d}}$. The \textit{r}-band best-fit surface brightness for the two components found in Section~\ref{section1} ($I_{\mathrm{b}}$ and $I_{\mathrm{d}}$) is thus decomposed into two factors: the dimensionless surface brightness that depends on the coordinates, $i_{\mathrm{b/d}}(\xi,\eta)$ and another that is a constant $L_{\mathrm{b/d}}$ , so that
\begin{equation}
	I_{\mathrm{b}}(\xi,\eta)=i_{\mathrm{b}}(\xi,\eta)\, \alpha\,L_{\mathrm{b}}
	\label{ib}
\end{equation}
and
\begin{equation}
	I_{\mathrm{d}}(\xi,\eta)=i_{\mathrm{d}}(\xi,\eta)\,\alpha\,L_{\mathrm{d}}
	\label{id}
\end{equation}
where $\alpha$ in (\ref{ib}) and (\ref{id}) is a conversion factor from luminosity to surface brightness. The two constant terms $L_{\mathrm{b}}$ and $L_{\mathrm{d}}$ are the \textit{r}-band bulge and disc template luminosities:
\begin{equation}
	L_{\mathrm{b}}=\sum_{i=1}^{M}\frac{a_{\mathrm{i,b}}}{\overline{T}_{\mathrm{i,b}}}L_{r\,\mathrm{i,b}}
	\label{gb}
\end{equation}
and
\begin{equation}
	L_{\mathrm{d}}=\sum_{i=1}^{N}\frac{a_{\mathrm{i,d}}}{\overline{T}_{\mathrm{i,d}}}L_{r\,\mathrm{i,d}}.
	\label{gd}
\end{equation}
the terms $a_{\mathrm{i,b/d}}$ are template coefficients of equations~(\ref{eq:templateb}) and (\ref{eq:templated}); the terms $\overline{T}_{\mathrm{i,b/d}}$ are values of flux counts of the templates averaged in the spectral direction, introduced to take into account that in the equations~(\ref{eq:templateb}) and (\ref{eq:templated}) the templates $T_{\mathrm{i,b/d}}$  are normalized; $L_{r\,\mathrm{i,b/d}}$ are the corresponding $r$ luminosities for each of the simple stellar population templates. As mentioned before, the dimensionless surface brightness that depends on the coordinates, $i_{\mathrm{b/d}}(\xi,\eta)$, are the same as in equations~(\ref{eq:gaub}) and (\ref{eq:gaud}), because we assume that the bulge-to-disc ratio has the same dependence on the coordinates at different wavelength.\\
The ratio between $i_{\mathrm{b}}/i_{\mathrm{d}}$, that is the relative contribution of bulge and disc in equations~(\ref{eq:gaub}) and (\ref{eq:gaud}) can thus be expressed as
\begin{equation}
	\frac{i_{\mathrm{b}}(\xi,\eta)}{i_{\mathrm{d}}(\xi,\eta)}=\frac{I_{\mathrm{b}}(\xi,\eta)}{I_{\mathrm{d}}(\xi,\eta)}\frac{L_{\mathrm{d}}}{L_{\mathrm{b}}}.
	\label{k}
\end{equation}
This ratio is known for each galaxy at every point after our photometric decomposition and the choice of the bulge and disc templates, so this fixes the amplitudes of the two gaussians in equations~(\ref{eq:gaub}) and (\ref{eq:gaud}). The average value of the ratio $L_{\mathrm{d}}/L_{\mathrm{b}}$ in equation~\ref{k}, which is the conversion factor between the $r$-band and our studied spectral range, is $\sim0.85$. This value means that the discs are bluer than the bulges and it is in agreement with the S0 conversion factor between bands similar to those investigated in this work \citep{gram08}. The values of the conversion factors for each galaxy of our sample are reported in the third column of Table \ref{tab:bsut}.

\subsection{Fitting procedure}\label{sec:fit}
\begin{figure}
	\includegraphics[width=\columnwidth]{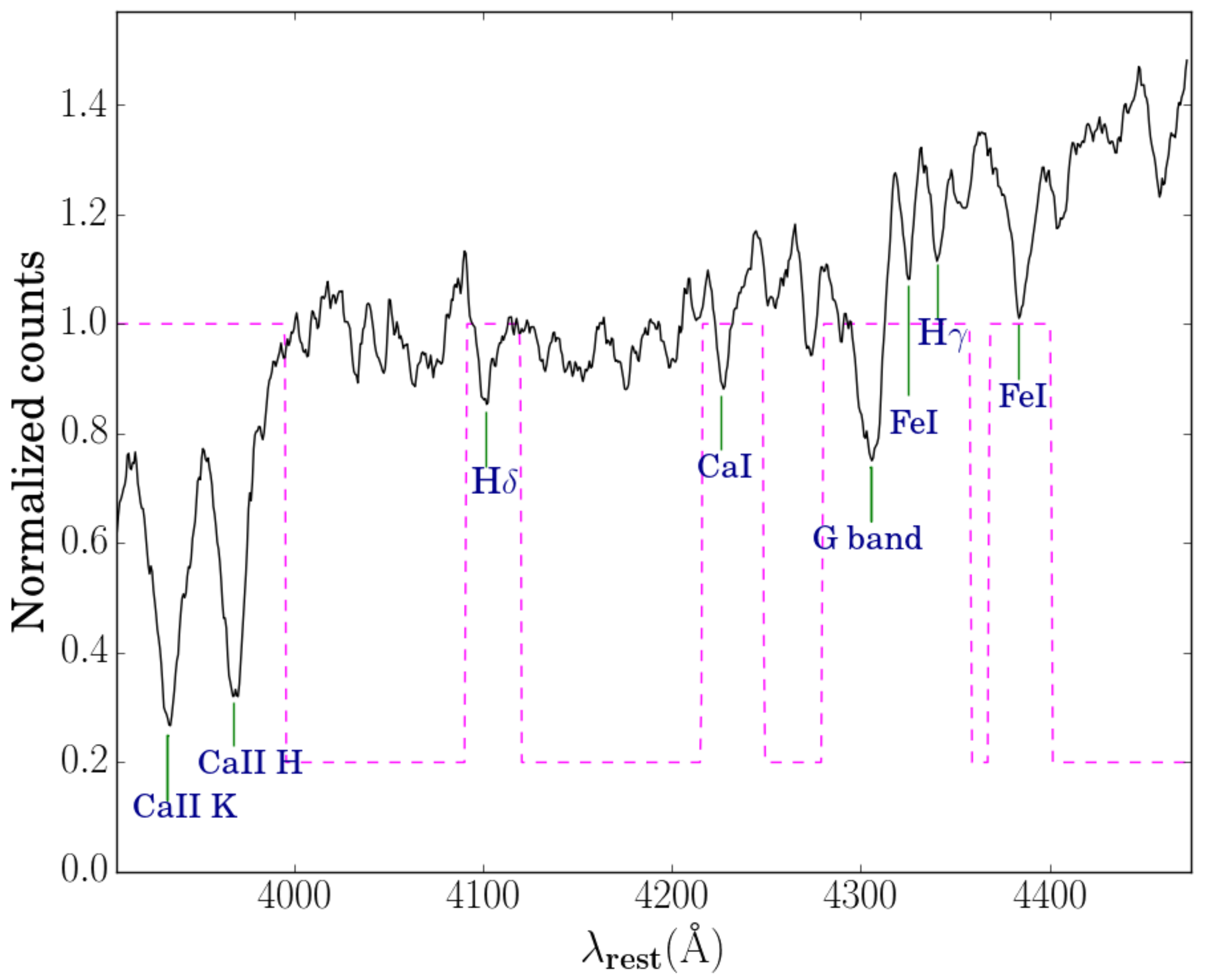}
	\caption{Spectrum extracted from the central parts of the galaxy NGC\,7025. The magenta dashed curve indicates the values of the fitting weight $w_{\mathrm{k}}$: it is equal to 1 in regions of absorption lines (green bars) considered in the fitting and 0.2 otherwise. }
	\label{spettrorighe}
\end{figure}

\begin{figure}
	\includegraphics[width=\columnwidth]{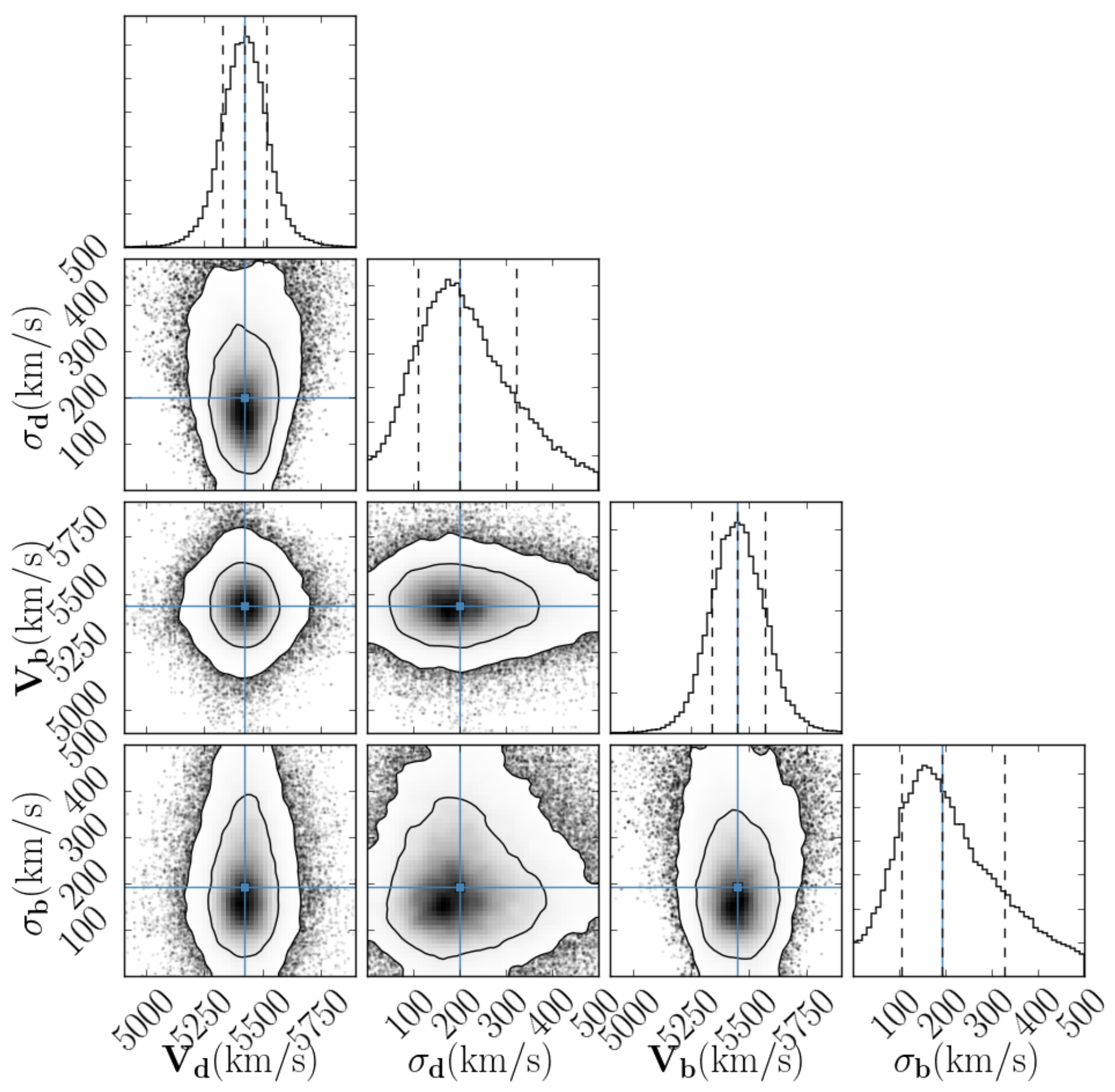}
	\caption{Corner plot for the distribution functions of the 4 parameters, the mean velocities $V_{\mathrm{b}}$ and $V_{\mathrm{d}}$ and the dispersions $\sigma_{\mathrm{b}}$, $\sigma_{\mathrm{d}}$, that define $F_{\mathrm{b}}$ and $F_{\mathrm{d}}$, see equations~(\ref{eq:bulge}) and (\ref{eq:disco}). The distribution functions, obtained using an MCMC sampler, are the results, in this example, of the fitting of a spatial pixel spectrum of the galaxy NGC\,5784. The contours in the 2D distributions are 68 and 95\% levels. The dashed lines in the 1D histograms show the 16th and the 84th percentiles, while the blue solid ones are 50th percentiles, corresponding to our adopted best-fit parameter values.}
	\label{mc}
\end{figure}

\begin{figure}
	\includegraphics[width=\columnwidth]{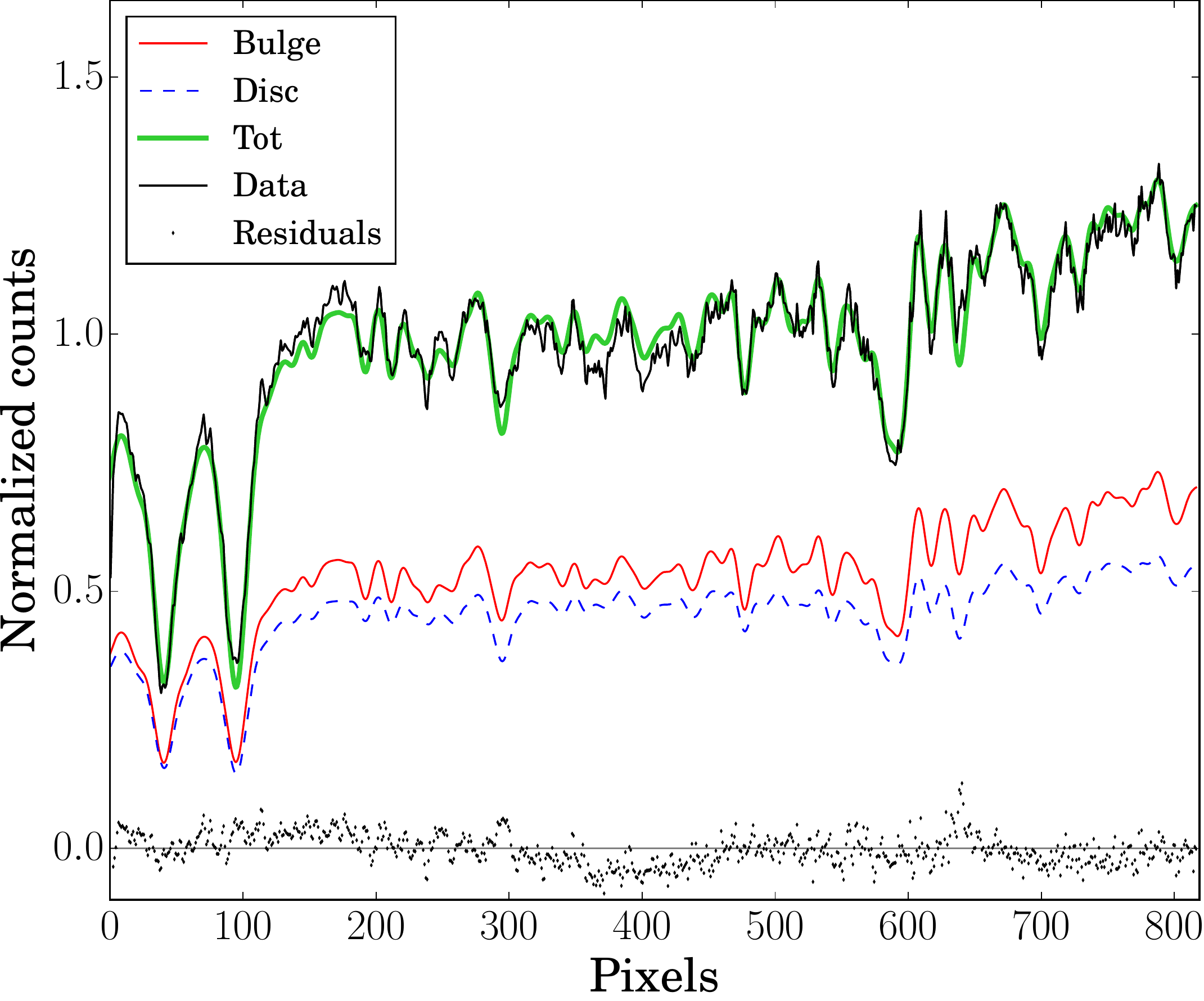}
	\caption{Spectrum (black solid line) for the same pixel shown in Fig.~\ref{mc}, in which the value of bulge-to-total ratio is 0.53. The spectrum covers the observed wavelength range [3975, 4552]\,\AA. The model spectra for bulge (red solid line) and disc (blue dashed line) are obtained after the convolution of disc and bulge template spectra with the LOSVDs defined by the best-fit parameter values shown in Fig.~\ref{mc}. The gray points represent the difference between observed and total model spectrum (green solid line). The gray solid line is the 0th level for reference.}
	\label{mcspettro}
\end{figure}

For each Voronoi bin in a galaxy we extract a spectrum that we model by using equation~(\ref{eq:2gaussiane}). Each model spectrum, $S_{\mathrm{mod}}$, is described by a set $\theta$ of four free parameters: $V_{\mathrm{b}}, V_{\mathrm{d}}, \sigma_{\mathrm{b}}, \sigma_{\mathrm{d}}$. The best parameters for the observed galaxy spectrum, $S_{\mathrm{obs}}$, are obtained by minimizing the weighted-absolute residuals\footnote{We made several experiments, using different kinds of residuals (in particular standard chi-square,  absolute residuals, absolute residuals weighted by the inverse of the sum of the data and the model) to choose the best quantity to minimize. The visual inspection of the fits led us to prefer the absolute residuals although the differences in the final results are minimal.}
\begin{equation}
	\sum_{\mathrm{k}}(|S_{\mathrm{obs,k}}-S_{\mathrm{mod,k}}|w_{\mathrm{k}})
	\label{eq:like}
\end{equation}
where the summation is over the set of spectral velocities corresponding to the wavelength range: [3899,4476]\,\AA. 
The coefficients $w_{\mathrm{k}}$ in equation~(\ref{eq:like}) are introduced to give higher weight in the fitting process at locations in the spectrum where the stellar lines are expected. We fix the weight to 1 in correspondence to the following absorption lines: CaII H and K (3968.5\,\AA, 3933.7\,\AA), G-band (4305\,\AA), H$\delta$ (4101.7\,\AA), CaI (4226\,\AA), FeI (4325\,\AA), H$\gamma$ (4340.5\,\AA), FeI (4383.5\,\AA), otherwise it is 0.2 (Fig.~\ref{spettrorighe}). The values of 1 and 0.2 are the result of tests made to optimize the fitting between the model and the data in the fixed spectral regions. The width of the spectral regions in which $w_{\mathrm{k}}=1$ (see Fig.~\ref{spettrorighe}) is such that our selected absorption lines are allocated more importance than the rest of the spectrum.\\
We use the Markov Chain Monte Carlo (MCMC) Python package EMCEE \citep{foreman} to explore the parameter space. MCMC is allowed to run, varying each of the four parameters in the range of [0, 500]\,km\,s$^{-1}$ for $\sigma_{\mathrm{b}}$ and $\sigma_{\mathrm{d}}$, and [$V_{\mathrm{est,sys}}-500, V_{\mathrm{est,sys}}+500$]\,km\,s$^{-1}$ for $V_{\mathrm{b}}$ and $V_{\mathrm{d}}$, where $V_{\mathrm{est,sys}}$ is an estimate of galaxy systemic velocity taken from NED.\\
In Fig.~\ref{mc} we show an example of the posterior distribution of each parameter from the Markov chains produced for the fitting of one spatial pixel of the galaxy NGC\,5784. This corner plot shows the distribution for each parameter in the panels along the diagonal and the joint distribution for each pair of parameters in the other panels. The kinematic parameters for bulge and disc components are reliable and well constrained for this pixel in which the relative contributions of bulge and disc flux are almost equal. We will discuss in Section~\ref{sec:exclbin} the exclusions of pixels in which the derived parameters are not reliable. The corresponding galaxy spectrum for this same spatial pixel is shown in Fig.~\ref{mcspettro} (black line). The solid red and dashed blue lines show the best-fit model spectra for bulge and disc components respectively. They are obtained as a convolution (see equations~(\ref{eq:gaub}) and (\ref{eq:gaud})) of the two LOSVDs, defined by the best-fit parameters obtained as explained above, with their respective bulge and disc template spectra, $T_{\mathrm{b}}$ and $T_{\mathrm{d}}$ (see equations~(\ref{eq:templateb}) and (\ref{eq:templated})). 

\section{From velocity fields to rotation curves}\label{rcurve}
The application of our software to bulge and disc kinematic decomposition on all bins of our sample of galaxies returns four maps, one for each fitted parameter: $V_{\mathrm{b}}$, $V_{\mathrm{d}}$, $\sigma_{\mathrm{b}}$, $\sigma_{\mathrm{d}}$. In this section we focus on the disc velocity fields that are analysed using the tilted-ring model \citep{begeman}. This divides a velocity map into a number of rings, from which the line-of-sight velocity is extracted and expanded in a finite number of harmonic terms. To first order and neglecting the radial motions, this expansion leads to the equation
\begin{equation}
	v_{\mathrm{los}}(R)=V_{\mathrm{sys}}+V_{\mathrm{rot}}(R)\cos \phi \sin\,i
	\label{vrot}
\end{equation}
where $R$ is the radius of a circular ring in the plane of the galaxy (or the semimajor axis length of the elliptical ring projected on the sky); $V_{\mathrm{sys}}$ is the systemic velocity; $\mathrm{\phi}$ is the anti-clockwise azimuthal angle measured from the projected major axis in the plane of the galaxy \citep{begeman}; $i$ is the inclination angle of the disc, related to the axial ratio $q_{\mathrm{d}}$ ($q_{\mathrm{d}}=\cos\,i$).\\
By fitting equation~(\ref{vrot}) to our ten disc velocity fields we obtain the disc rotation curves, which are one-dimensional representations of the circular velocities as a function of radius. This fit is performed using ROTCUR, which is a task of GIPSY \citep{van} that makes use of the Levenberg-Marquardt method. In the fitting process we consider rings with width of 2 arcsec rather than of 1 arcsec, which is the dimension of each spaxel, because the fibers of the CALIFA IFU have a diameter of 2.7 arcsec. Note, however, that even with our choice of 2 arcsec, there is an oversampling if we consider the loss of spatial resolution caused by the Voronoi binning. In the fitting process the axis ratio, $q_{\mathrm{d}}$, and the position angle, $PA_{\mathrm{d}}$, are held fixed to the values used for the kinematic decomposition (e.g. see derivation of bulge-to-disc ratio, Section~\ref{bsutsec}), which are those found by the photometric decomposition (Section~\ref{section1}). Our fitting strategy is the following: initially we leave $V_{\mathrm{rot}}(R)$ and $V_{\mathrm{sys}}(R)$ free to vary. We focus on the estimated value of $V_{\mathrm{sys}}(R)$, fixing it to the mean value along the rings. Then we run ROTCUR twice to fit the approaching and receding sides separately.\footnote{We decided to fit the approaching and receding sides separately, and not both sides simultaneously, because this fitting strategy allowed us to estimate the errors on the rotation velocities (see Section~\ref{sec:errv}). We verified that the rotation curves obtained considering the two fitting processes (approaching and receding fit or both sides) are fully compatible within our errors.}   When applicable, we consider as final rotation velocities,$V_{\mathrm{rot}}(R)$, those obtained as averages of the approaching and receding sides. In two cases (IC\,1652 and NGC\,7025) this approach is not applicable at all radii because of the exclusion of regions contaminated by external stars (see Section~\ref{resgal}).

\subsection{Error estimates}\label{sec:errv}
We estimate the errors on the rotation velocities considering the following three contributions.
\begin{enumerate}
\item $\delta_{\mathrm{Q}}$: as mentioned in Sections~\ref{section1} and \ref{rcurve}, we assume that the inclination of the discs, $i$, is given considering $\cos\,i$ as equal to the observed disc axis ratio $q_{\mathrm{d}}$ and the rotation velocities are derived under this assumption, equation~(\ref{vrot}). However, this is valid only for an infinitely thin disc. To take into account that the discs could have finite thickness, they are often \citep[e.g][]{gram08, cortese16} modelled as oblate spheroids with an intrinsic short-to-long axis $Q$. In this case, the inclination $i$ of the galaxy is given by \citep{weij}:
\begin{equation}
	\cos\,i=\sqrt{\frac{q_{\mathrm{d}}^{2}-Q^{2}}{1-Q^{2}}}.
	\label{angolo-inc}
\end{equation}
In general, the value of the intrinsic axis ratio $Q$ is highly uncertain and it could vary within the range $\sim$0.1-0.25 \citep[e.g.][]{giova, weij, lambas}. To estimate the errors that we make under our thin disc approximation, we run ROTCUR in the same manner as described above, but with an inclination $i$, given by equation~(\ref{angolo-inc}), with a value of $Q=0.2$ \citep{padilla}. Since equation~(\ref{angolo-inc}) gives inclination angles larger than those obtained with $Q=0$ (thin disc approximation), the rotation velocities obtained are lower than those in the thin disc approximation, as expected because of the dependence as 1/$\sin\,i$, equation~(\ref{vrot}). For each galaxy, we define an error $\delta_{\mathrm{Q}}$ as the largest difference (in the radial range considered in the fitting process) between the rotation velocities obtained in the thin disc approximation and those obtained deriving $i$ from \ref{angolo-inc} with $Q=0.2$. The typical values of $\delta_{\mathrm{Q}}$ obtained are $\sim$|5-8| km\,s$^{-1}$. For the reason explained above, these errors give a contribution only towards lower values of the velocities.
\item $\delta_{\mathrm{i}}$: The kinematic decomposition is performed fixing the geometrical parameters (see Section~\ref{sec:coordinates} and \ref{bsutsec}) of the bulge and disc components to those obtained by the photometric decomposition (Section~\ref{section1}). However, it is possible that these geometrical parameters are different from those obtained by the fitting of the velocity fields. In other words, it is possible that the photometric values of the position angles $PA_{\mathrm{d}}$ and of the inclinations $i=\arccos\,q_{\mathrm{d}}$ are different from their kinematic estimates, obtained letting these two parameters be free during the fitting of rotation velocities with equation~(\ref{vrot}). To take into account the errors that we make by fixing the inclination, we run ROTCUR in the same manner as described above, but letting the inclination angles be free in a first run. Then we average the values of the inclination angles along the fitted rings. The values of the kinematic inclination angles obtained are larger (for 2 galaxies: NGC\,7671 and NGC\,6081) and smaller (for the others) by $\lesssim$5\% than the photometric ones. For two galaxies, NGC\,7025 and NGC\,0528, in contrast, the kinematic inclination angles are larger by $\sim$16\% and smaller by $\sim$14\% respectively than their photometric counterparts. Finally, for each galaxy we run ROTCUR to estimate the values of rotation velocities, while holding the inclination angle fixed to this kinematic value. The comparison of the rotation velocities obtained in this way, with those assumed as our best fit, allows us to estimate another contribution to the errors, which we define as $\delta_{\mathrm{i}}$. We estimate it as the largest difference (in the radial range considered in the fitting process) between our best-fit rotation velocities and those obtained with the kinematic inclination angle. For three galaxies (NGC\,7671, NGC\,6081 and NGC\,7025) these errors give a contribution towards lower values of the velocities, $\delta_{\mathrm{i-}}$ while for the others towards larger values $\delta_{\mathrm{i+}}$. The typical values of $|\delta_{\mathrm{i-/+}}|$ obtained are $\sim$4 km\,s$^{-1}$, except for NGC\,7025 and NGC\,0528 for which we have $|\delta_{\mathrm{i-/+}}|\sim$ 15 km\,s$^{-1}$.
\item $\delta_{\mathrm{v}}$: We estimate the standard uncertainties, $\delta_{\mathrm{v}}$, on the best-fit rotation velocities considering the prescription of \citet{swater99} who assumed that the 1$\sigma$-errors on the rotation velocities are given by adding quadratically measurement and asymmetry errors. The asymmetry errors derive from non-circular motions and asymmetry of the galaxy and can be estimated as one fourth of the difference in rotation velocity between approaching and receding sides. For measurement errors we assume the fit-parameter uncertainties given by ROTCUR, which are computed from the covariance matrix produced during the least-square minimisation algorithm.
\end{enumerate}
Finally, we combine the three errors: $\delta_{\mathrm{Q}}$, $\delta_{\mathrm{i}}$ and $\delta_{\mathrm{v}}$ using the following expressions:
\begin{equation}
	\Delta_{\mathrm{v+}}=\delta_{\mathrm{i+}}+\delta_{\mathrm{v}}
	\label{deltav+}
\end{equation}
\begin{equation}
	\Delta_{\mathrm{v-}}=\sqrt{\delta_{\mathrm{i-}}^{2}+ \delta_{\mathrm{Q}}^{2}}+\delta_{\mathrm{v}}.
	\label{deltav-}
\end{equation}
The values $\delta_{\mathrm{Q}}$, $\delta_{\mathrm{i}}$ are not summed quadratically with $\delta_{\mathrm{v}}$ as in the standard error combinations because they are not the result of random fluctuations but are systematic errors due to our ignorance about the true inclinations of our discs \citep{martinelli}.

\subsection{Exclusion of bins}\label{sec:exclbin}
\begin{figure}
	\includegraphics[width=\columnwidth]{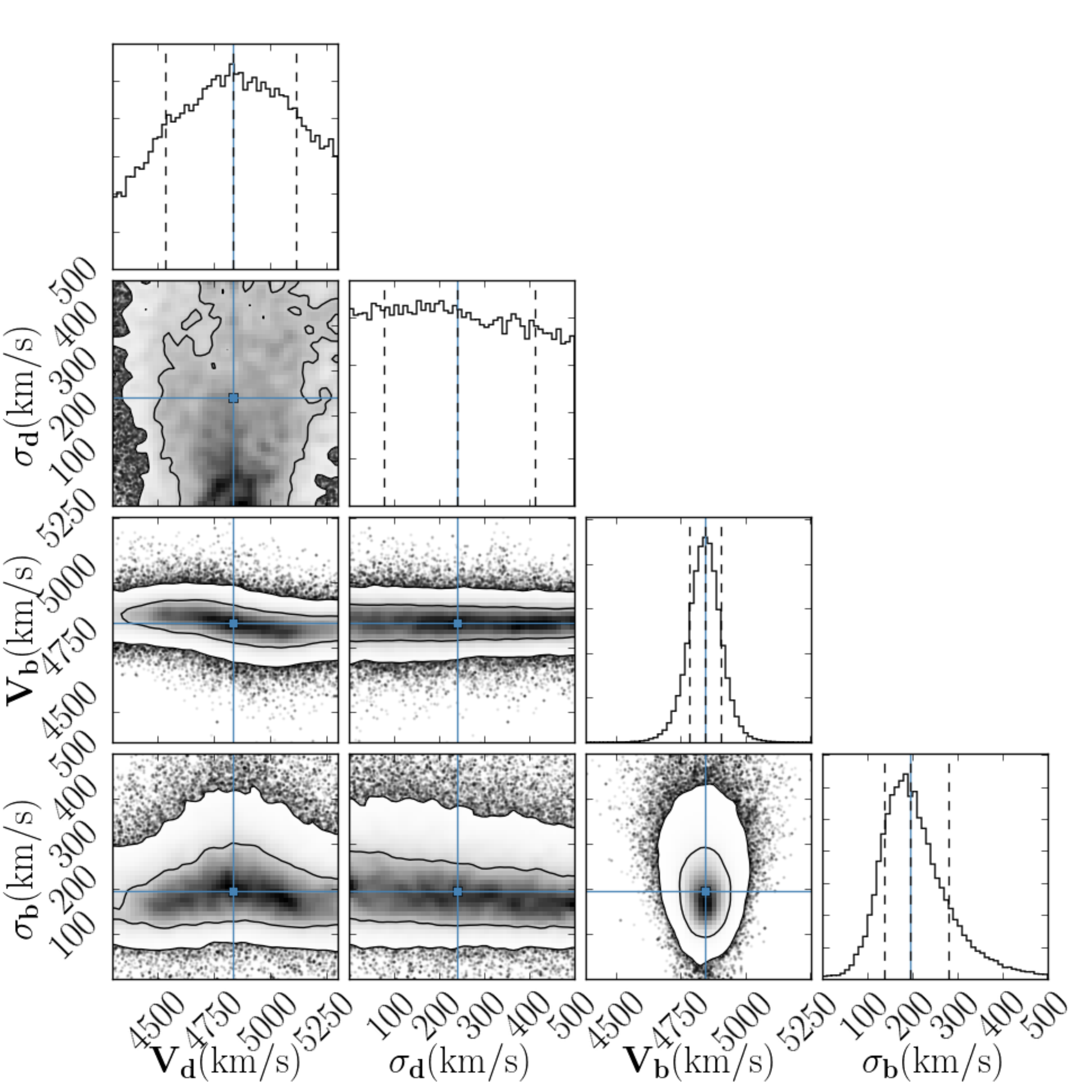}
	\caption{Example of corner plot extracted from a location where the bulge dominates (bulge-to-total ratio is 0.87) for the galaxy NGC\,0528. This location is excluded from the disc tilted-ring fitting process. The two criteria described in Section~\ref{sec:exclbin} are not fulfilled: the distribution of $\sigma_{\mathrm{d}}$ is similar to the prior uniform distribution for this parameter, while the 1$\sigma_{+/-,V_{\mathrm{d}}}$ uncertainties for the posterior probability distribution obtained for $V_{\mathrm{d}}$ are larger than the difference between the extremes of the interval in which the values of $V_{\mathrm{d}}$ are allowed to vary during the MCMC sampling and the best fit $V_{\mathrm{d}}$ (blue solid line). Note that the parameters of the bulge are extremely well constrained.}
	\label{pix-escl}
\end{figure}

\begin{figure*}
	\includegraphics[width=18cm]{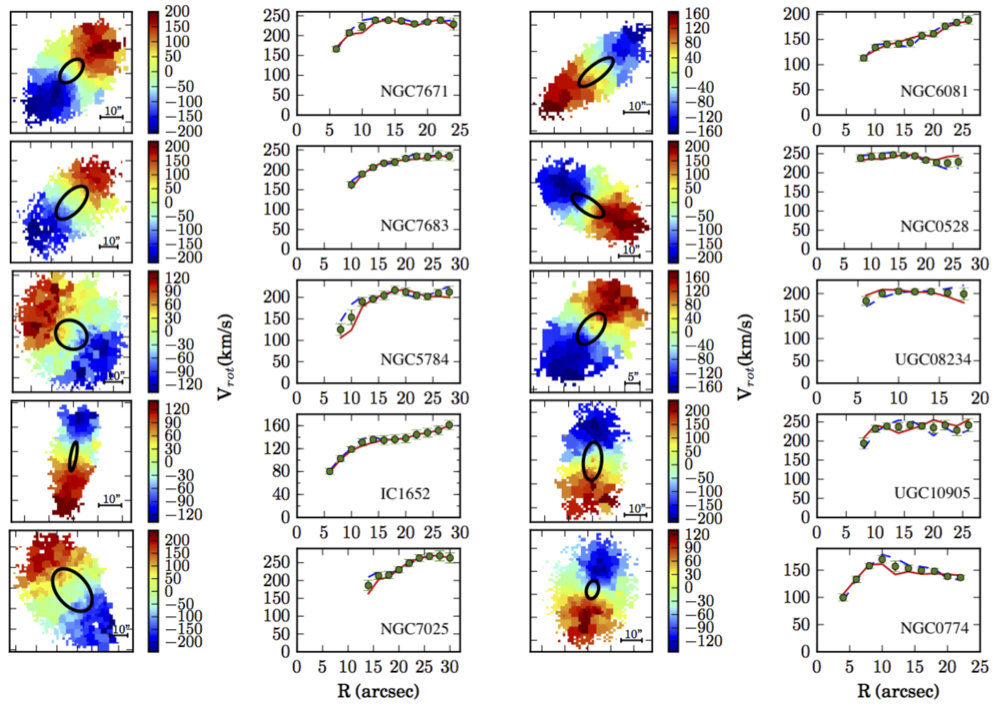}
	\caption{Disc velocity fields and respective deprojected rotation curves for (first and second columns, from top to bottom) NGC\,7671, NGC\,7683, NGC\,5784, IC\,1652, NGC\,7025 and (third and fourth columns, from top to bottom) NGC\,6081, NGC\,0528, UGC\,08234, UGC\,10905, NGC\,0774. The points within the black ellipse in the velocity maps are excluded from the tilted-ring fitting because the disc parameters are not well constrained (see Section~\ref{sec:exclbin}). The units of the colour-bars are in km\,s$^{-1}$. The dashed blue and red solid curves in the rotation velocity profiles show the approaching and receding sides from the radius at which the disc parameters are well constrained; the green points represent their average. The error-bars represent the 1$\sigma$ standard uncertainties, $\delta_{\mathrm{v}}$, calculated as explained in Section~\ref{sec:errv}, which do not take into account the systematic uncertainties.}
	\label{fig:vel10}
\end{figure*}

The fitting process described in the previous section is applied to all radii, but in the successive analysis we consider the rotation velocities obtained in rings where the disc parameters of $F_{\mathrm{d}}$ are better constrained. In the regions where the values of bulge-to-total flux ratio are $\gtrsim$0.7-0.8 we obtain values of $V_{\mathrm{d}}$ and $\sigma_{\mathrm{d}}$ that are not reliable (see e.g. the corner plot of Fig.~\ref{pix-escl}). For this reason, the bins that do not satisfy the following criteria are excluded:
 \begin{enumerate} 
 \item $1\sigma_{+,V_{\mathrm{d}}}<(V_{\mathrm{est,sys}}+500-V_{\mathrm{best-fit,d}})$ and \\
 \hspace*{0.7cm} $1\sigma_{-,V_{\mathrm{d}}}<(V_{\mathrm{best-fit,d}}-V_{\mathrm{est,sys}}+500)$\\
1$\sigma_{+/-,V_{\mathrm{d}}}$ uncertainties for the posterior probability distribution obtained for $V_{\mathrm{d}}$ must be smaller than the difference between the extremes of the interval in which the values of $V_{\mathrm{d}}$ are allowed to vary during the MCMC sampling ($[V_{\mathrm{est,sys}}-500, V_{\mathrm{est,sys}}+500$]\,km\,s$^{-1}$) and the best fit $V_{\mathrm{d}}$;
\item 1$\sigma_{+/-,\sigma_{\mathrm{d}}}$ uncertainties for the posterior probability distribution obtained for $\sigma_{\mathrm{d}}$ must be lower than 144.33 km\,s$^{-1}$, that is the standard deviation of a uniform distribution defined in an interval [0, 500] \footnote{The standard deviation of a uniform distribution defined in the interval [a, b] is given by: $\mathrm{|a-b|/\sqrt{12}}$.}, equal to those over which the values of $\sigma_{\mathrm{d}}$ were allowed to vary during the MCMC sampling. 
 \end{enumerate} 
 The exclusion of these internal bins (see the black ellipses in the first and third columns of Fig.~\ref{fig:vel10}) is not a problem for the derivation of the specific angular momentum, because this quantity is very insensitive to the values of velocity at these small radii (see Section~\ref{sec:jind}). Similarly, the effect of the beam smearing is also negligible in the derivation of the specific angular momentum. In order to quantify the effects of the beam smearing we consider model galaxies with different velocity gradients in the rising part of the rotation curves and with a resolution typical of our observations. Even if the main effect of the beam smearing is to reduce the velocity gradient in the inner parts of the rotation curve \citep{barolo}, the resulting specific angular momentum is scarcely not influenced (at most ~2\%), because it receives the greatest contribution from the outer regions. \\
Furthermore, we introduce another criterion on the quality of the spectral fit, because some spectra, with low S/N, give unreliable values of $V_{\mathrm{best-fit,d}}$ and have to be excluded from the tilted-ring fitting process. In particular, this criterion is based on the computation of the standard deviation, E, of the residuals $\sum_{j}(S_{\mathrm{obs,j}}-S_{\mathrm{mod,j}})$ (grey points in Fig.~\ref{mcspettro}), where the summation is over the set of spectral velocities. We exclude all Voronoi bins that have pixels with values of E$\gtrsim$0.13 counts, while the typical values of this quantity in the other pixels is $\lesssim$0.06 counts. The typical number of excluded Voronoi bins for each galaxy is about 10 and they are located in the most external regions.

\subsection{Results on individual galaxies}\label{resgal}
In this section we show the rotation curves obtained under the assumptions and using the techniques described above. The second and fourth columns panels of Fig.~\ref{fig:vel10} show the approaching and receding rotation velocities (blue and red solid curve) for the radii at which the disc parameters are well constrained (see Section~\ref{sec:exclbin}), while the green points show the average of approaching and receding sides. For a better visualisation, we show only the uncertainties $\delta_{\mathrm{v}}$, estimated as explained in Section~\ref{sec:errv}. The velocity fields from which these rotation velocities are derived are shown in the adjacent panels. We can see that for most galaxies of our sample the rotation curves become asymptotically constant out to the radii investigated in our analysis, like the typical rotation curve of spiral galaxy discs. In Table~\ref{tab:v} we report the characteristic disc rotation velocities for all galaxies, $\bar{V}_{\mathrm{flat}}$. These are obtained as average of the values of rotation velocities taken from the radius at which the curve becomes flat out to the outermost data point. The determination of this radius can be easily done for most galaxies, except for NGC\,1652, NGC\,6081 and NGC\,0774. The first two galaxies have rotation curves that are still rising at the outer reaches of the data and so, to be conservative, we average the last three points of their rotation curves. For NGC\,0774 we average the last two points of its rotation curve, because the approaching side is flat, while the receding side is declining.\\  
The velocity fields of NGC7671 and NGC0528 are comparable to those obtained with a different method by \citet{tabor} for the same galaxies. Indeed as discussed in Sections \ref{ngc7671_photo}  and \ref{ngc0528_photo} the parameters that describe the disc components are comparable. The differences in the bulge parameters do not affect the extraction of the kinematics of the disc component, especially in the regions in which the luminosity contribution of the disc is dominant. \\
An interesting question is whether performing a full kinematic bulge-disc decomposition as done in this paper makes a significant difference with respect to using a one-component model. We tested this by fitting one single velocity and velocity dispersion to the absorption lines in our data and compare the resulting velocity fields with the one obtained for the disc alone. For UGC\,08234, which has the largest B/T ratio, the value of the one component $\bar{V}_{\mathrm{flat}}$ is 165 km\,s$^{-1}$, so the two values differ by $\sim$20\% (2.4$\sigma$) . For galaxies for which the bulge contribution is less prominent, e.g. NGC\,7025, the one-component $\bar{V}_{\mathrm{flat}}$ is about 1.5$\sigma$ lower than the value obtained for the disc alone. In general, the residuals of the fit reveal that the two-component model reproduces the spectra more accurately with respect to the single component model. We can conclude that performing a one-component fit results in a systematic underestimate of the specific angular momentum by about 15-20\%.\\
In Table~\ref{tab:v} we report also the values of the systemic velocities, which are the result of the tilted-ring fitting process (see discussion in Section~\ref{rcurve}). The values found for $V_{\mathrm{sys}}$ are in agreement within 1$\sigma$ both with the values reported in NED and with those reported in the header of each datacube (keyword: MEDVEL, see \citet{garcia15}). For two galaxies, IC\,1652 and NGC\,7025, we cannot apply the same method as for the others. These two galaxies are indeed strongly contaminated by an external star (see Fig.~\ref{n1652} and \ref{n7025}), so that these regions had to be completely excluded from the fitting. For this reason, their rotation velocities are the result of the average of approaching and receding velocities out to 16 arcesc (for IC\,1652) and 26 arcsec (for NGC\,7025), while for larger radii only the rotation velocities on one side are considered. In this case, the errors are assumed equal to the maximum error obtained in the region where the average is performed. The fitting process for the two problematic galaxies NGC\,5784 and NGC\,0774 is discussed in Appendix \ref{app:n5784} and \ref{app:n0774}. 

\begin{table}
	\centering
 	\caption{For each galaxy in our sample (column 1) we report the values of systemic velocity $V_{\mathrm{sys}}$ (column 2) found after the application of the tilted-ring fitting method and of the average of the rotation velocities, $\bar{V}_{\mathrm{flat}}$ (column 3), and their respective errors $\Delta_{\mathrm{v+/-}}$, calculated as explained in Section~\ref{sec:errv}. In column 4 we show the maximum radius for which we have CALIFA data.}
	\label{tab:v}
	\begin{tabular}{l l l l}
	\hline
 	Galaxy & $V_{\mathrm{sys}}$ (km\,s$^{-1}$) & $\bar{V}_{\mathrm{flat}}$ (km\,s$^{-1}$) &$R_{\mathrm{max}}$ (kpc)\\
 	\hline
	NGC\,7671 & 3900$\pm$11 & $235^{+15}_{-19}$ &  6.9\\
	NGC\,7683 & 3703$\pm$14  & $233^{+17}_{-17}$ & 7.3\\
	NGC\,5784 & 5426$\pm$15  & $210^{+42}_{-42}$ & 10.8\\
	IC\,1652 & 5029$\pm$17  & $149^{+12}_{-14}$ & 10.1\\
	NGC\,7025 & 4927$\pm$17  & $267^{+14}_{-32}$ & 10.5 \\
	NGC\,6081 & 5064$\pm$11  & $186^{+8}_{-16} $ & 9.5\\
	NGC\,0528 & 4788$\pm$16  & $228^{+25}_{-14}$ & 8.7 \\
	UGC\,08234 & 8026$\pm$15  & $203^{+18}_{-16}$ &10.1\\
	UGC\,10905 & 7658$\pm$14  & $237^{+18}_{-21}$ & 14.1\\
	NGC\,0774 & 4601$\pm$11  & $138^{+18}_{-15}$ & 6.9\\
 	\hline
  	\end{tabular}
\end{table}

\section{Specific angular momentum and stellar masses}\label{sec:jandm}
\subsection{Estimates of the specific angular momentum}\label{sec:angularmomentum}
The general expression for stellar specific angular momentum is:
\begin{equation}
	\vec{j}_{\star}= \frac{\vec{J}_{\star}}{M_{\star}}=\frac{\int(\vec{x}\times\vec{\bar{v}})\,\rho(\vec{x})\,d^{3}\vec{x}}{\int\rho(\vec{x})\,d^{3}\vec{x}}.
	\label{eq:jt} 
\end{equation}
where $\vec{x}$ and $\vec{\bar{v}}$ are the position and mean velocity\footnote{The three-dimensional mean velocity vector is defined as: \[\vec{\bar{v}}=\frac{M_{\star}}{\rho(\vec{x})}\int \vec{v} f(\vec{x},\vec{v}) d^{3}\vec{v}\] where $\vec{v}$ is the velocity vector relative to the galactic centre and $f(\vec{x},\vec{v})$ is the phase-space density.} vectors with respect to the centre of mass of the galaxy. These two quantities in a galaxy are fully specified by the phase-space density $f(\vec{x},\vec{v})$. The quantity $\rho(\vec{x})$ is the three-dimensional stellar mass density, which is expressed in terms of $f(\vec{x},\vec{v})$ and of the total stellar mass $M_{\star}$ as:
\begin{equation}
	\rho(\vec{x})=M_{\star} \int f(\vec{x},\vec{v}) d^{3}\vec{v}.
\end{equation}
However, all these three-dimensional quantities are not accessible directly and the usual adopted assumptions to obtain $j_{\star}$ from observations are the following:
\begin{enumerate}
\item Galaxies are axisymmetric and have cylindrical rotation specified by the same symmetry axis as the density distribution;
\item Adopting the cylindrical galactic coordinates ($R$,\,$z$,\,$\phi$), $\vec{\bar{v}}$ becomes $v_{\mathrm{rot}}(R)\hat{\phi}$, $\rho(\vec{x})$ becomes $\rho(R,z)$ because there is no dependence on $\phi$ and equation~(\ref{eq:jt}) becomes
\begin{equation}
	j_{\star}=\frac{\int \Sigma_{\star}(R)v_{\mathrm{rot}}(R)R^{2}\,dR}{\int \Sigma_{\star}(R)\,R\,dR}
 	\label{eq:jcil}
\end{equation}
where $\Sigma_{\star}(R)$ is the surface mass density of a galaxy viewed face-on: $\Sigma_{\star}(R)=\int \rho(R,z) dz$;
\item There is no radial variation of the stellar mass-to-light ratio across the disc and the surface mass density profile $\Sigma_{\star}(R)$ becomes equivalent to the surface brightness profile because the conversion factor in equation~(\ref{eq:jcil}) between these two quantities cancels out.
\end{enumerate}
In order to compare the distribution of our S0 discs in the $j_{\star}$-$M_{\star}$ plane to the relations found by \citetalias{rf12} for all morphological types, we have first estimated the values of $j_{\star}$ using their angular momentum estimator and their assumptions (Section~\ref{sec:jref}). Subsequently, we have also used our independent estimator (Section~\ref{sec:jind}). We note that in \citet{fall13} the values of disc $j_{\star}$ are unchanged with respect to \citetalias{rf12}, while the stellar masses, in particular for discs, are smaller because these were derived using $M_{\star}/L$ derived from the colour-$M_{\star}/L$ relation. However, since in \citet{fall13} the updated values of the stellar masses were not given, we decided to carry out our comparison with \citetalias{rf12}. As explained in Section \ref{sec:jmdiagram} our results are not affected by this assumption.

\subsubsection{RF angular momentum estimator}\label{sec:jref}
The values of $j_{\star}$ used by \citetalias{rf12} to construct the $j_{\star}$-$M_{\star}$ relation, are not the results of a calculation using the equation~(\ref{eq:jcil}). For their galaxy sample they used an angular momentum estimator, $j_{\star,\mathrm{RF}}$, introduced to obtain a quick estimate of $j_{\star}$ without the need to evaluate the full integrals in (\ref{eq:jcil}).\\
The expression for $j_{\star,\mathrm{RF}}$ is based on the assumption that there is a constant value of $v_{\mathrm{rot}}$ that when substituted in (\ref{eq:jcil}) gives the same value of $j_{\star}$ as that obtained using the true rotation curve. Furthermore, if in (\ref{eq:jcil}) the velocity has the constant value $v_{\mathrm{s}}$, the dependence of $j_{\star}$ on $v_{\mathrm{s}}$ and on the surface brightness parameters, i.e. the S\'ersic index $n$ and the effective radius $R_{\mathrm{e}}$, becomes:
\begin{equation}
	j_{\star,\mathrm{RF}}=k_{\mathrm{n}}\,v_{\mathrm{s}}\,R_{\mathrm{e}}
	\label{jk}
\end{equation}
where $k_{\mathrm{n}}$ is a dimensionless function of S\'ersic index $n$. For exponential profiles, the full integral (\ref{eq:jcil}) with a constant rotation velocity has the simple solution: $2\,v_{\mathrm{s}}\,R_{\mathrm{d}}$, so that the expression equivalent to (\ref{jk}), valid for the discs is: 
\begin{equation}
	j_{\star,\mathrm{RF}}= 2\,v_{\mathrm{s}}\,R_{\mathrm{d}}.
	\label{jkd}
\end{equation}
In principle each galaxy has its radius $R_{\mathrm{s}}$, which depends on its velocity profile, where the value of $v_{\mathrm{s}}$, substituted in (\ref{jk}) and (\ref{jkd}), returns the correct value of $j_{\star}$.  However, \citetalias{rf12}, using both models and real galaxies found that a good approximation for $j_{\star}$ can be obtained substituting $v_{\mathrm{s}}$  with the value of the rotation velocity taken at $2\,R_{\mathrm{e}}$ both in (\ref{jk}), which is valid for a generic S\'ersic profile, and in (\ref{jkd}), for exponential profile.\footnote{For an exponential profile the effective radius $R_{\mathrm{e}}$ is linked to the scale radius by the relation $R_{\mathrm{e}}=1.68\,R_{\mathrm{d}}$, so that $2\,R_{\mathrm{e}}=3.36\,R_{\mathrm{d}}$.}
In our galaxy sample, for the value of v at $R_{\mathrm{s}}$ we use that obtained by extrapolating the rotation beyond the outermost measured data point assuming a constant, as done by \citetalias{rf12}. In practice, the constant values of velocity are assumed equal for each galaxy to the values of $\bar{V}_{\mathrm{flat}}$ found in Section~\ref{rcurve} (see Table~\ref{tab:v}). The values of $j_{\star,\mathrm{RF}}$ for each disc galaxy of our sample are shown in column 2 of Table~\ref{tab:jem}.\\
For NGC\,0774, which is the galaxy characterized by two disc components, we report the value of $j_{\star,\mathrm{RF}}$ obtained using the proxy for luminosity weight average of the two scale radii of the best-fit model (see Appendix~\ref{ph:n0774}), $R_{\mathrm{d}}\sim6.33$\,arcsec, as indicated with suffix D in Table~\ref{tab:jem}. Furthermore, we also report the values of $j_{\star,\mathrm{RF}}$ obtained considering the internal disc, characterized by $R_{\mathrm{d}}\sim4.42$\,arcsec (indicated with suffix D1) and the external disc that has $R_{\mathrm{d}}\sim15.15$\,arcsec (indicated with suffix D2). In this case, it is difficult to establish which of the three values of $j_{\star,\mathrm{RF}}$ reported (D, D1, D2) should be compared with those of the other galaxy sample, and adopting a unique value among the three could be overly simplistic (see also discussion in Section~\ref{sec:discussion}).

\begin{table*}
	\caption{For each galaxy in our sample (column 1) we report the values of the angular momentum estimator, $j_{\star,\mathrm{RF}}$ (column 2), as defined by \citetalias{rf12}, and computed as explained in Section~\ref{sec:jref}. In column 3 we show the values of disc $j_{\star}$, calculated, as explained in Section~\ref{sec:jind} using equation~(\ref{eq:jsum}), in columns 4 and 5 the values of disc stellar masses, calculated under the assumptions of \citetalias{rf12} (Section~\ref{sec:mref}) and under our assumptions (Section~\ref{sec:mind}).}
	\label{tab:jem}
	\begin{tabular}{l l l l l}
	\hline
	Galaxy & $j_{\star,\mathrm{RF}}$ & $j_{\star,2R_{\mathrm{d}}}$ & $\log(M_{\star \mathrm{d,RF}}/M_{\odot})$ & $\log(M_{\star \mathrm{d}}/M_{\odot})$ \\
 & ($\mathrm{km\,s^{-1}\,kpc}$) & ($\mathrm{km\,s^{-1}\,kpc}$)\\
	\hline
	NGC\,7671 &       $1359^{+120}_{-140}$  &$806^{ +45}_{-54}$     & 10.91$\pm$0.05 & 10.72$\pm$0.07\\
	NGC\,7683 &       $2048^{+186}_{-183}$ &$1164^{+89 }_{-87}$    & 10.81$\pm$0.03 & 10.67$\pm$0.07\\
	NGC\,5784 &       $1408^{+283}_{-283}$ & $692^{ +55}_{-55}$     & 11.22$\pm$0.02 & 10.98$\pm$0.09\\
	IC\,1652    &        $1003^{+81}_{-95}$     & $513^{+39 }_{ -46}$     & 10.63$\pm$0.10 & 10.55$\pm$0.06\\
	NGC\,7025 &       $3824^{+304}_{-512}$ & $2025^{+118 }_{-256}$& 11.30$\pm$0.08 & 11.12$\pm$0.08\\
	NGC\,6081&        $1979^{+113}_{-188}$ & $855^{ +76}_{ -106}$   & 11.03$\pm$0.02 & 10.81$\pm$0.09\\
	NGC\,0528 &       $1470^{+166}_{-100}$ & $925^{+95}_{-61}$       & 10.78$\pm$0.07 & 10.63$\pm$0.07\\
	UGC\,08234 &     $1833^{+167}_{-149}$ & $1034^{+66}_{-56}$     & 10.97$\pm$0.27 & 10.95$\pm$0.07\\
	UGC\,10905 &     $3284^{+346}_{-378}$ & $1835^{+145}_{-164}$ & 11.18$\pm$0.26 & 10.94$\pm$0.10\\
	NGC\,0774-D &   $535^{+101}_{-92}$     & $385^{ +59}_{-52}$      & 10.92$\pm$0.02 & 10.79$\pm$0.11\\
	NGC\,0774-D1 & $362^{+48}_{-40}$       & $273^{+48}_{-44}$\\
	NGC\,0774-D2 & $1242^{+194}_{-169}$ & $549^{ +77}_{-65}$ \\
	\hline
	\end{tabular}
\end{table*}

 \subsubsection{Our estimate of the angular momentum} \label{sec:jind}
\citetalias{rf12} used the angular momentum estimator described in the previous section in order to obtain a quick and easy quantity to calculate for their large sample of galaxies. In this work in addition to estimate the angular momentum in the same manner as \citetalias{rf12} (see Section~\ref{sec:jref}), we also compute a more direct estimation of $j_{\star}$.\\
Our estimation of the values of $j_{\star}$ is based on the equation~(\ref{eq:jcil}), in which the integral is substituted with a summation extended out to a radius $R_{\mathrm{i}}=2\,R_{\mathrm{d}}$:
\begin{equation}
	j_{\star,2R_{\mathrm{d}}}=\frac{\sum_{i}\Sigma_{\star}(R_{\mathrm{i}})V_{\mathrm{rot,i}}R_{\mathrm{i}}^{2}}{\sum_{i}\ \Sigma_{\star}(R_{\mathrm{i}})\,R_{\mathrm{i}}}
	\label{eq:jsum}
\end{equation}
where:
\begin{enumerate}
\item $V_{\mathrm{rot,i}}$ is the value of the rotation velocity, found in Section~\ref{rcurve}, at radius $R_{\mathrm{i}}$;
\item $\Sigma_{\star}(R_{\mathrm{i}})$ represents the disc surface brightness, defined for each galaxy by the parameters found in Section~\ref{section1}, calculated at $R_{\mathrm{i}}$.
\end{enumerate}
The choice of $2\,R_{\mathrm{d}}$ as the outer radius of our summation is justified for the following reasons: 
\begin{enumerate}
\item We have values of the rotation velocities out to $\sim2\,R_{\mathrm{d}}$ for all galaxies of our sample (see details below), so we can use measured data without any assumptions. 
\item If the discs have rotation-velocity profiles that are not too dissimilar (e.g.\,flat rotation velocity profiles) the total angular momentum $j_{\star}$ will be a well defined multiple of $j_{\star,2R_{\mathrm{d}}}$. This can be well understood considering the simple case of a constant rotation velocity curve: in this case the ratio between the angular momentum calculated out to $2\,R_{\mathrm{d}}$ (equation~(\ref{eq:jcil}) with integral defined in the interval [0, $2\,R_{\mathrm{d}}$]) and the total angular momentum (equation~(\ref{eq:jcil}) with integral defined in the interval [0, $\mathrm{\infty}$]) is equal to $\sim0.55$.
\item The rotation curve of an exponential disc reaches its flat part beyond about two disc scale lengths \citep{swater99}.
\end{enumerate}
In our sample, we have values of the rotation velocities out to $2\,R_{\mathrm{d}}$ for eight galaxies, while for NGC\,7683 the velocity data extend out to $\sim1.6\,R_{\mathrm{d}}$ and for NGC\,7025 out to $\sim$1.4$R_{\mathrm{d}}$. In these two last cases we adopt the same extrapolation as in the previous section, which is to assume a constant velocity profile equal to $\bar{V}_{\mathrm{flat}}$. In the inner regions, where the values of our disc velocities are not well constrained (see Section~\ref{sec:exclbin}) we assume values of $V_{\mathrm{rot,i}}$ equal to that of the first radius for which we have reliable rotation velocity. We verified, using model velocity profiles that this assumption causes an overestimation in the value of stellar specific angular momentum of at most 2\% with respect to the case in which there is a linear growth of the velocity from $R=0$ to $R=R_{\mathrm{in}}$.\\
The values of $j_{\star,2R_{\mathrm{d}}}$ calculated in this way are reported in the third column of Table~\ref{tab:jem}.\\
The mean ratio, for our galaxy sample, between the angular momentum computed using the extended summation out to $2\,R_{\mathrm{d}}$ and the $j_{\star,\mathrm{RF}}$, calculated in the previous section, is $\sim0.55$. As mentioned this value is also that obtained as the ratio between the angular momentum calculated out to $2\,R_{\mathrm{d}}$ and the total angular momentum for galaxy models characterized by a constant velocity profile, with no trend in this ratio depending on the disc stellar mass. Therefore, this result confirms that $j_{\star,\mathrm{RF}}$ is a good estimator of the total angular momentum. Four galaxies, NGC\,5784, IC\,1652, NGC\,0528, NGC\,08234, have rotation curve data going beyond $2\,R_{\mathrm{d}}$. For NGC\,5784 we have data out to 3.3$\,R_{\mathrm{d}}$, while for the others the data extend out to $\sim 2.5 R_{\mathrm{d}}$. As a further check, for these galaxies we calculate the ratio between $j_{\star}$ obtained using equation (\ref{eq:jsum}) out to the most external radius and $j_{\star,\mathrm{RF}}$. The resulting values for this ratio: 0.86 for NGC\,5784 and $\sim 0.75$ for the others, confirm that the values of $j_{\star,\mathrm{RF}}$ better approximates $j_{\star}$ the more extended the rotation curve. \\
The errors on the values of $\j_{\star,2R_{\mathrm{d}}}$ are estimated using the following expressions:
 \begin{equation}
	\Delta_{j+}=\sqrt{[j(v+\Delta_{v+})-j(v)]^{2}+[\delta j(\delta_{R_{\mathrm{d}}},\delta_{D})]^{2}}
\end{equation}
\begin{equation}
 	\Delta_{j-}=\sqrt{[j(v-\Delta_{v-})-j(v)]^{2}+[\delta j(\delta_{R_{\mathrm{d}}},\delta_{D})]^{2}}
\end{equation}
where $j(v+/-\Delta_{v+/-})$ is the value of $j_{\star,2R_{\mathrm{d}}}$, calculated using equation~(\ref{eq:jsum}) in which each $V_{\mathrm{rot,i}}$ is replaced with the value of the velocity increased/reduced by its 1$\sigma$ uncertainty, whose estimation is explained in Section~\ref{sec:errv}. This is a conservative choice that allows us to take into account that the asymmetric errors on the velocities, as discussed in Section~\ref{rcurve}, are not just the result of random fluctuations but also of systematic errors due to our ignorance about the true inclinations of our discs. The values obtained in this way are then summed quadratically with those obtained considering the formula of error propagation that takes into account uncertainties on $R_{\mathrm{d}}, \delta_{R_{\mathrm{d}}}$, and on distances $D, \delta_{D}$.\\
For NGC\,0774 we compute, as in the previous section, the values of $j_{\star}$ considering the summation of equation~(\ref{eq:jsum}) out to: the luminosity weighted average scale radius (suffix D in Table~\ref{tab:jem}), the scale radius of the internal disc (suffix D1) and that of the external disc (suffix D2).

\subsection{Estimate of stellar masses}  
The estimate of galaxy stellar mass is one of the most challenging tasks in astrophysics, because it requires the knowledge of the stellar mass-to-light ratio, $M_{\star}/L$, which depends on the initial mass function and on the age and metallicity of the composite stellar population that characterizes a galaxy. Similarly to what was done in the previous section we estimate the stellar masses both by making the same assumptions as \citepalias{rf12}, in order to compare with their results, and using more appropriate assumptions to our sample in order to obtain our revised $j_{\star}$-$M_{\star}$ relation.

\subsubsection{RF stellar masses} \label{sec:mref}
We first calculate the total stellar mass, $M_{\star\,\mathrm{tot,RF}}$ under the same assumption of \citepalias{rf12} as:
\begin{equation}
 	M_{\star\,\mathrm{tot,RF}}=M_{\star\,\mathrm{d,RF}}+M_{\star\,\mathrm{b,RF}}=f_{\mathrm{d}}(M_{\star}/L)_{\mathrm{d}}L_{K}+f_{\mathrm{b}}(M_{\star}/L)_{\mathrm{b}} L_{K}
 	\label{mstar}
	\end{equation}
where:
\begin{enumerate}
\item $L_{K}$ is the total \textit{K}-band luminosity obtained from values of apparent magnitude provided by the 2MASS catalogue \citep{2mass}. To convert these apparent magnitudes into absolute ones we use, as \citetalias{rf12}, distances derived from redshifts with a Hubble parameter h=0.73 and corrected for Virgo, Shapley and Great Attractor which differ from those in Table~\ref{nome}\footnote{The distances used in this section, corrected for infall into the Virgo, Shapley and the Great Attractor, are taken from NED.} (column 2 of Table~\ref{tab:mtot}).
\item $f_{\mathrm{b}}$ and $f_{\mathrm{d}}$ are the bulge and disc luminosity fractions. Similarly to RF12, these two quantities are taken from a 2-component decomposition in \textit{r}-band, neglecting the difference between these fractions moving from the $r$ to the \textit{K}-band. Therefore the quantity $f_{\mathrm{b}}$ corresponds to the inclination-corrected bulge-to-total ratio, B/T, reported in column 5 of Table~\ref{tab:bsut}, and $f_{\mathrm{d}}$ is (1-B/T).
\item We adopt, as \citetalias{rf12}, values of $(M_{\star}/L)_{\mathrm{d}}=(M_{\star}/L)_{\mathrm{b}}=1$ in \textit{K}-band.
\end{enumerate}
The values of total stellar masses, $M_{\star\,\mathrm{tot,RF}}$, are reported in column 3 of Table~\ref{tab:mtot}. For comparison,  we report in column 4 the analogous quantities, $M_{\star\,\mathrm{tot,SED}}$, provided by CALIFA \citep{walcher2014} and obtained through SED fitting of five optical bands\footnote{The CALIFA stellar mass catalogue was computed from the SDSS \textit{ugriz} bands using the code of \citet{walcher2008} for the SED fitting and the following cosmological parameters: $H_{0}=70$\,km\,s$^{-1}$Mpc$^{-1}$, $\Omega_{\mathrm{M}}=0.3$, $\Omega_{\mathrm{\Lambda}}=0.7$.}. Since these last quantities can be considered more reliable, we can conclude that the assumptions of \citetalias{rf12} cause a systematic overestimation of stellar masses, as discussed in \citet{fall13}. However, a proper comparison between $M_{\star\,\mathrm{tot,RF}}$ and $M_{\star\,\mathrm{tot,SED}}$ should take into account that these two quantities are calculated considering different cosmological parameters and therefore different distances. The masses $M_{\star\,\mathrm{tot,RF}}$ calculated considering the same distances used to obtain $M_{\star\,\mathrm{tot,SED}}$ are in fact larger only by a factor $\sim1.1$. The values of disc stellar masses, $M_{\star\,\mathrm{d,RF}}$, obtained under the above mentioned assumptions are reported in column 4 of Table~\ref{tab:jem}. The uncertainty estimates are calculated considering the error propagation on the quantities that enter in equation~(\ref{mstar}). The dominant contribution to these errors derives from bulge-to-total ratio uncertainties (column 3 of Table~\ref{tab:bsut}), which have typical relative values of 10\%, reaching up to 25-30\% for UGC\,08234 and UGC\,10905. This corresponds to an uncertainty of 60\% in the disc mass of these two galaxies.

\begin{table*}
	\caption{For each galaxy in our sample (column 1) we report the values of distances (column 2) taken from NED, using a Hubble parameter h=0.73 and correcting the Hubble flow for infall into Virgo Shapley and Great Attractor. These values differ from those reported in Table~\ref{nome} and used for the rest of this work because of different cosmological parameters and different infall corrections. In column 3 we show the values of total stellar masses, $M_{\star\,\mathrm{tot,RF}}$, estimated using the same assumptions of \citetalias{rf12}, as explained in Section~\ref{sec:mref}, in order to directly compare our results with theirs. In column 4 we report, for comparison, the values of total stellar masses from the CALIFA catalogue \citep{walcher2014}. These are derived from SED fitting of optical bands. In column 5 we show the total stellar masses, derived under our assumptions, as explained in Section~\ref{sec:mind}.}
	\label{tab:mtot}
	\begin{tabular}{l l l l l}
	\hline
 	Galaxy & $\mathrm{D_{h=0.73}}$&$\log(M_{\star\,\mathrm{tot,RF}}/M_{\odot})$ &$\log(M_{\star\,\mathrm{tot,SED}}/M_{\odot})$ & $\log(M_{\star\,\mathrm{tot}}/M_{\odot})$\\
& (\textit{Mpc})\\
	 \hline
	NGC\,7671 & 56.4 & 11.09$\pm$0.06 & $10.95^{+0.09}_{-0.09}$  & 10.97$\pm$0.05\\
	NGC\,7683 & 51.0 & 11.05$\pm$0.07 & $11.02^{+0.10 }_{- 0.10}$ & 11.06$\pm$0.04\\
	NGC\,5784 & 80.1 & 11.42$\pm$0.08 &$11.22^{+0.10}_{-0.11}$    & 11.28$\pm$ 0.06\\
	IC\,1652     & 69.4 & 10.82$\pm$0.07 & $10.61^{+0.08}_{-0.09}$   & 10.84$\pm$ 0.04\\
	NGC\,7025 & 70.7 & 11.50$\pm$0.07 & $11.53^{+0.17}_{-0.18}$   & 11.41$\pm$ 0.08\\
	NGC\,6081 & 80.9 & 11.22$\pm$0.07& $11.12^{+0.07}_{-0.09}$    & 11.07$\pm$ 0.06\\
	NGC\,0528 & 64.5 & 11.10$\pm$0.06 & $10.87^{+0.09}_{-0.09}$   & 11.06$\pm$ 0.05\\
	UGC\,08234 & 116.1 & 11.39$\pm$0.07 & $11.13^{+0.11}_{-0.10}$& 11.55$\pm$ 0.08\\
	UGC\,10905 & 114.1& 11.53$\pm$0.08 & $11.61^{+ 0.20}_{-0.05}$& 11.47$\pm$ 0.09\\
	NGC\,0774 & 61.8 & 10.97$\pm$0.07& $10.92^{+0.07}_{-0.09}$    & 10.87 $\pm$ 0.10\\
	\hline
	\end{tabular}
\end{table*}

\subsubsection{Our estimate of the stellar masses} \label{sec:mind}
The use of the same value of stellar-mass-to light ratio for bulges and discs implies that we are assuming that these two components are composed of the same stellar populations. However, as derived in Section~\ref{template} and Appendix \ref{app:b}, the bulges and discs of our sample appear described by different stellar populations. For this reason, in order to obtain a more accurate estimate of the disc stellar masses, which is one of our main goals, we proceed as follows.\\
We use a colour-$M_{\star}/L$ relation, CMLR, based on stellar population synthesis models, to convert the luminosities into stellar masses. Although infrared luminosity would be the best choice to estimate this quantity because it is only slightly affected by recent star formation and dust extinction, we use the \textit{r}-band luminosity, derived from our previous photometric decomposition. The use of SDSS images allow us to take advantage of the deep \textit{r}-band images, as necessary to trace the external surfaces of the discs. To perform this conversion from bulge and disc $L_{r}$ to corresponding stellar masses we use the linear equation of \citet{into} that relates the ($g-i$) colour to log($M_{\star}/L$) in \textit{r}-band with the following expression:
\begin{equation}
	\log{M_{\star}/L}=1.005\,(g-i)-0.652
	\label{color}
\end{equation}
The numerical coefficients in (\ref{color}) are the result of CMLR fitting of population synthesis models with an exponential Star Formation History and with a Kroupa IMF. We select the relation between log($M_{\star}/L$) and ($g-i$), because this colour is considered as the most robust optical $M_{\star}/L$ indicator \citep[e.g.][]{gallazzi, into}. In principle, the application of equation~(\ref{color}) with different ($g-i$) for bulges and discs would allow us to obtain different log($M_{\star}/L$) for these two components. However, since we do not have colours measured separately for the bulge and the disc components, we assume values of ($g-i$) typical of bulges and discs of S0 galaxies, equal to 1.19$\pm$0.04 and 1.06$\pm$0.04 respectively. We obtained these two values as averages of the colours of bulges and discs of 20 S0s in the CALIFA sample \citep{mendez15}, derived by applying a 2D photometric decomposition to SDSS-DR7 images. To make sure that we are not biased by low number statistics we check that they are in agreement with values found by different authors on a more extended S0 sample of $\sim\,200$ galaxies \citep[e.g.][]{head14}. To summarize, the application of equation~(\ref{color}) with ($g-i$) typical of S0 bulges and discs, combined with our measurement of $L_{r}$ for the two components, allows us to obtain bulge and disc stellar masses. The disc masses are shown in Table~\ref{tab:jem} (column 5), compared to those calculated under \citetalias{rf12} assumptions, $M_{\star\,\mathrm{d,RF}}$ (column 4). The values of total stellar masses, calculated as the sum of those of bulges and discs, are reported in column 5 of Table~\ref{tab:mtot}. For further discussion about the values of our stellar masses see Appendix \ref{app:mass}.\\

\subsection{$j_{\star}$-$M_{\star}$ diagram: comparison with other morphological types}\label{sec:jmdiagram}
\begin{figure}
	\includegraphics[width=\columnwidth]{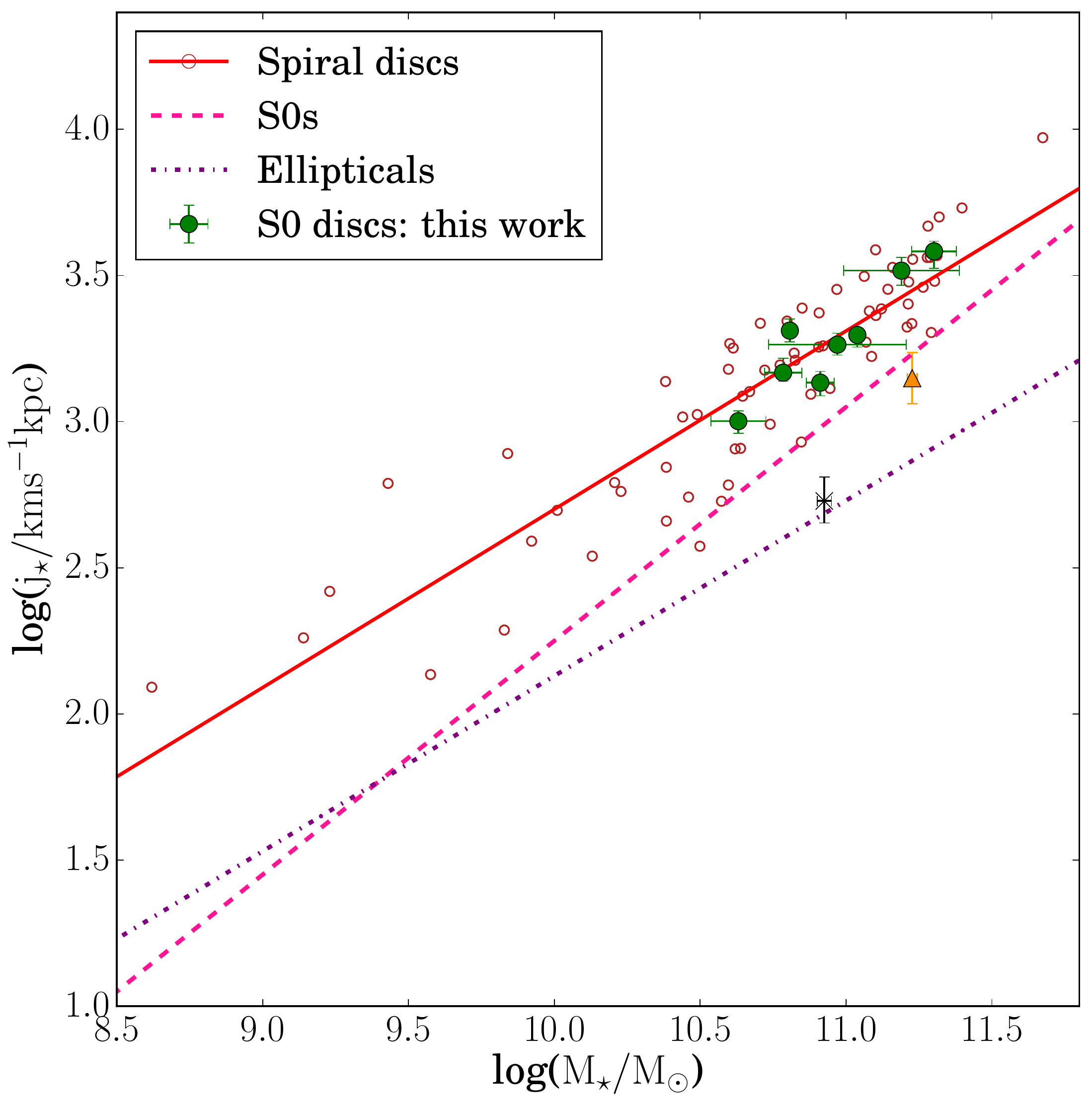}
	\caption{Distribution of the ten S0 discs of our sample in the $j_{\star}$-$M_{\star}$ log-space. Eight of our galaxies are represented by green circles, while for the two problematic cases we use an orange triangle (NGC\,5784) and a black cross (NGC\,0774). For NGC\,0774 we show the value of $j_{\star,\mathrm{RF}}$ indicated with suffix D in Table~\ref{tab:jem} (see discussion in Section~\ref{sec:jind}). The values of $j_{\star}$ are those obtained using the quantities $j_{\star,\mathrm{RF}}$ (column 3 of Table~\ref{tab:jem}, for explanation of their derivation see Section~\ref{sec:jind}), while for the stellar disc masses we refer to $M_{\star\,\mathrm{d,RF}}$ (column 4 of Table~\ref{tab:jem}, for explanation of their derivation see Section~\ref{sec:mind}). The lines represent the best-fit relation for $j_{\star}$-$M_{\star}$ found by \citetalias{rf12}. The red line, with a slope of $\sim 0.6$, refers to all spiral discs (Sa, Sb, Sc, Sd, Sm); the pink dashed line, with a slope of $\sim 0.8$, was found for S0s (without separation of their bulge and disc components), while the purple dot-dashed line, with a slope of $\sim 0.6$, is the best-fit for elliptical galaxies. The red empty circles represent the spiral discs from \citetalias{rf12}.}
	\label{ref}
\end{figure}

The derivation of $j_{\star,\mathrm{RF}}$ (Section~\ref{sec:jref}) and of disc stellar masses ($M_{\star\,\mathrm{d,RF}}$, Section~\ref{sec:mref}) allows us to compare the distribution of our S0 discs with the relation between these two quantities derived by \citetalias{rf12} for different morphological types. In Fig.~\ref{ref} we show the best-fit found by these authors for different subtype spiral discs, S0 and elliptical galaxies. The relation shown for spirals (red line) in Fig.~\ref{ref} refers to their disc components, while the dashed pink and dot-dashed purple lines for S0s and ellipticals refer to the best-fits for total $j_{\star}$-$M_{\star}$, without taking into account the decomposition in disc and bulge. Even if there is a general consensus that a significant fraction of early type galaxies consist of a spheroid and a disc-like component \citep[e.g.][]{cap16}, an extended analysis of them and of their position in the $j_{\star}$-$M_{\star}$ plane is not present in the literature. A rough analysis of a sample of ellipticals was reported in \citetalias{rf12}, but the uncertainties in the decomposition did not allow those authors to draw a clear conclusion. From Fig.~\ref{ref} we can see that eight of our ten S0 discs have a distribution in the $j_{\star}$-$M_{\star}$ plane that is in full agreement with those of spiral discs. Furthermore, their scatter is fully compatible with the intrinsic scatter $\sigma_{\log j_{\star}}=0.17$ found by \citetalias{rf12}, as showed also by the red points in Fig.~\ref{ref} that represent the spiral discs in \citetalias{rf12}. This result implies that spiral and S0 discs are dynamically similar, differing more in their morphological features mostly related to the presence or absence of star formation activity.\\
We note that that these results are unchanged if we consider $M_{\star}/L$ ratios smaller than 1, as done in \citet{fall13}. To verify this we compute the S0 disc stellar masses using a range of $M_{\star}/L$ values, from $M_{\star}/L=0.62$, typical of Sa/Sab discs \citep[e.g.][]{fall13, portinari04}, to an extreme value of 0.84. This later was obtained as a median for elliptical galaxies by \citet{fall13} and it should be considered an upper limit for S0 discs.  With the values of the stellar masses obtained using this range of $M_{\star}/L$, we compute the respective values of $j_{\star}$, from the best-fit parameters of \citet{fall13}  spiral discs $j_{\star}$-$M_{\star}$ relation. The relative residuals between these latter and the values of  $j_{\star}$ found in this work are smaller than 7\% for the eight normal discs, while for the outliers the relative residuals reach a value of 20\%, confirming the results discussed previously.\\
In Section~\ref{sec:discussion} we discuss the implication of this result for S0 formation scenarios, which are briefly overviewed in Section~\ref{sec:mech}. Our interpretation of the reasons for which two galaxies, NGC\,5784 (orange triangle in Fig.~\ref{ref}) and NGC\,0774 (black cross in Fig.~\ref{ref}), show a clear incompatibility with the rest of our galaxy sample will also be given in Section~\ref{sec:discussion}. 

\section{Discussion}
\label{sec:discussion}
\subsection{How different mechanisms change the position of discs in $j_{\star}$-$M_{\star}$ plane}\label{sec:mech}
As already discussed in the introduction a number of mechanisms have been proposed to explain the transformation of spirals into S0s.  Some of these (e.g. minor or major merging, \citet{que, falcon15} tidal encounters, \citet{bekki11}), cause a partial or total disruption of the stellar discs, while others act mostly on the gas, removing it and causing little or no disruption of the stellar discs. Among these second types, the traditional commonly-invoked processes are ram-pressure stripping \citep[e.g.][]{que}, which removes the cold gas from the galaxy's disc, or starvation \citep[e.g.][]{boselli06} which removes the hot halo gas reservoir. These mechanisms should be less important for S0 formation in low density environments, as studied in this work, so tidal encounters, mergers \citep{que} or the effects of internal processes (e.g. active galactic nuclei, weak magnetic field, \citet{vandenbergh09}) are considered the most likely process for S0 formation in galaxy groups and in the field. However, to date, it is unclear whether or not a galaxy with a morphology and properties typical of an S0 could be the result of merger processes \citep[e.g.][]{laur10, bellstedt}. \\
The analysis of the position of S0s in the $j_{\star}$-$M_{\star}$ diagram could reveal the relative importance of the different processes considered plausible for their formation, because each of these is associated with a characteristic change of $j_{\star}$ and/or $M_{\star}$. The major contribution to the change of $j_{\star}$ should be caused by a change in the rotation velocity and not to a change in the structure of the disc: \citet{laur10} showed that the scaling relations between the sizes of bulges and discs in S0s are similar to those of the spirals. The average disc scale length of the discs of our sample, $\sim$4.3 kpc, is comparable to that  of spirals studied in \citetalias{rf12}, $\sim$4.2 kpc.\\
In this section we describe how the three mechanisms belonging to the violent scenario (tidal interaction, major and minor merger between spirals) lead to a change of the position of a disc in the $j_{\star}$-$M_{\star}$ plane.\\

\begin{figure}
	\includegraphics[width=\columnwidth]{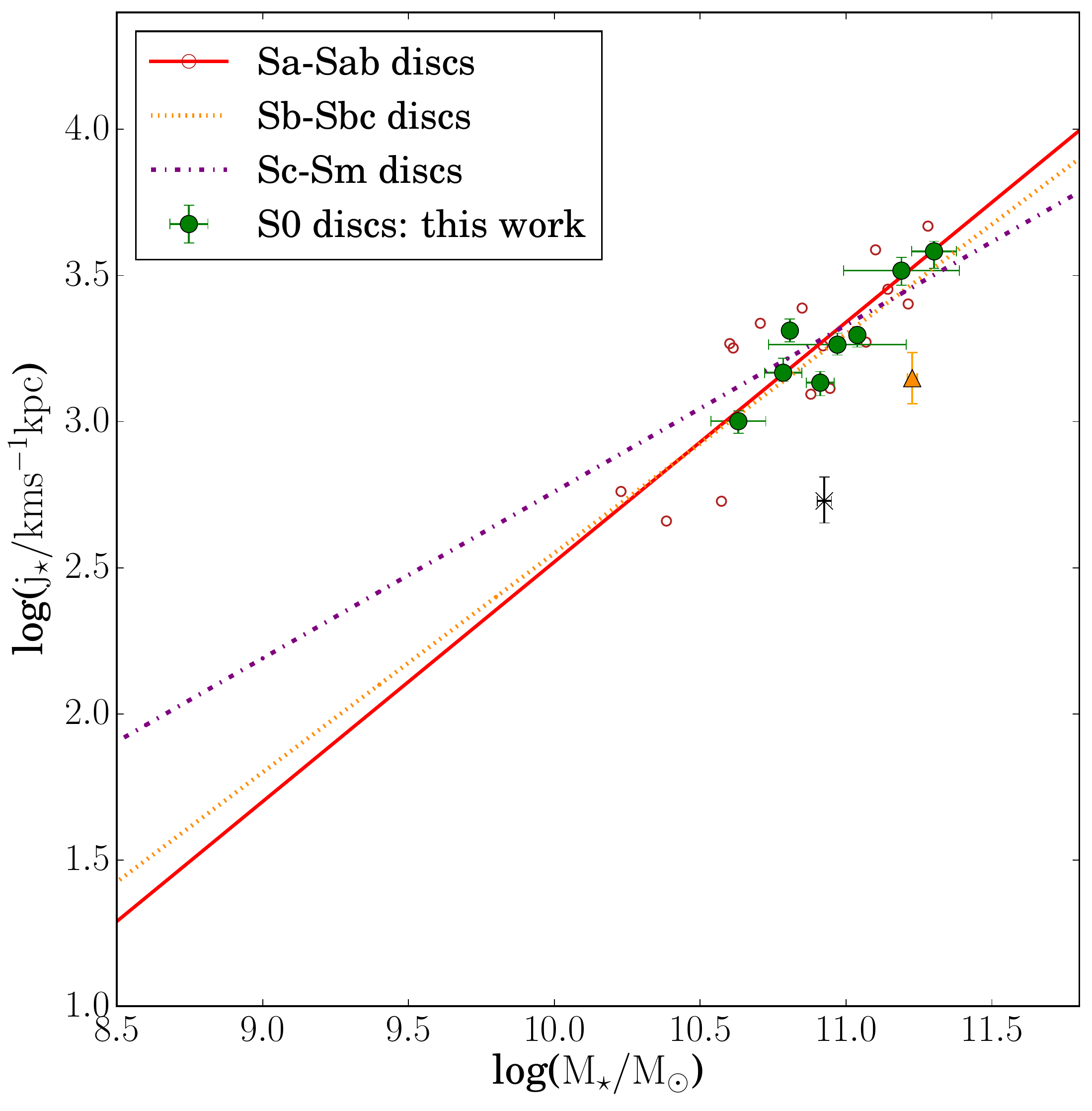}
	\caption{The same as Fig.~\ref{ref}, but now we show the best-fit relation between disc specific angular momentum and disc stellar mass for different spiral subtypes. The red line, with a slope of $\sim$0.82, corresponds to Sa and Sab discs; the orange, with a slope of $\sim$0.75, is for Sb and Sbc discs, while the purple line, with a slope of $\sim$0.57, corresponds to Sc and Sm discs. The red empty circles represent the Sa and Sab discs from \citetalias{rf12}.}
	\label{ref_sa}
\end{figure}

\begin{enumerate}
\item \textbf{Tidal interactions}\\
Simulations \citep[e.g.][]{bekki11} showed that S0s formed by tidal interactions in groups have kinematical properties that are significantly different from those of spirals. Their discs show maximum rotation velocity up to 50\% lower than those of their spiral progenitors, owing to the tidal heating. The mass of these S0 discs is not larger than the parent spiral discs: although gas is converted efficiently into stars during S0 formation, a fraction (10\%-30\%) of the initial disc stars is stripped, so that the balance between these two processes results in a negligible mass change. Therefore we expect that if S0s were the result of tidal interaction, they should have values of disc $j_{\star}$ systematically lower than those of spirals at fixed stellar mass.
\item \textbf{Major mergers}\\
It is not fully understood how discs survive or re-form after major mergers: some hydrodynamical simulations showed that major mergers are capable of strongly altering the morphologies of discs, transforming them into remnants with typical properties of elliptical galaxies \citep[e.g.][]{bois11, moody}. Other simulations \citep[e.g.][]{governato07, rob, moody} suggested that even major mergers can leave remnants with a disc structure. In the attempt to reproduce rotationally supported galaxies, akin to the observed systems, these studies used different feedback prescriptions, such as star-formation feedback \citep{rob} or galactic wind feedback \citep{governato07}. This latter, enables, for example, disc survival by redistributing gas out to large radii, preventing in this way a total angular momentum loss and so allowing the formation of discs after mergers. Other works \citep{hopkins09} suggested that in the case of gas-rich major mergers it is possible to obtain a disc remnant, irrespective of feedback processes. \citet{hopkins09} showed that during a merger between two gas-rich disc galaxies, the stellar and gas component are subjected to different effects: the stellar disc is partially or totally destroyed as a result of violent relaxation in the collision. The fraction of the initial gas that has not been subjected to internal torques driven by the galaxy interaction sees the central potential rapidly relax and having conserved its angular momentum, rapidly cools and reforms a thin, rotationally supported disc. The gas in the inner region falls into the galaxy centre and transforms into stars. Irrespective of the detailed mechanisms that shape the disc-remnant, at the end of this violent process, the galaxy will have lower $j_{\star}$ than its spiral progenitors, because the stellar disc is destroyed and only a small fraction of gas, whose quantity depends on the details of the relative orbits of the progenitors \citep[e.g.][]{bois11, hopkins09}, retains its initial value of angular momentum.\\
A recent study \citep{lagos2} found that both a dry and a wet major merger can reduce the value of $j_{\star}$  in the outer regions (that contribute more to the value of $j_{\star}$) by $40\%$.
\item \textbf{Minor mergers}\\
Minor mergers do not destroy entirely the discs \citep{bour05}, but according to $\Lambda$CDM, they are more frequent than major mergers. A sequence of minor mergers, each characterized by a random vector change in the value of specific angular momentum, cumulatively has an effect similar to or larger than that of a major merger, both in terms of the final angular momentum \citepalias{rf12} and of the destruction of the stellar disc \citep{bour07}. A recent study \citep{lagos2} found that wet and dry minor mergers have different results on the $j_{\star}$ of the remnants. Dry mergers reduce$j_{\star}$ by $30\%$, while wet mergers can increase $j_{\star}$. However this increase of $j_{\star}$ influences the central regions (bulge) while the outer regions are not influenced in the case of co-rotating mergers or are subjected to a reduction of $j_{\star}$ in the case of perpendicular or counter-rotating mergers. Since the co-rotating mergers are less common than the perpendicular and counter-rotating mergers and the probability to have a  sequence of only co-rotating mergers is lower than randomly-oriented mergers, we can conclude that the cumulative effect of minor mergers leads to a reduction of the disc $j_{\star}$. We also note that in the above studies \citep{bour05, lagos2} the merger stellar mass ratios have values between 0.1 and 0.4.
\end{enumerate}
The indication that we derive from the distribution of our S0 discs in the $j_{\star}$-$M_{\star}$ plane (Fig.~\ref{ref}) is that all the above mentioned violent processes are not the dominant mechanisms that transform spirals into S0s. The position of our S0 discs in $j_{\star}$-$M_{\star}$ diagram, indeed, is fully compatible with the kinematic properties of the discs of spiral galaxies. There is no systematic shift towards lower values of $j_{\star}$, caused by tidal interactions or mergers. Eight of our ten S0s can be considered as faded spirals, namely as a result of the quenching of their star formation and of the consecutive fading of the spiral arms \citep{carollo}. To our knowledge, this passive fading is the only mechanism that does not change the disc $j_{\star}$ and $M_{\star}$ values. Only two galaxies, NGC\,5784 and NGC\,0774, are located systematically below the others. By inspecting their SDSS images and/or bulge-disc decomposition we could have expected a similar discrepancy for these two galaxies with respect to the rest of our sample. The image of NGC\,5784 reveals, indeed, that it is in an interaction phase, probably with a companion visible in Fig.~\ref{n5784_tot} and in addition its dust lanes reveal signs of recent star formation. NGC\,0774 is the galaxy for which the addition of a second disc to a simple two component decomposition was necessary. As discussed in Section \ref{sec:fitting process} this upbending profile is usually explained as a result of recent interactions \citep{erwin05}.\\
Thus, it is possible that these two galaxies have indeed become S0s as a consequence of some violent process. Alternatively, their kinematics/morphology could be too disturbed to allow for an accurate determination of $j_{\star}$.\\
The mechanisms for the quenching of the star formation or the interruption of the gas supply are still a matter of debate. Cosmological models predict that galaxies with virial masses $M_{\mathrm{vir}} \gtrsim 10^{12} \mathrm{M}_{\sun}$ at $z$\,$\lesssim1$ should accrete gas through the so-called hot-mode accretion \citep{keres, dekel}. This accretion mode generates an extended corona \citep{miller} that can eventually feed the star formation in the disc through cooling \citep[e.g][]{fraternali17}. However, \citet{arm} claimed that the cooling of a galactic corona becomes less efficient for increasing virial mass of the dark matter halo $M_{\mathrm{vir}}$. Their model suggest that all disc galaxies characterized by $M_{\mathrm{vir}}\gtrsim3\times10^{12}$, typical of Sa types, could be considered as plausible progenitors for S0s. The shut down of the cold gas supply could be explained also if, for example, the progenitor spiral is an early type (Sa-Sb) that has less cold gas \citep{catinella} with respect to the other spirals. The stellar masses of our sample are indeed typical of Sa-Sb galaxies \citep{gonzalez15}. \\
The average $r$-band bulge-to-total ratio for the S0s in our sample is $\sim0.41$, value typical of Sa galaxies. However, the comparison of luminosity B/T ratio between S0 and spiral galaxies should not be considered as a 1:1 relation. The fading of the disc once star formation ceases increases the luminosity B/T of the quenched galaxy with respect to the star-forming one, even if the masses of the disc and the bulge do not change. \citet{carollo} showed, for example, that the fading of a disc that has been quenched 1 Gyr ago leads to galaxies with initial B/T$\sim0.24$ to become B/T$\sim0.44$ in $I$ bands. Considering that $r$-band and $I$-band have similar wavelength range, we can speculate that a quenched galaxy with B/T$\sim0.4$ is the result of a spiral with B/T$\sim0.2$ (typical of an Sb galaxy).
In principle, these scenarios could be checked against the position of our S0 discs with respect to the different subtypes of spiral discs (Fig.~\ref{ref_sa}). However given that there is a negligible distinction, in particular in the mass range studied, between the$j_{\star}$-$M_{\star}$ relations of the Sa/Sab, Sb/Sbc and Sc-Scm discs, we cannot conclude that our distribution is more compatible with the Sa-Sab distribution (Fig.~\ref{ref_sa}) than with the general relation for spiral discs (Fig.~\ref{ref}).

\subsection{Comparison to other work}
\citet{math} found previous kinematic evidence of a strong link between S0s and spirals: their S0 sample follows a Tully-Fisher relation (TFR), which is offset from the relation for normal spiral galaxies by the amount that one would expect if star formation had been simply shut off a few Gyrs ago. \citet{laur10} found a tight scaling relation between the sizes of bulges and discs in S0s, similar to those of spirals, and interpreted them as the result of passive evolution, arguing that mergers would destroy that structural coupling. A recent comparison between kinematic properties (e.g. $V/\sigma$, local spin parameter $\lambda_{R}(R)$)  of a sample of S0 galaxies with those from different simulated galaxies \citep{bellstedt} showed that S0s resemble the spiral progenitors more than the merger remnants. This is interpreted as an indication that S0s are not the result of mergers, but of the fading of spiral galaxies. \\
Nevertheless, our result is in contrast with those found by other observational and theoretical (N-body numerical simulations) studies and based on a comparison of spirals and S0s in the $\lambda_{R}$ vs concentration ($R_{90}/R_{50}$) plane \citep{falcon15, que}\footnote{The parameter $\lambda_{R}$, \citep{ems07} is a measure of rotational support of galaxies in the regions inside the effective radius, while ($R_{90}/R_{50}$) is a measure of \textit{r}-band light concentration and it can be considered as a proxy for the relative flux in spheroid and disc component.}. The conclusion of these two studies is that S0s cannot be considered as simply faded spirals that have had their gas removed, because there is an incompatibility between their positions in the $\lambda_{R}$-$(R_{90}/R_{50})$ plane. However, these authors emphasized that, in contrast to other spirals, many Sa galaxies have a distribution similar to S0s in this plane.\\
The discrepancy between our result and those found by these studies could be ascribed to the fact that the $\lambda_{R}$ parameter does not allow one to measure scales that are important for a true estimate of the physical angular momentum, as opposed to $j_{\star}$. A clear indication of this comes from the fact that the distribution of slow rotators (which are characterized by low values of $\lambda_{R}$) is not different in the $j_{\star}$-$M_{\star}$ plane from fast rotators (which are characterized by high values of $\lambda_{R}$, \citetalias{rf12}). In addition, the conclusion of \citet{falcon15} and \citet{que} about the difference between spirals and S0s is based on a measure of the parameter $\lambda_{R}$ obtained for the whole galaxy and not on a measure of the dynamical properties of the single subcomponents which is the main novelty of the present work. Finally their results are based on a difference in terms of concentration. Regarding this last quantity, we expect a difference between spirals and S0s in the passive evolutionary scenario described above, because a simple fading in surface-brightness of star-forming discs, once star-formation ceases, causes an increase of the relative importance of bulge flux, without the need for any substantial bulge mass growth caused by merging \citep{carollo}. In addition, the differential fading between discs and bulges cause a change of the flux-weighted $\lambda_{R}$ parameter. If we studied the distribution of our S0 sample in the $\lambda_{R}$-$(R_{90}/R_{50})$ plane we would likely find the same results as \citet{falcon15} and \citet{que}.\\
We conclude by mentioning the study of \citet{cortesi13} who studied a sample of 6 S0s from different environments, analysing both the TFR and the $V/\sigma$ values of their discs. They found that their S0s follow a TFR that is offset to fainter magnitudes and the $V/\sigma$ values are systematically slightly lower ($\sim$4.2) than the typical values of spirals ($\sim$5.2). This latter result led \citet{cortesi13} to the conclusion that S0s formed neither through simple fadings of spirals nor through mergers, but other mechanisms must be invoked. The Tully-Fisher offset could be explained as a result of simple fading, but a stellar mass TFR would be a better answer to this issue. Further investigations with larger samples are required to solve this potential discrepancy with our results.

\subsection{Our independent fit}
\begin{figure}
	\includegraphics[width=\columnwidth]{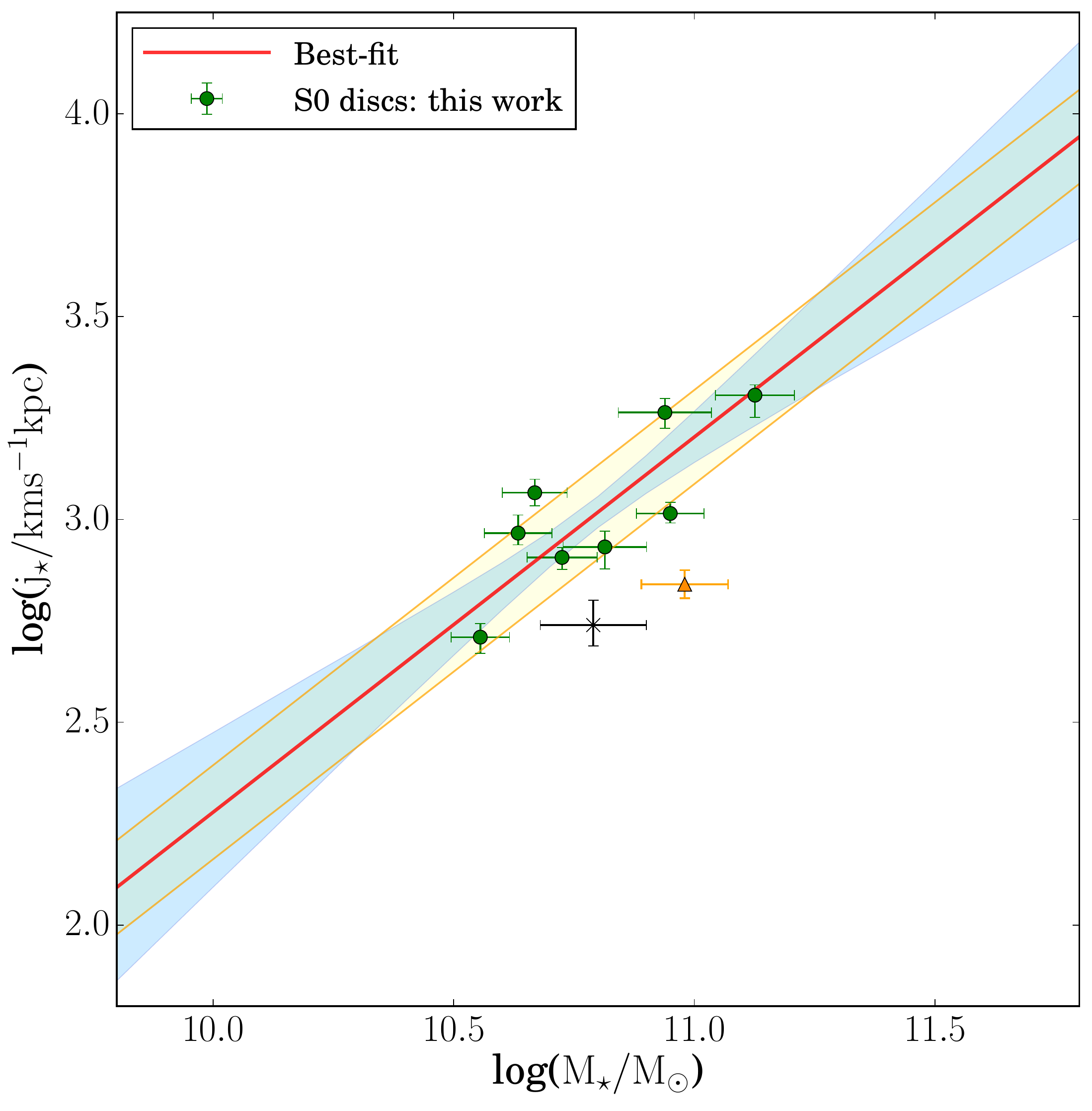}
	\caption{Distribution of the ten S0 discs of our sample in the $j_{\star}$-$M_{\star}$ log-space. Only eight galaxies, those shown with green circles, are considered in the fitting process. The reason for the exclusion of two galaxies, NGC\,0774 (black cross) and NGC\,5784 (orange triangle), is explained in the main text. The values of $j_{\star}$ are those obtained using the equation~(\ref{eq:jsum}) (column 3 of Table~\ref{tab:jem}, for explanation of their derivation see Section~\ref{sec:jind}), while the stellar disc masses are those derived using a colour mass to light relation (column 5 of Table~\ref{tab:jem}, for explanation of their derivation see Section~\ref{sec:mind}). The red line represents the best-fit relation, while the blue band shows the 16th and 84th percentiles; the yellow band represents the best fit relation $\pm$ our estimate of the intrinsic scatter.}
	\label{fig:ref-mc}
\end{figure}

In this section we show the best-fit parameters for the linear relation between disc $j_{\star}$ and $M_{\star}$ in log-space: $\log(\,j_{\star}/\mathrm{kpc\,km\, s^{-1}})=a(\log\,(M_{\star}/\mathrm{M}_{\sun})-10.7) +b$, in the case in which these two quantities are estimated under our assumptions. Therefore we consider the specific angular momentum, $j_{\star,2R_{\mathrm{d}}}$, calculated as described in Section~\ref{sec:jind}, using the equation~(\ref{eq:jsum}) out to $2\,R_{\mathrm{d}}$, and the disc stellar masses, $M_{\star,\mathrm{d}}$, estimated in Section~\ref{sec:mind}, using a colour-mass-to-light relation. We recall that our $j_{\star,2R_{\mathrm{d}}}$ is not the total angular momentum, but can be considered as a scaled version of it (see discussion in Section~\ref{sec:jind}). The two galaxies, NGC\,5784 and NGC\,0774, that have clear signs of interaction, are not considered in the fitting process. \\
The best-fit slope and intercept, found using MCMC Python package EMCEE \citep{foreman} are respectively: $0.91^{+0.23}_{-0.24}$ and $2.93^{+0.04}_{-0.04}$ (red line in Fig.~\ref{fig:ref-mc})\footnote{The likelihood function used to find the slope and intercept has the following expression: $-\frac{1}{2}\sum_{\mathrm{i}}\Big[\frac{(\log{j_{\star \mathrm{i}}}-a\,\log{M_{\star \mathrm{i}}}-b)^{2}}{s_{\mathrm{i}}^{2}}+\ln(2\pi s_{\mathrm{i}}^{2})\Big]$, with $s_{\mathrm{i}}^{2}=\sigma_{\log{j_{\star \mathrm{i}}}}^{2}+b^{2}\sigma_{\log{M_{\star \mathrm{i}}}}^{2}+f^{2}$. The terms $\sigma_{\log{M_{\star \mathrm{i}}}}$ and $\sigma_{\log{j_{\star \mathrm{i}}}}$ take into account the uncertainties on the values of $\log{M_{\star}}$ and $\log{j_{\star}}$. The term $f$ represents an extra scatter, which is introduced to take into account either an underestimation of observational errors or an intrinsic scatter.}. The value of the slope is different from that of $\sim0.6$ found by \citetalias{rf12} as the best-fit value for all spiral discs. It is in agreement with both the value of $0.82\pm0.08$ found by \citetalias{rf12} for Sa-Sab and with the two slope values of $0.96\pm0.07$ and $0.80\pm0.09$  found for Sa/Sb and Sbc/Sd respectively by \citet{cortese16} in the case in which $j_{\star}$ is calculated within one effective radius using SAMI. However, we cannot exclude that the slope value that we obtained could be influenced by the small number of fitted data points and by the small range of stellar masses covered by our sample, so this result has to be confirmed using a larger sample. \\
The blue band in Fig.~\ref{fig:ref-mc} shows the 16th and 84th percentiles obtained from the combination of the posterior probability distribution of the slope and intercept. The yellow band in Fig.~\ref{fig:ref-mc} shows the region of the best-fit relation summed and subtracted to the value of the best-fit intrinsic scatter, $f=0.1$. However, we test that the intrinsic scatter is underestimated in the case of a small sample, as that used in this work.\\
The position of the two galaxies NGC\,0774 and NGC\,5784 in the $j_{\star}$-$M_{\star}$ diagram with respect to the rest of the sample validates our results and interpretation given in the previous sections. They are the two S0s for which one of the mechanisms of the violent scenario could be responsible for their formation. In particular, this appears likely for NGC\,0774, given its position in Fig.~\ref{fig:ref-mc} (black cross). This is the galaxy for which we computed three values of the specific angular momentum (Table~\ref{tab:jem}) because it is characterized by two discs (the third value takes into account a single disc, obtained as a luminosity average of the other two, see Section~\ref{sec:fitting process}). Although the black cross in Fig.~\ref{fig:ref-mc} represents an upper limit, because it shows the value of $j_{\star,2R_{\mathrm{d}}}$-D2 (using the $R_{\mathrm{d}}$ value of the external disc, see Section~\ref{sec:jind}), it is systematically lower than the other S0s with similar mass. We can speculate that the double disc structure is the result of a merger process: the inner one could be the remnant of the old disc, while the outer one is the result of the rebuilding driven by the gas accretion (see Section~\ref{sec:mech}). This would confirm that it is possible to have a disc remnant as a result of a merger process, but its kinematic properties will be different from those of its progenitors.

\section{Concluding remarks}
\label{sec:conclusion}
In this work we studied the stellar kinematics of a sample of ten S0 galaxies. This morphological Hubble types are characterized by properties that encompass multiple distinct classes of galaxies \citep[e.g.][]{capps0, k12}. Their main features are that they have a bulge and a disc, like spiral galaxies, but lack cold gas and, as a consequence, star formation.\\
The formation of S0 galaxies in clusters is well understood and it is explained taking into account for environment-driven processes (e.g. ram pressure stripping, \citet{gunn}, starvation, \citet{larson80}) that remove their gas and quench their star-formation activity. There is not, instead, a general consensus on the physical mechanisms that lead to the formation of S0s in galaxy groups and in the field. This is the main motivation that prompted us to study group and field S0s.\\
In low density environments, the numerous S0 formation mechanisms belong to two main categories: passive, in which the cold gas of the progenitor disc galaxy is slowly removed because of star formation \citep[e.g.][]{vandenbergh09, will, laur10, bellstedt}, or violent, in which, as a result of mergers \citep[e.g.][]{falcon15, que} or tidal interactions \citep[e.g.][]{bekki11}, the galaxy cold gas is rapidly consumed. These two scenarios cause distinctive changes to the properties of a spiral galaxy, in addition to quenching its star formation. Therefore, some of the structural properties that characterize S0 galaxies could provide a record of the physical mechanisms which have triggered their evolution.\\
Most previous work has focused on structural and morphological properties of S0s \citep{laur10, head14}. In this paper we have, instead, focused on the kinematic properties of the stellar components of S0s, using the ten unbarred S0 galaxies from DR2 CALIFA Survey \citep{garcia15}. In particular we studied the stellar specific angular momentum and mass of their discs (Section~\ref{sec:angularmomentum}). This analysis was possible thanks to a development of software that performs a kinematic bulge-disc decomposition (Section~\ref{sec:kinematic}) and by taking advantage of a photometric decomposition (S\'ersic+exponential profiles) of these two components (Section~\ref{section1}). The same technique could be applied in the future to a sample of spiral galaxies.\\
In our analysis, we focused on the distribution of the S0 discs in the plane of specific angular momentum ($j_{\star}$) vs stelar mass ($M_{\star}$). Observations obtained with a number of techniques and instruments \citepalias{rf12} show that different morphological types follow different relations between their $j_{\star}$ and $M_{\star}$. This allowed us to compare the distribution of our S0 discs in the $j_{\star}$-$M_{\star}$ plane with the relations and distributions that characterize spiral discs and ellipticals. This comparison gives important clues to the formation of S0s, since the position of a galaxy in this plane is subject to a characteristic change with respect to its progenitors depending on the physical mechanisms associated to the two formation scenarios (passive or violent). \\
Our main result is that eight out of the ten S0 discs of our sample have a distribution in the $j_{\star}$-$M_{\star}$ plane that is fully compatible with that of spiral discs, both in terms of the relative trends between the two quantities and their scatters. Two S0 discs have values of $j_{\star}$ systematically lower than those of the discs of spiral galaxies. These two outliers are the only two galaxies that show signs of interaction/merger. \\
The implications of our results are that violent processes, such as mergers or tidal interactions, can be ruled out as the dominant processes in S0 formation in low-density environments. Our results on the kinematics of S0 discs suggest that a passive scenario for the transformation from spiral galaxies to S0s is the most plausible. In this passive evolutionary scenario a spiral galaxy is transformed into an S0 because of the consumption of the cold gas reservoir. The exhaustion of the fuel for sustaining the star formation causes the fading and reddening of the disc, as well as the disappearance of its spiral arms. As a result, this quenched spiral galaxy is classified as an S0 and ends up on the red sequence.

\section*{Acknowledgements}
\addcontentsline{toc}{section}{Acknowledgements}
We thank an anonymous referee for a very careful report and for his/her comments. We thank Michele Cappellari for making his code public and for helpful suggestions. We thank also Luca Cortese, Miguel Querejeta and Eric Emsellem for providing stimulating discussions and comments. This study makes use of the data provided by the Calar Alto Legacy Integral Field Area (CALIFA) survey and of the Sloan Digital Sky Survey data. The research has made use of the NASA/IPAC Extragalactic Database (NED) which is operated by the Jet Propulsion Laboratory, California Institute of Technology, under contract with the National Aeronautics and Space Administration.

\bibliographystyle{mnras.bst}
\bibliography{bibliography_nb}

\appendix
\section{From $\lambda$ to $j_{\mathrm{DM}}$}\label{app:lambda}
The spin parameter for a DM halo is expressed in terms of its total energy $E$, angular momentum $J$ and total mass $M$ using the expression \ref{eq:lj}.\\
We consider a virialized dark halo with a Navarro, Frenk and White density profile given by
\begin{equation}
\rho(r)=\rho_{\mathrm{crit}}\frac{\delta_{0}}{(r/r_{\mathrm{s}})(1+r/r_{\mathrm{s}})^{2}}
\end{equation}
with \begin{equation}
\delta_{0}=\frac{200}{3}\frac{c^{3}}{\ln(1+c)-c/(1+c)}
\end{equation}
$c=r_{\mathrm{vir}}/r_{\mathrm{s}}$ is the halo concentration factor, $r_{\mathrm{s}}$ is a scale radius, $\rho_{\mathrm{crit}}=3H^{2}/8\pi G$ is the critical density. $r_{\mathrm{vir}}$ is the limiting radius of a virialized halo, assumed to be that within which the mean mass density is $\Delta_{\mathrm{vir}}$ times the critical density. Under the assumption that all particles are on circular orbits and using the virial theorem \citep{mo10, vandenbosch}, the total energy $E$ of the truncated halo can be expressed in terms of the kinetic energy $K$ as
\begin{equation}
E=-K=-4\pi\int_{0}^{r_{\mathrm{vir}}}\frac{\rho(r) V_{\mathrm{c}}^{2}(r)}{2}r^{2}dr=-\frac{M V_{\mathrm{vir}}^{2}}{2}f_{c}
\label{etot}
\end{equation}
with $V_{\mathrm{vir}}=(GM/r_{\mathrm{vir}})^{1/2}$ and 
\begin{equation}
f_{\mathrm{c}}=\frac{c}{2}\frac{[1-1/(1+c)^{2}-2\ln(1+c)/(1+c)]}{[c/(1+c)-\ln(1+c)]^{2}.}
\end{equation}
Since we have defined the mass of the dark halo so that the average overdensity of the halo is $\Delta_{\mathrm{vir}}$ times the critical density, we can find a relation between $M$ and $V_{\mathrm{vir}}$,
\begin{equation}
M=\frac{4 \pi}{3} (r_{\mathrm{vir}})^{3} \Delta_{\mathrm{vir}} \rho_{\mathrm{crit}}=\frac{4 \pi}{3} \left(\frac{GM}{V_{\mathrm{vir}}^{2}}\right)^{3}\Delta_{\mathrm{vir}} \frac{3H^{2}}{8\pi G}, 
\end{equation}
so that the expression for $V_{\mathrm{vir}}$ is
\begin{equation}
V_{\mathrm{vir}}=\left(M\,\sqrt{\frac{\Delta_{\mathrm{vir}}}{2}}GH\right)^{1/3}.
\end{equation}
Finally we express the total energy in \ref{etot} in terms of $M$ as:
\begin{equation}
E=-M^{5/3}\frac{f_{c}}{2}\Big(\frac{\Delta_{\mathrm{vir}}}{2})^{2/3} (GH)^{2/3}
\end{equation}
Substituting this expression in \ref{jfin} the dependence of $j_{\mathrm{DM}}=J/M$ on $M$ becomes:
\begin{equation}
j_{\mathrm{DM}}=\frac{\lambda G M^{3/2}}{|E|^{1/2}}=\lambda M^{2/3} \Big(\frac{2}{f_{c}}\Big)^{1/2}\Big(\frac{2}{\Delta_{\mathrm{vir}}}\Big)^{1/3}\Big(\frac{G^{2}}{H}\Big)^{1/3}=\lambda M^{2/3}\alpha
\label{jfin}
\end{equation}
where $\alpha$ contains the constant terms, which enter in equation~(\ref{jfin}).

\section{Photometric decomposition: results on individual galaxies}\label{app:a}
\begin{figure*}
	\includegraphics[width=16cm]{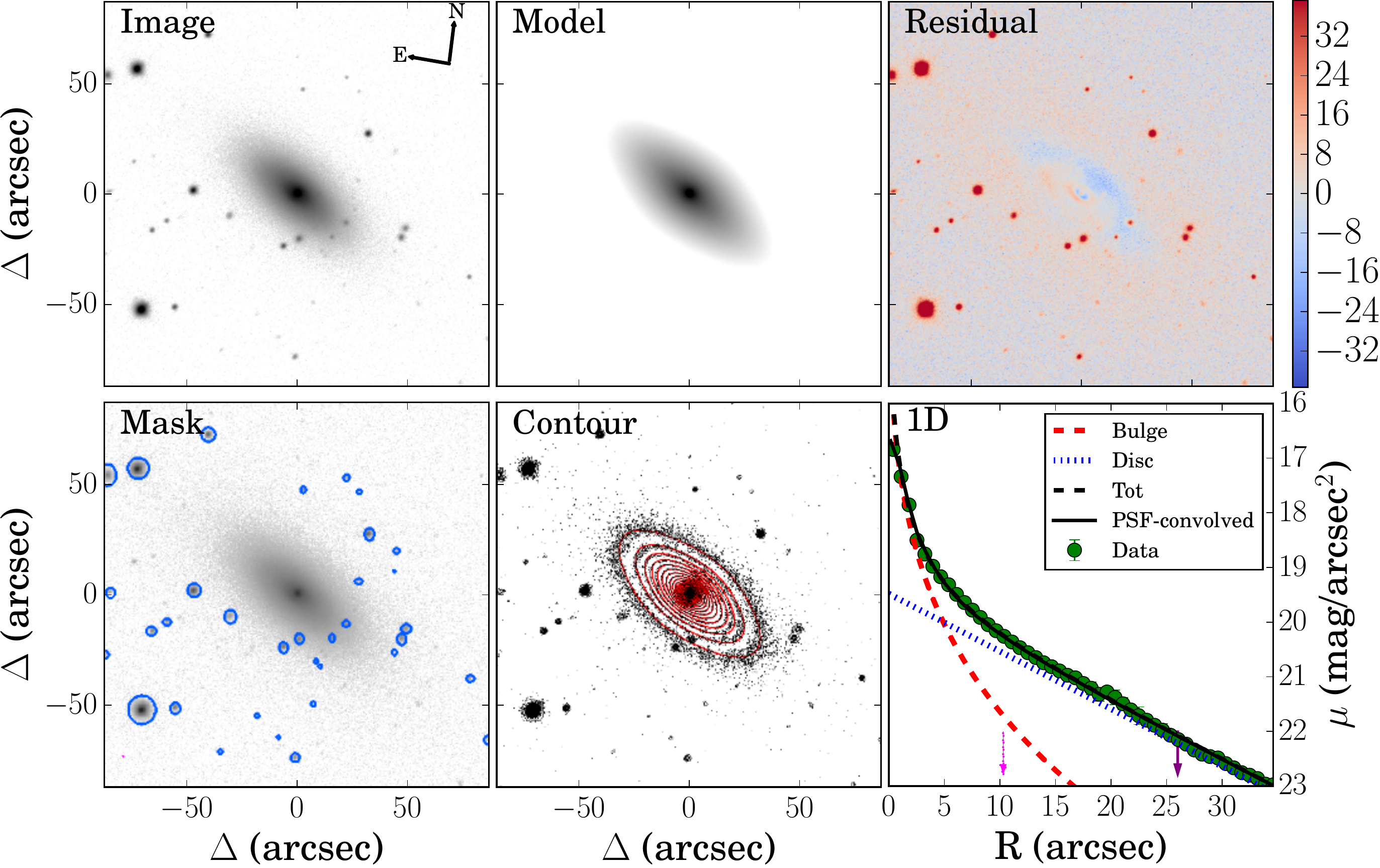}
	\caption{NGC\,0528: same as in Fig.~\ref{n08234}. The north direction is at -7.5$^\circ$. }
	\label{n0528}
\end{figure*}

In this section we show the bulge-disc photometric decomposition for each galaxy of our sample. The results of this analysis, obtained with the technique described in Section~\ref{section1}, are shown in Fig.~\ref{n7683}-\ref{n0774}. In each figure the upper panels show (from left to right) the SDSS image, the model obtained with GALFIT, and a relative residual map. The latter is obtained considering the quantity (data$-$model)/data$\cdot$100 for each pixel. The lower panels show (from left to right) the mask, created as explained in Section~\ref{sec:maskphoto}, the superimposition of contour levels of data and model and the 1D surface brightness profile. In this plot the points represent the mean of the fluxes extracted from the opposite sides with respect to the centre of the galaxy along the major axis of the disc component. This extraction is done in a wedge of 10$^\circ$ around the major axis. We show indicative error bars calculated as one fourth of the flux difference between these two sides, expressed in magnitudes. These error bars can be considered as a measure of the uncertainties due to asymmetries along the major axis and so they should not be confused with those used by GALFIT to perform the fitting. The coloured lines are profiles of individual components, while the black dashed line is the sum of them. All of these lines represent seeing-free profiles. The black solid line is obtained from the total model that fits the observed profile. The flux for this profile, which is a convolution of the intrinsic profile with the PSF, is extracted using ELLINT in the 2D model and imposing the same geometric conditions applied to extract the data. For every galaxy we show only a fraction of the 1D profile, out to a distance of $3.36\,R_{\mathrm{d}}$. The reason for this choice is discussed in Section~\ref{sec:angularmomentum}. We indicate the disc scale length with a dashed magenta arrow in the 1D profile, while the largest radius for which we have kinematic data is shown with a purple arrow. 

\subsection{NGC\,7671} \label{ngc7671_photo}
\begin{figure*}
	\includegraphics[width=16cm]{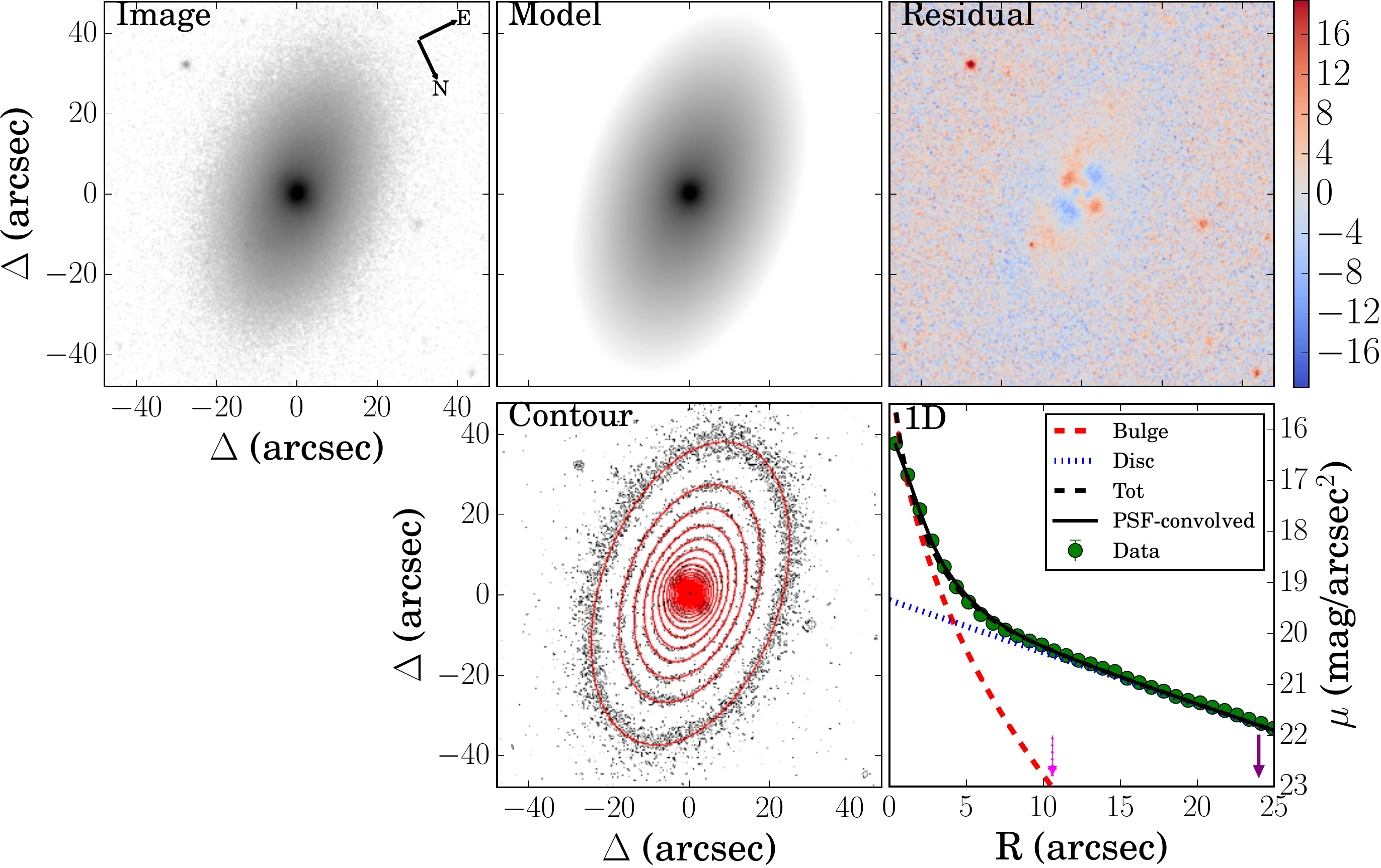}
	\caption{NGC\,7671: same as in Fig.~\ref{n08234}. The north direction is at -153.7$^\circ$. In this galaxy there is no need for a mask. }
	\label{n7671}
\end{figure*}

In this galaxy we fix the position angle of bulge and disc to the same value after verifying that this constraint does not affect the parameters of the two components with respect to the model in which these are let free to vary. The residuals show a symmetric pattern in the inner region (see upper-right panel of Fig.~\ref{n7671}) and this is a clue, as suggested by \citet{peng2002}, to the presence of other subcomponents, as for example a bar. Using a more complicated model, obtained adding a Sérsic profile to two component model, the residuals do not seem to have a systematic appearance. Nevertheless, the interpretation of the two Sérsic profiles in terms of a bulge and a bar is not straightforward since a mix of typical features of both is present in the parameters which describe the two Sérsic functions. Comparing the structural parameters of the two and three-component model we find that those of the bulge are more affected by the presence of subcomponents because the bulge model in the simpler fit is a compromise between the various subcomponents of the galaxy and it reflects neither one perfectly. On the other hand, the parameter values of the disc are not significantly different from those of the simpler model. For this reason we assume the two-component one as the best-fit model for the surface brightness of this galaxy, aware that in this case the bulge parameters are affected by model-dependent measurement uncertainties. The change of the parameters due to the use of different models is suggested as a method to estimate errors \citep{peng2010}. However, to follow a consistent method for all galaxies in Table~\ref{tab:photo} we show the errors due to the sky value uncertainty.\\
This galaxy was photometrically decomposed into its bulge and disc components by \citet{tabor}. The values of the recovered parameters for the disc are consistent with those found in our analysis. However there is a difference in the value of the bulge Sérsic index, equal to 0.99 in \citet{tabor}, while we found a value of 2.34. To test this discrepancy we run GALFIT fixing the values of the geometrical parameters, the effective radius and the Sérsic index for the bulge and the scale length for the disc to those reported in \citet{tabor} and leaving free only the surface brightnesses $\mu_{\mathrm{e,b}}$ and $\mu_{\mathrm{0,d}}$. The resulting reduced chi-square is increased from 1.215 to 1.953, and the visual inspection of the contours levels reveals a clear worsening of the fit, especially in the outer regions. We can ascribe this difference of the recovered values of the bulge to a different wavelength range in which the decomposition is performed. In particular the low value of the Sérsic index could indicate a possible contamination of the blue light from the disc; indeed for this galaxy \citet{tabor} found a bulge older than its disc. Another cause of discrepancy between the values of the parameters recovered could be the smaller field of view in \citet{tabor} and a different PSF with a FWHM$\sim$2.2 arcsec in \citet{tabor} and $\sim$1.3 arcsec in this work. As shown in \citet{johnston16} both a smaller field of view and a poorer spatial resolution can affect the values of the Sérisic index and the effective radius. However, this difference does not influence our results given that the disc parameters are comparable in the two studies, as well as the derived disc velocity fields (Section~\ref{resgal}).

\subsection{NGC\,7683}
\begin{figure*}
	\includegraphics[width=16cm]{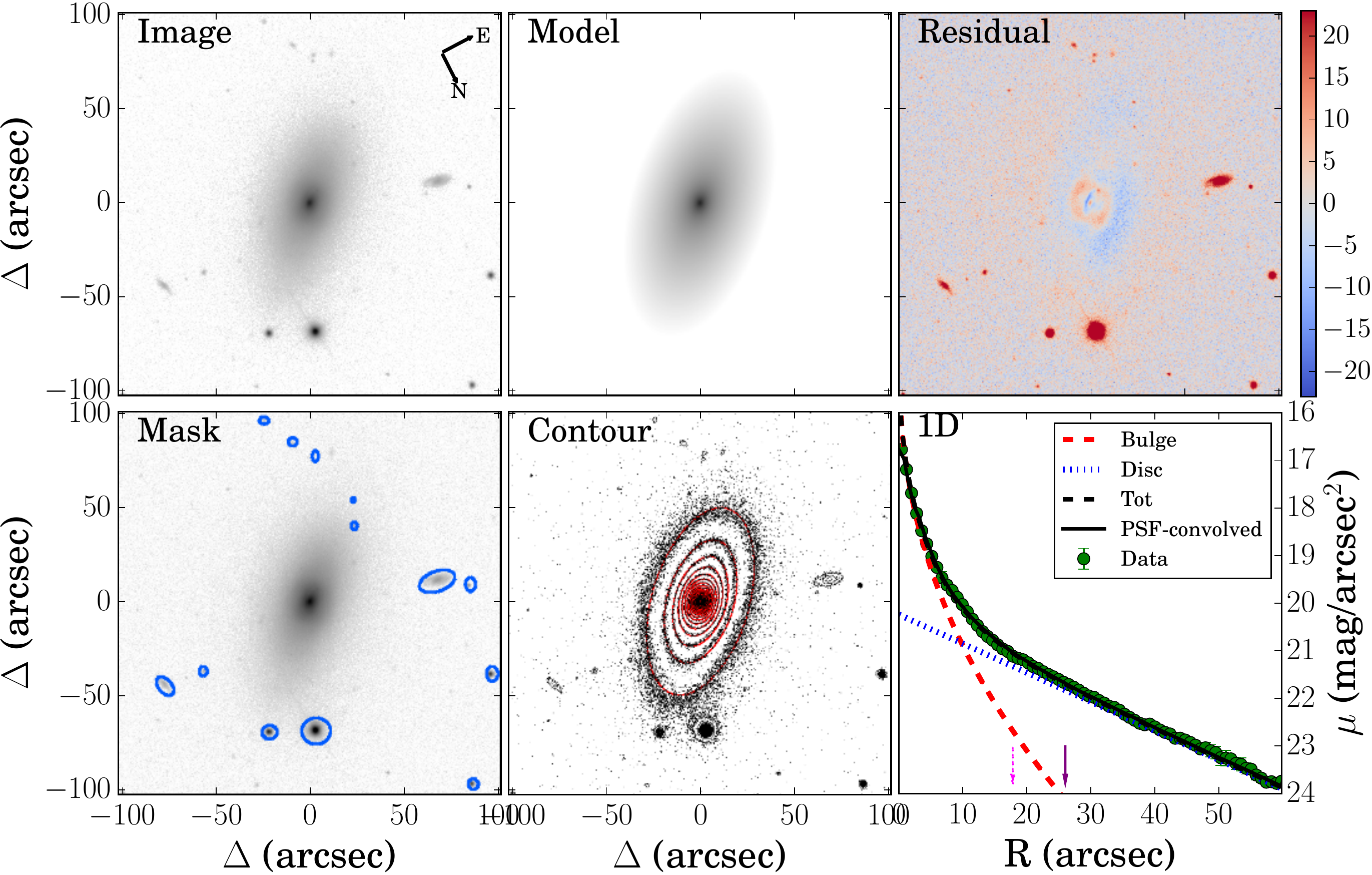}
	\caption{NGC\,7683: same as in Fig.~\ref{n08234}. The north direction is at -153.8$^\circ$.}
	\label{n7683}
\end{figure*}

The image of this galaxy is contaminated by stars and/or background galaxies. For this reason we had to mask them prior to the fit (lower left panel of Fig.~\ref{n7683}). Pixels contained in the mask are not included in the $\chi^{2}$ calculation. After a first run in which the position angle of the bulge and disc are let free to vary, we obtain that the difference between these geometric quantities is less than 1$^\circ$ in the 2 components. For this reason in the successive runs we keep the two position angles fixed at the value of the disc. The relative residuals reach values of $\sim$10\% at the centre where dust lanes are present, but the fit is very good further out. 

\subsection{NGC\,5784}\label{sec:n5784}
\begin{figure}
	\includegraphics[width=\columnwidth]{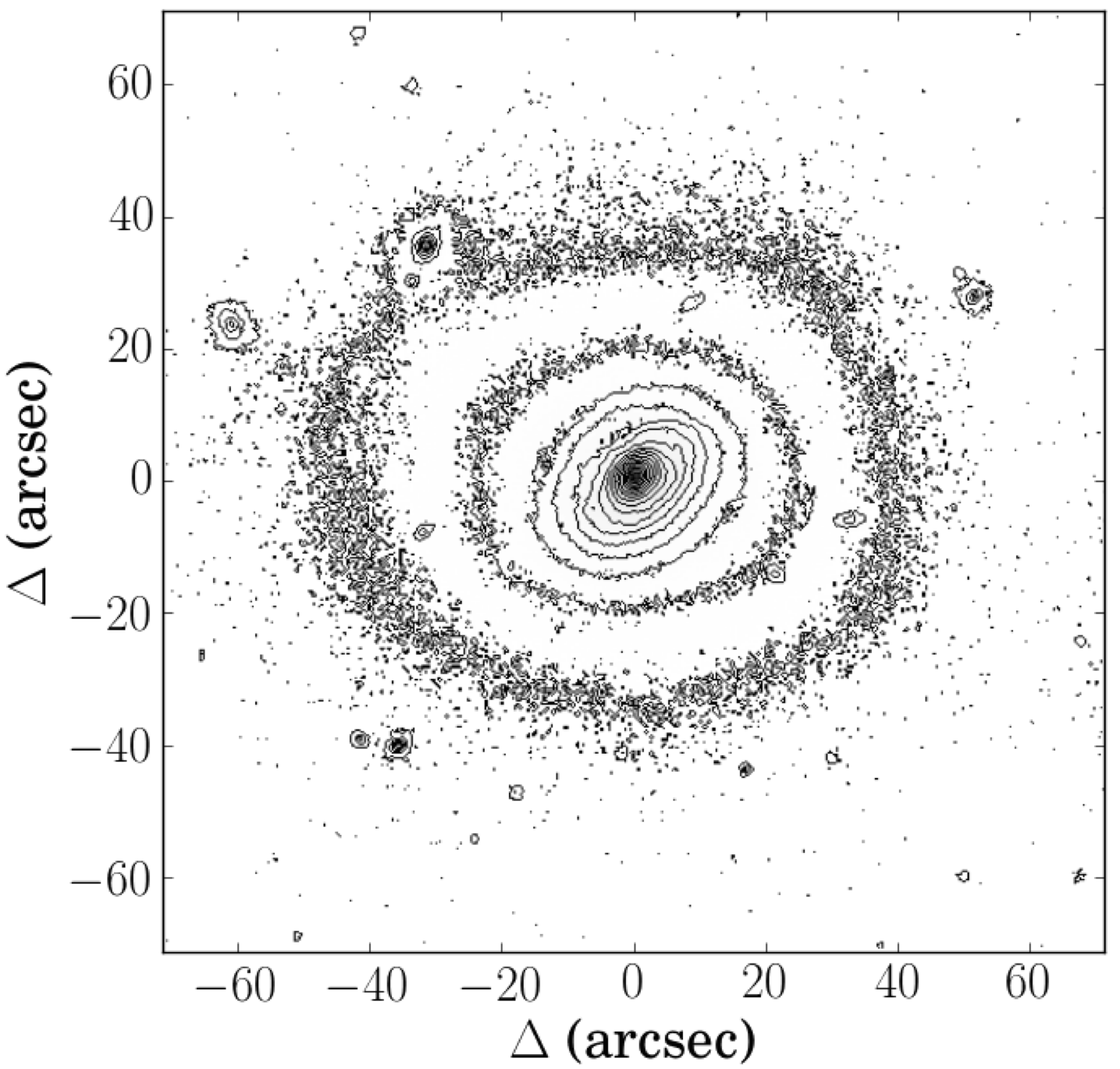}
	\caption{Map of contour levels for the full image of the galaxy NGC\,5784. The distortion of the external isophotes respect to the inner ones is clearly visible.}
	\label{n5784_tot}
\end{figure}

\begin{figure*}
	\includegraphics[width=16cm]{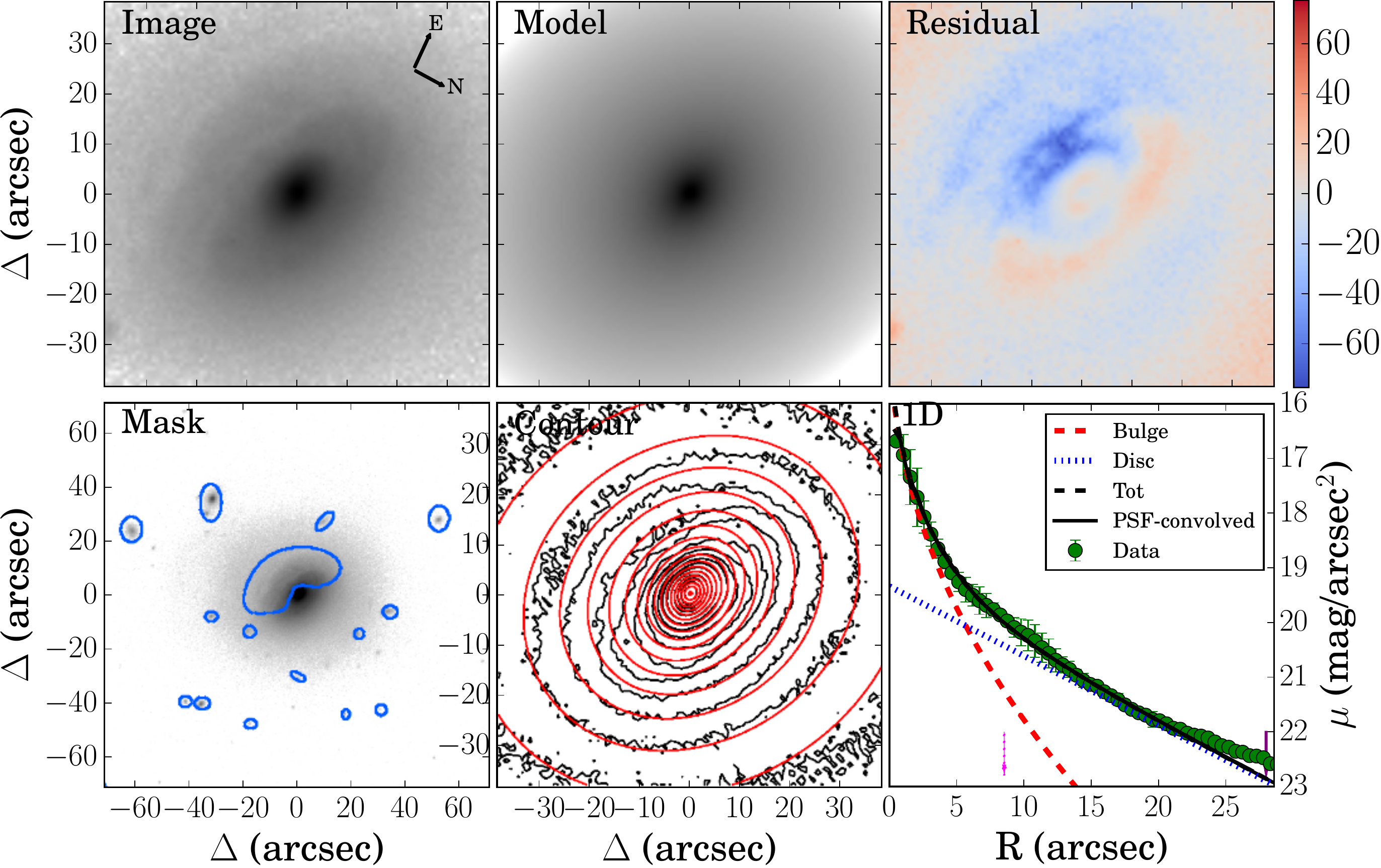}
	\caption{NGC\,5784: same as in Fig.~\ref{n08234}. The north direction is at -118.2$^\circ$. The panels show only the region considered in the fitting process (see Section~\ref{sec:n5784}). The lower left panel shows the mask applied to the full image of this galaxy (see Fig.~\ref{n5784_tot}).}
	\label{n5784}
\end{figure*}

This galaxy is contaminated not only by numerous stars and background galaxies, but also by a dust lane. To exclude the pixels influenced by these effects a mask is created (see lower left panel of Fig.~\ref{n5784}). For the dust regions, we try to build a mask with a symmetrical shape with respect to the minor axis. Thus, in the inner region only half of the galaxy is considered in the fitting process. The two-component fit for this galaxy is characterised by relative large residuals, mostly due to the dust. The outer contour levels show a distortion in comparison to the inner one (see Fig.~\ref{n5784_tot}). This feature can be viewed as an indication of disturbance, due to an interaction with a companion or a post merger. In order not to be influenced in the fitting process by external distorted regions, we fit only the region where we have kinematical data (upper left panel of Fig.~\ref{n5784}). The reason for this choice is due to the fact that in the successive analysis we consider the relative weight of the bulge and disc components only in these regions. The shape of the residuals in the upper right panel of Fig.~\ref{n5784} reflects the pattern of the dust lane.

\subsection{IC\,1652}\label{sec:ic1652}
\begin{figure*}
	\includegraphics[width=16cm]{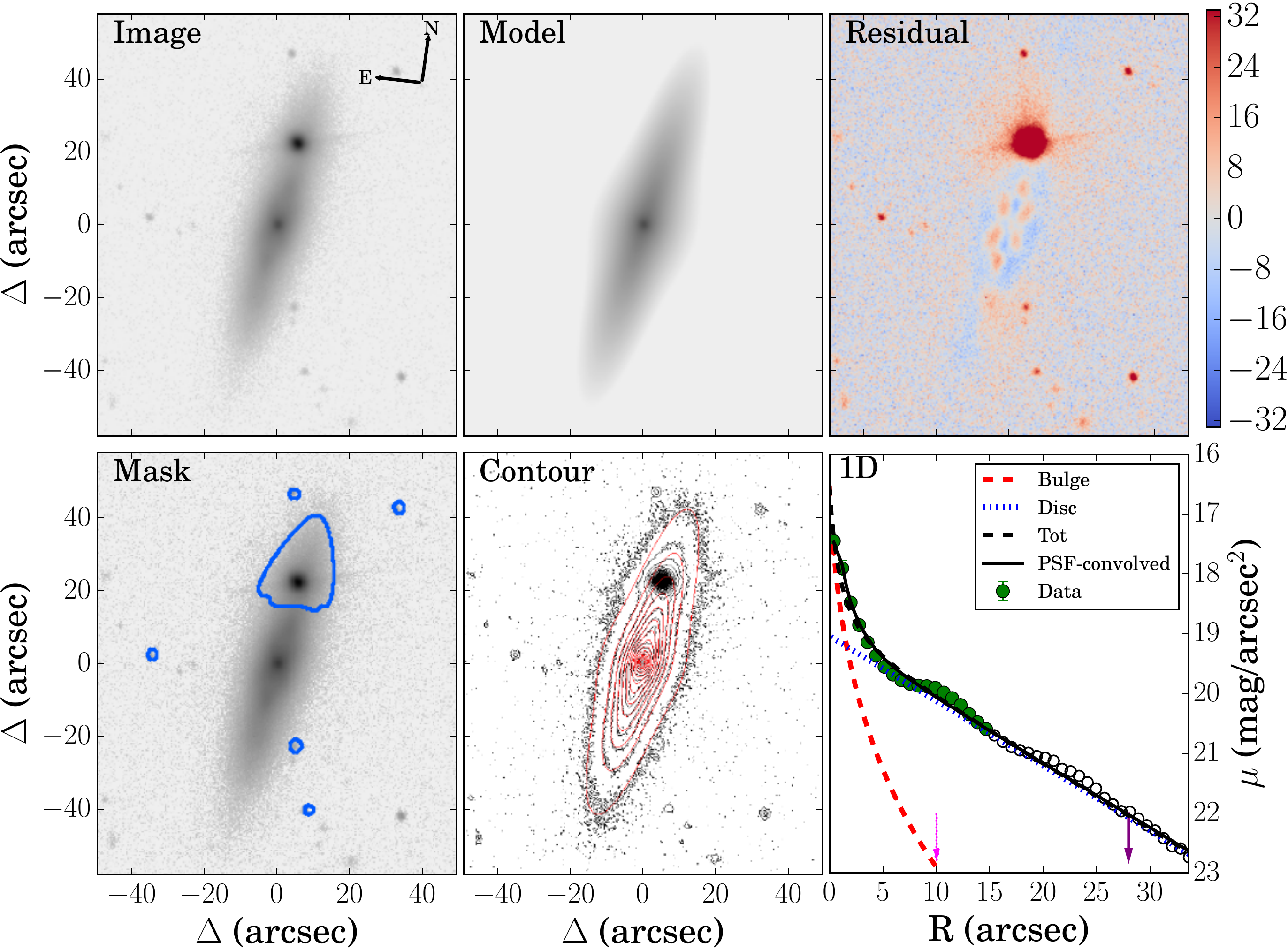}
	\caption{IC\,1652: same as in Fig.~\ref{n08234}. The north direction is at -7.4$^\circ$. For explanation of the different colours of the points in the lower right panel see the main text (Section~\ref{sec:ic1652}).}
	\label{n1652}
\end{figure*}

Half of the image of this galaxy is contaminated by a star. In order to not consider the pixels influenced by its light during the fitting process and to be sure that the fitting parameters would not be influenced by spurious effects we create a mask which covers not only the star but also the regions of the galaxy around it (Lower left panel of Fig.~\ref{n1652}). The residuals of the two-component model have a quadrupole symmetry, signal of the presence of other components (Upper right panel of Fig.~\ref{n1652}). The over-subtraction along the major axis is probably due to a contamination of the star. Adding other components to the two-component model makes the cross pattern of the residuals almost disappear while those along the major axis remain almost unchanged. However, all this complications do not influence our analysis: comparing the parameters of the disc for the two-component model with that of a more complicated one we find that the structure of disc is well constrained also by the simpler model. In Table~\ref{tab:photo} we report the parameter uncertainties based on our method (2$\sigma$ sky variation, see Section~\ref{incertezze}), aware that for this galaxy those relative to the bulge could be underestimated. From a distance of 15 arcsec from the centre the data for the 1D profile (lower right panel of Fig.~\ref{n1652}) are extracted only from the side of galaxy not influenced by the star and they are represented with empty black circles.

\subsection{NGC\,7025}\label{sec:n7025}
\begin{figure*}
	\includegraphics[width=16cm]{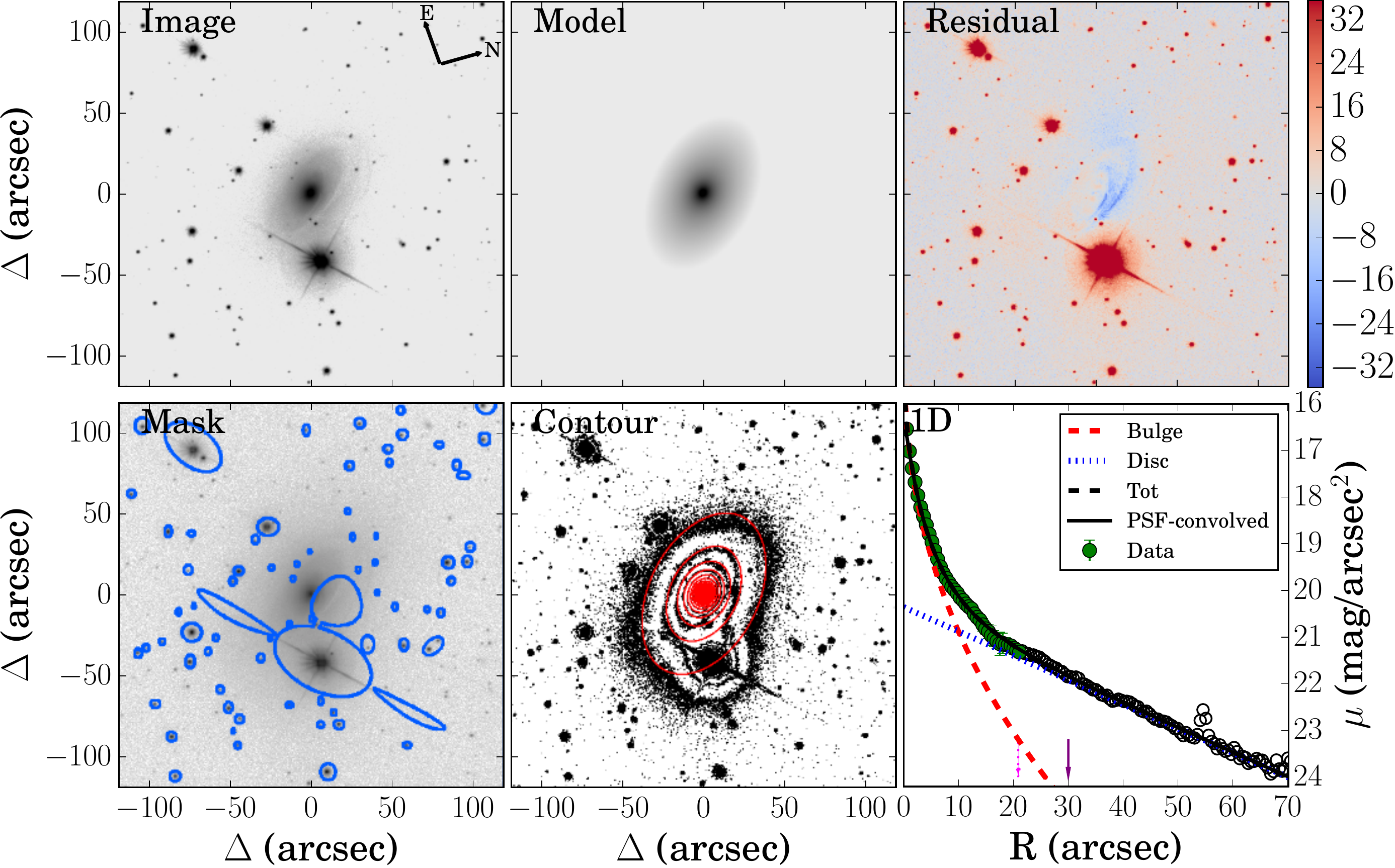}
	\caption{NGC\,7025: same as in Fig.~\ref{n08234}. The north direction is at -71.4$^\circ$. For explanation of the different colours of the points in the lower right panel see the main text (Section~\ref{sec:n7025}).}
	\label{n7025}
\end{figure*}

This galaxy is strongly influenced by foreground stars, in addition it has also a clear dust lane. The mask created to fit only the good regions is shown in the lower left panel of Fig.~\ref{n7025}. This galaxy is well described by a two-component model: the parameters are well constrained and there is no evidence of systematic residuals due to the choice of the model. The outer contour level depicted in the lower middle panel of Fig.~\ref{n7025} shows a discrepancy with the model, which would seem to suggest a slightly different position angle for the disc. On the other hand, if we force the disc model to follow this trend, stronger residuals are produced in the more inner regions. The residuals along the minor axis (upper right panel of Fig.~\ref{n7025}) are due only to the presence of dust in those regions. The position of data points below the model in the 1D profile in the region around 15 arcsec is due to the influence of dust. Starting from 22 arcsec (corresponding to $(\Delta x,\Delta y)\sim (-9,-20)$ arcsec in the lower left panel of Fig.~\ref{n7025}) the data are extracted only on the side of galaxy not covered by the mask and they are represented as empty black circles. The peak in the extracted profile visible in the lower right panel of Fig.~\ref{n7025} at $\sim$55 arcsec is due to a star covered by the mask.

\subsection{NGC\,6081}\label{sec:n6081}
\begin{figure*}
	\includegraphics[width=16cm]{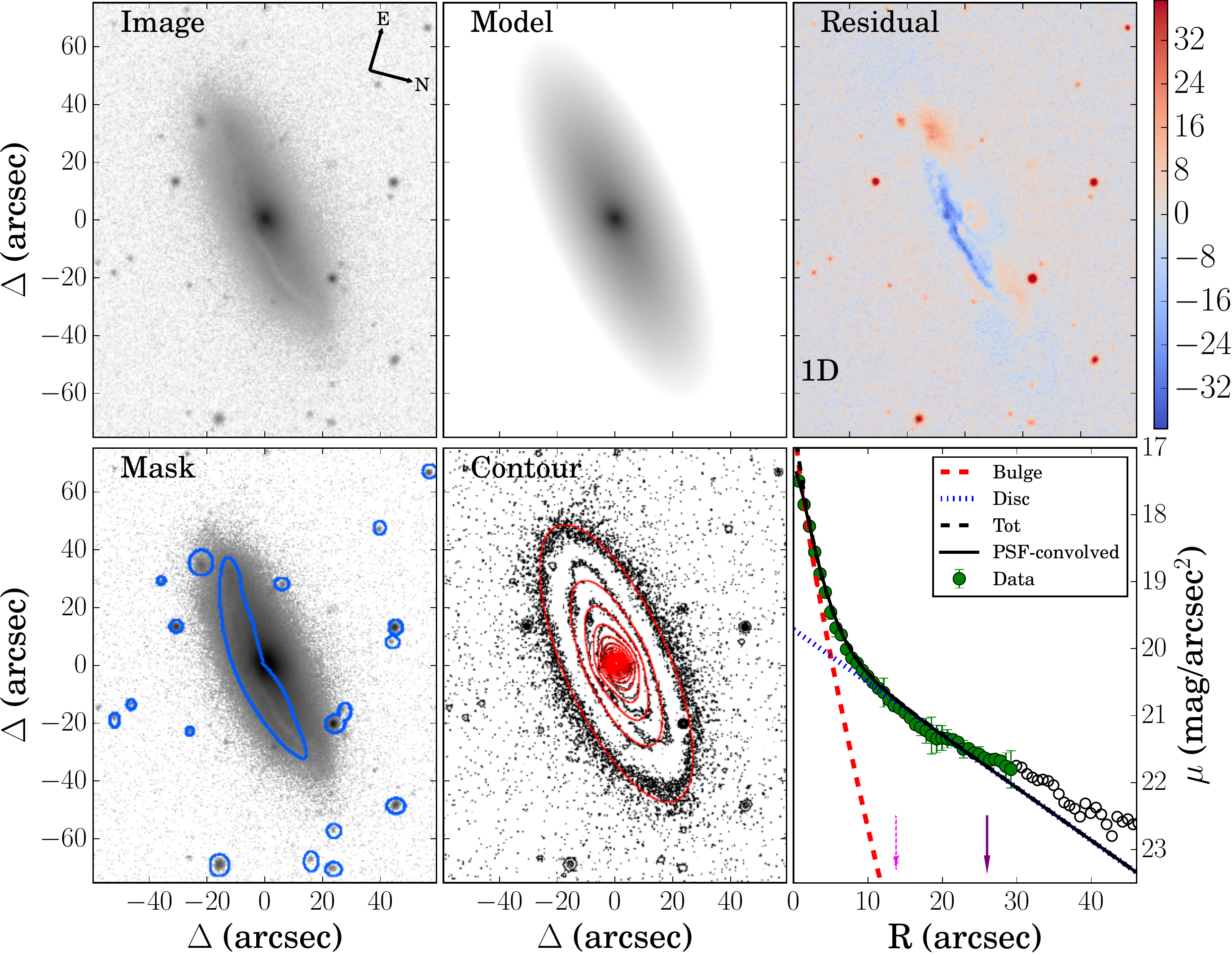}
	\caption{NGC\,6081: same as in Fig.~\ref{n08234}. The north direction is at -104.7$^\circ$. For explanation of the different colours of the points in the lower right panel see the main text (Section~\ref{sec:n6081}). }
	\label{n6081}
\end{figure*}

A mask is created for this galaxy to exclude the contamination of dust and stars. The extension of the masked regions covers not only the clear dust lane visible in the lower left panel of Fig.~\ref{n6081} as the darkest regions, but also the more inner regions. The inspection of contour levels reveals, indeed, that also these central regions are influenced by dust. This is the galaxy with the lowest Sérsic index among the ten in our sample. This Sérsic index is well constrained as it is subject to a relative variation lower than 1\% against a variation in the sky level of 2 standard deviation. The data points are extracted in a wedge of 5$^\circ$ with the disc major axis as first segment, in order to avoid regions covered by the mask. The green empty points in the lower right panel of Fig.~\ref{n6081} are extracted in a boundary region between the masked and fitted galaxy area. Starting from 20 arcsec the black empty points show data extracted only along the side of the major axis at positive value of $\Delta x$ and negative of $\Delta y$.

\subsection{NGC\,0528}\label{sec:n0528} \label{ngc0528_photo}
This is another example in which we had to mask the stars that influenced the image of the galaxy prior to the fit. This galaxy is characterised by a great difference between the position angle of the bulge and the disc, so we cannot fix them to the same value. Letting all the photometric parameters free the fit returns an unphysical model in which the effective radius for bulge is twice that for disc and the bulge Sérsic index is 6.7. However, the values of both $n$ and effective radius are strongly dependent on the mean sky estimate. For these reasons we fix the bulge Sérsic index to the value of 4. Although this choice can be considered arbitrary, it reflects the more traditional, conservative and simple way used in the literature to describe the surface brightness profile of a bulge \citep{carollo98, moll}. The parameters of the two components with this assumption on $n$, are those reported in Table~\ref{tab:photo}. The relative residuals show an over-subtraction of the model along one side of the minor axis. We verify that it is an intrinsic effect (probably dust absorption) and it is not due to a wrong estimate of the centre of the galaxy. This galaxy was photometrically decomposed into bulge and disc components by \citet{tabor}. As for NGC7671 the values of the recovered disc parameters are consistent with those found in our analysis, while for the bulge there is a discrepancy that could be explained taking into account the difference of the data features (FOV, PSF; wavelength range) used in the two studies (see discussion in Section \ref{ngc7671_photo}).

\subsection{UGC\,08234}
The image of this galaxy is contaminated by a few stars that we exclude with the mask present in the lower left panel of Fig.~\ref{n08234}. The fitting runs in which the position angle of the two components are let free to vary, give a difference less than 1$^\circ$ and so we fix them to the same value. The parameters of the bulge are strongly dependent on the sky estimate in this case. A variation of 2 standard deviation in the sky mean estimate causes a variation of 12\% and 11\% on the value of effective radius and Sérsic index respectively. This is due to the fact that the bulge Sérsic index is larger than 4. This means that in the external regions the relative importance of the bulge respect to the disc becomes important given the extended outer wings typical of Sérsic profile with $n\ge$4. On the other hand, the most important disc parameters vary by only $\leqslant3\%$.

\subsection{UGC\,10905}
\begin{figure*}
	\includegraphics[width=16cm]{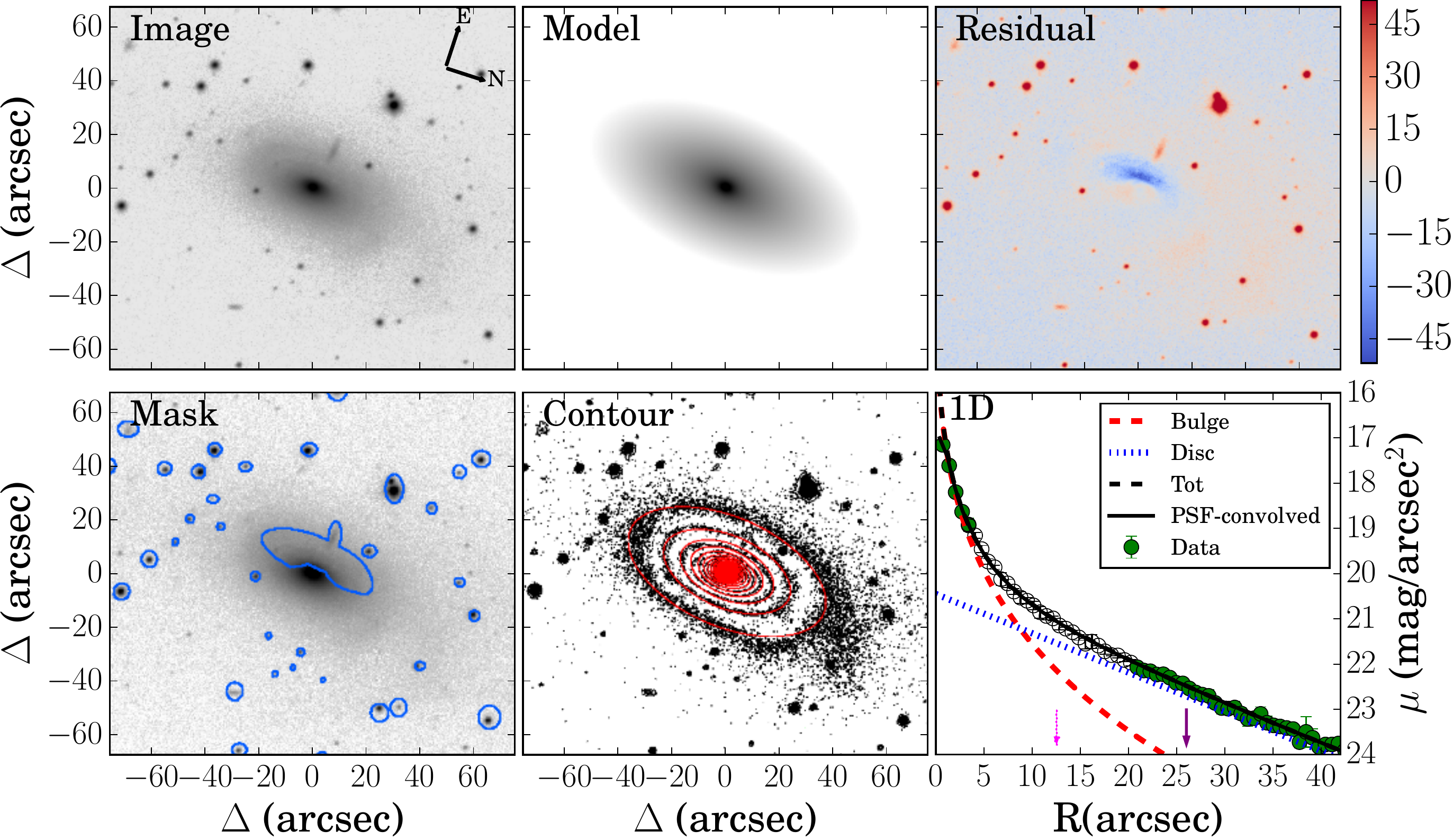}
 	\caption{UGC\,10905: same as in Fig.~\ref{n08234}. The north direction is at -108.2$^\circ$. }
	\label{n10905}
\end{figure*}

The image of this galaxy is polluted by numerous stars. We build a mask (lower left panel of Fig.~\ref{n10905}) in the central regions to cover both the contamination of a dust lane and a stream-like structure that appears perpendicular to the plane of the galaxy. The most external contour level in the lower middle panel of Fig.~\ref{n10905} appears asymmetric along the major axis. After the first runs, the position angle of the bulge is kept fixed at the same value of the disc, because the run in which they are letting free, returns a difference $\lesssim1^\circ$ between these angles. The negative values of the residuals, under the application of this model, are due to the regions influenced by dust in the inner regions. In the region around 5 and 20 arcsec the data of the 1D profile (lower right panel of Fig.~\ref{n10905}) are extracted in a wedge which has an amplitude of $5^\circ$ and has the disc major axis as first segment to avoid the area contaminated by dust. The points extracted under this geometrical assumption are represented as black empty circles.

\subsection{NGC\,0774}\label{ph:n0774}
\begin{figure*}
	\includegraphics[width=16cm]{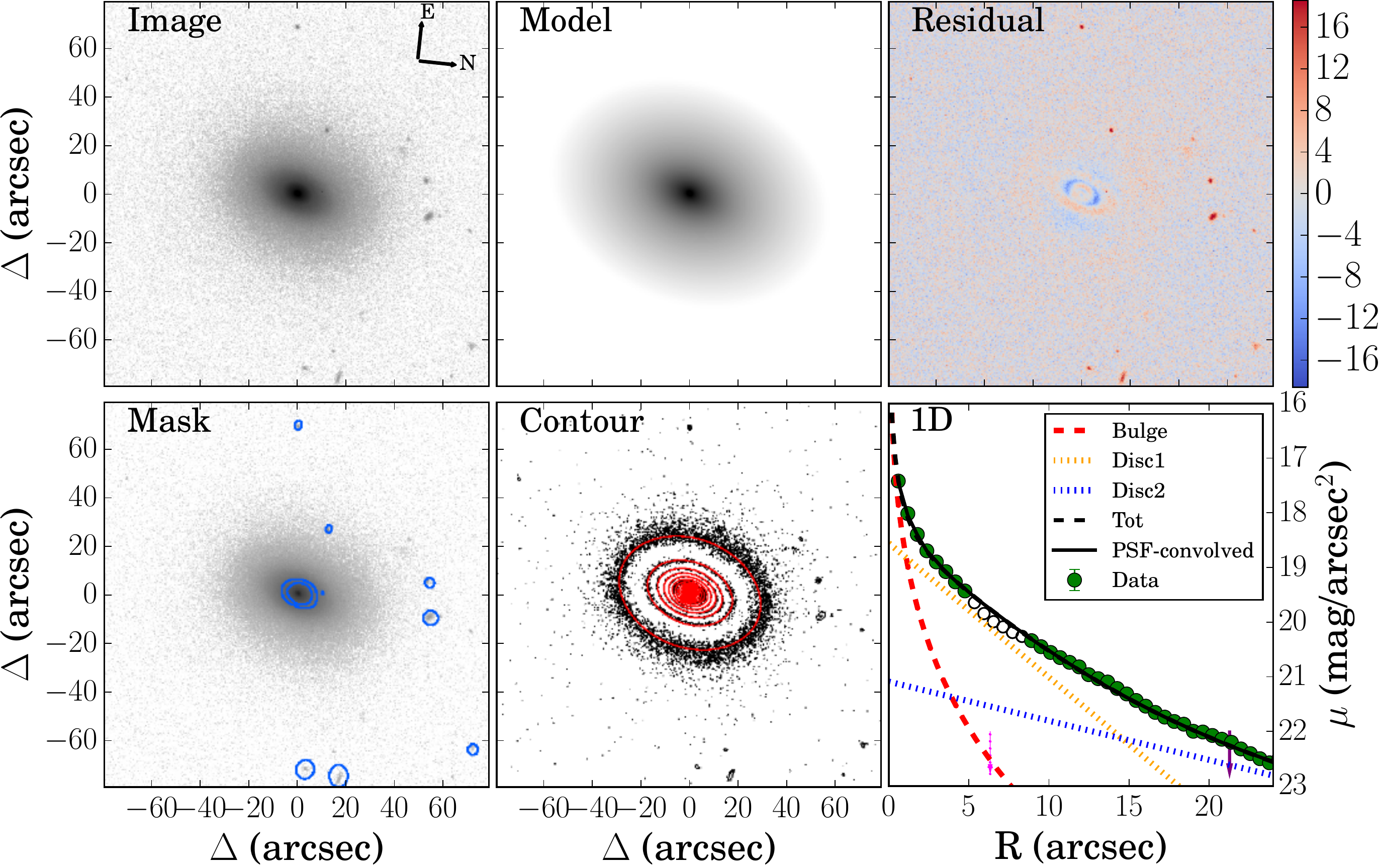}
	\caption{NGC\,0774: same as in Fig.~\ref{n08234}. The north direction is at -96.1$^\circ$. The two arrows in the lower right panel refer to a luminosity weight average of the scale radii of the two discs (see the main text, Section~\ref{ph:n0774}). For explanation of the different colours of the points in the lower right panel  see the main text (Section~\ref{ph:n0774}). }
	\label{n0774}
\end{figure*}

A small dust ring fainter than the surrounding regions is present in the image of this galaxy. For this reason a ring mask which covers the pixels influenced by this effect is created (lower left panel of Fig.~\ref{n0774}). For this galaxy, the fit in which the photometric parameters are let free returns an unphysical model with the effective radius of the bulge twice that of the disc. Moreover, this model is characterised by large relative residuals in the external regions and for this reason we add another disc component. The inspection of the contour level map (middle panel of Fig.~\ref{n0774}), as well as of the residuals (upper right panel of Fig.~\ref{n0774}) reveals a good agreement between this three-component model and the galaxy data. In Table~\ref{tab:photo} we indicate with the suffix D1 the parameters of the inner disc and with D2 those of the outer disc. The ring shape in the residuals is due to the dusty regions covered by the mask, which are represented by empty black circles in the 1D profile of the lower right panel of Fig.~\ref{n0774}.\\
As discussed in the main text, this is the only galaxy of our sample for which the addition of a third component was necessary. Since in our successive analysis we want to have a unique value of disc scale length we consider also that obtained by a two component fit with bulge parameters fixed at the values obtained with three-component model. The value of 6.33 arcsec for the disc scale length, obtained under this assumption, could be considered as a luminosity-weight average of the two scale radii of the best-fit model. The other parameters for this disc are shown in Table~\ref{tab:photo} and indicated with suffix D. The dashed magenta arrow in the lower right panel of Fig.~\ref{n0774} is referred to this luminosity-weight scale radius. 


    

\section{Kinematic decomposition}\label{app:b}
\subsection{Description of bulge and disc spectral templates}
In Section~\ref{sec:kinematic} we describe our fitting process developed to perform a kinematic bulge-disc decomposition. Two of the quantities used in this fitting are the bulge and disc spectral templates, $T_{\mathrm{b}}$ and $T_{\mathrm{d}}$. These two templates are found applying the routine pPXF \citep{cap-ppxf} to our representative disc and bulge spectra. As described in Section~\ref{template}, pPXF finds the best linear combination of spectral templates (see details for them in the main text) that are able to best reproduce the observed spectra. In Table~\ref{tab:template} we report both the properties of simple stellar population templates that fit our representative disc and bulge spectra and the coefficients $a_{\mathrm{i}}$ assigned to each of them.  A linear combination, see equations~(\ref{eq:templateb}) and (\ref{eq:templated}) of these SSP spectral templates, defined by their age (column 3 and 6) and metallicity (column 4 and 7) with their respective coefficients $a_{\mathrm{i}}$ allows us to create bulge and disc spectral templates for each galaxy of our sample. As mentioned in the main text we consider only bulge templates with t\,$\geqslant10$ Gyr, we verify that releasing this constraint results in meaningful weights given to the younger templates. Comparing, for each galaxy, the disc and bulge templates with the largest weights, we can see that there is an indication that discs are younger and more metal poor than their respective bulges, as expected by the common scenario of inside-out galaxy formation, which predicts that discs formed later than bulges and that star formation continues for a longer time in discs than in bulges. This scenario is confirmed also by the large range of ages and metallicities covered by the disc templates. These results on the higher metallicities and older ages of the bulges with respect to the discs are in contrast with the results of e.g. \citet{john14, bekki11}. We could ascribe this difference to the different environment of the studied galaxies: the Virgo cluster in \citet{john14} and field/group in this work. However, as mentioned in Section~\ref{template}, the choice of an optimal template for the bulges and discs has no effects on the main results of the present work.

\begin{table*}
	\caption{For each galaxy of our sample (column 1) we report the templates able to reproduce bulge and disc spectra. Each of these templates is identified by age and metallicity (columns\,3-4 and columns\,6-7 for bulge and disc respectively), as well as by a coefficient $a_{\mathrm{i}}$ (columns\,2 and 5), through which a linear combination of a bulge, equation~(\ref{eq:templateb}), and a disc template, equation~(\ref{eq:templated}), is built for each galaxy.}
	\label{tab:template}
	\begin{tabular}{c| c c c |c c c }
\hline
 Galaxy &\multicolumn{3}{c}{Bulge} &\multicolumn{3}{c}{Disc}\\
&  $a_{\mathrm{b,i}}$ & Age(Gyr) & [M/H]  & $a_{\mathrm{d,i}}$ & Age(Gyr) & [M/H] \\
\hline
NGC\,7671 &  0.81 &13 & 0.40 &  0.46 & 3.25 & 0.26 \\
&  0.19 & 12 & 1.49&0.37 & 9.5 & -0.35 \\
&&&&0.12 & 5.5&-0.66\\
&&&& 0.03 & 14 & 0.26 \\
\hline
 NGC\,7683  & 0.52 & 10.5 & 0.26& 0.54& 5.5 & -0.66\\
  & 0.26 & 10.5 & -1.26& 0.42& 11.5 & 0.26\\
  & 0.21 & 14.0 & 0.26& 0.025 & 0.2 & -0.35\\
 \hline
 NGC\,5784 & 0.35& 14.0&0.26 & 0.48&9.5&-0.35\\
 &0.34&12.0 &-1.49& 0.20&5.5&-0.66\\
 &0.21&10.5 &0.26&0.14 &1.25&0.4\\
 &0.11 & 10.5 &-1.26 &0.06&0.03&0.26\\
 &&&& 0.06&0.5 &-0.96\\
 &&&&0.03&7.5&-0.96\\     
  \hline
 IC\,1652 & 0.46 & 10.5 &0.26& 0.89& 9.5 & -0.35\\
 & 0.28 & 10.5 & -1.26&0.05& 0.2& -0.35\\
 & 0.19 &10.0& -0.25& 0.04& 0.5 &0.40\\
& 0.05 & 11.5 & -0.66& 0.01&7.5 & -0.96\\
\hline
NGC\,7025&0.62&13.0&0.26&0.31&10.5&-1.26\\
&0.17&12.0&-1.49&0.18&13.0&0.40\\
&0.11&10.5&-1.26&0.16&12.5&0.06\\
&0.06&14.0&0.26&0.16&1.25&0.40\\
&0.03&11.5&-1.49&0.08&3.25&0.26\\
&&&&0.06&4.0&-1.49\\
\hline
 NGC\,6081 &0.50& 14.0 & 0.26& 0.66&5.5 &-0.66\\
 &0.23 & 11.5 & -0.66&0.24 & 14.0 & 0.26\\
 &0.14&10.5 & -1.26& 0.04 & 0.06 & 0.06\\\
 &0.11&12.0& -1.49&0.03& 0.10& -0.96\\
 &&&&0.02& 0.03 & -2.27\\
 \hline
 NGC\,0528   & 0.56 & 13.0 & 0.40  & 0.6& 5.5 & -0.66\\
& 0.32 & 10.5 & -1.26 &0.26&14.0 & 0.26\\
 &0.08 & 12.0 & -1.49 &0.05&0.03 & -2.27\\\
 & 0.03 & 14.0 & 0.26 \\
 \hline
 UGC\,08234  &0.45 & 11.5 & -1.49& 0.78 & 7.5 & -0.96\\
 & 0.37 & 11.5 & -1.26& 0.12 & 0.2 & -0.35\\
 & 0.17& 10.0 & 0.26&0.08& 5.5 & -0.66 \\
 &&&&0.01 & 0.5 & 0.96\\
 \hline
 UGC\,10905  &0.55& 10.5 & 0.26 & 0.64& 7.0& -0.66\\\
 & 0.25 & 10.5 & -1.26 & 0.17& 3.25 & 0.26\\
 &0.18&11.0 & -1.26 &0.09 &5.5& -0.66\\
 &0.01&12.0 & -1.49 &0.06 & 11.5 & -0.25\\\
 &&&&0.01& 1.25 &0.40\\
 \hline
 NGC\,0774  & 0.29&10.5 & 0.26 & 0.43 & 5.5 &-0.66\\
 &0.27&14.0& 0.26&0.31 &12.5 & 0.06\\
& 0.26 & 10.0 & -0.66&0.19 & 7.5 & -0.96\\
 &0.08&11.5 & -1.49& 0.05 & 0.03 &-2.27\\
 &0.07& 11.5&-0.66\\
 \hline
	\end{tabular}
\end{table*}

\subsection{Discussion on the kinematic fitting of the two problematic galaxies}
\subsubsection{NGC\,5784} \label{app:n5784}
\begin{figure}
	\includegraphics[width=\columnwidth]{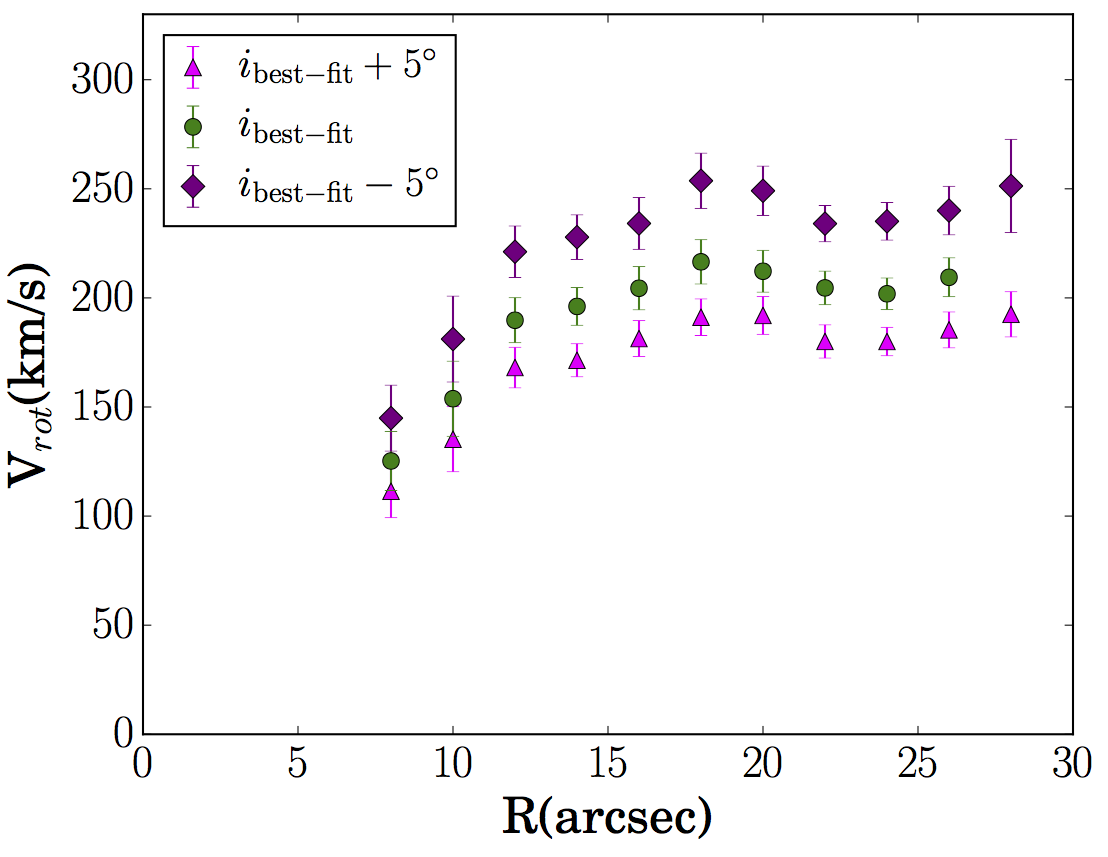}
	\caption{Rotation curves derived under different inclination angles for NGC\,5784. The green points are the same as in Fig.~\ref{fig:vel10}, while magenta triangles and purple diamonds represent rotation velocities derived with a value of $i=i_{\mathrm{best-fit}}+5^\circ$ and $i=i_{\mathrm{best-fit}}-5^\circ$ respectively, with $i_{\mathrm{best-fit}}$ the value derived from photometry.}
	\label{confronto5784}
\end{figure}

This is the most face-on galaxy between those studied in this work. It has a value of $q_{\mathrm{d}}=0.85$, corresponding to an inclination i$\sim32^{\circ}$. Since the rotation velocity has a dependence on the inclination angle $i$ that goes as 1/sin\,$i$, see equation~(\ref{vrot}), the same uncertainty on the  angle $i$ corresponds to a stronger variation of the velocity in a nearly face-on galaxy. Aware of this, we derive the rotation curve not only for the best-fit value of $i$ found by photometric decomposition, but also for inclinations that deviate of $5^{\circ}$ from the best-fit value, that is in general a reasonable uncertainty for this parameter.\footnote{An error of $5^{\circ}$ of the inclination angle corresponds to an error of $\sim$0.04 (equal to the maximum uncertainty on this parameter among those found for our sample, see Table~\ref{tab:photo}) in the value of $q_{\mathrm{d}}$, as given by the expression of error propagation: $i=\arccos\,q_{\mathrm{d}} \rightarrow \Delta\,i=(180/\pi)\Delta\,q_{\mathrm{d}}/\sqrt{1-q_{\mathrm{d}}^{2}}$.} In Fig.~\ref{confronto5784} we show both the rotation velocity profiles (magenta triangles and purple diamonds) found under these assumptions and the profile found with the best-fit value of $i$ for comparison. As we expect there is a significant variation in the values of rotation velocities, that reach a maximum relative difference of 18\% in the case of a difference of $-5^{\circ}$ from the best fit. The maximum relative difference for $+5^{\circ}$ difference is of 13\%, more similar to those of $\sim$10\% (i.e.\,see the first green point in Fig.~\ref{confronto5784}), obtained with our uncertainty estimation method. Therefore, the uncertainties shown as error-bars in figure are underestimated and for this reason we consider a conservative relative error of 20\% for the rotation velocities derived for this galaxy. The errors reported in Table~\ref{tab:v} are the result of this assumption.

\subsubsection{NGC\,0774} \label{app:n0774}
\begin{figure}
	\includegraphics[width=\columnwidth]{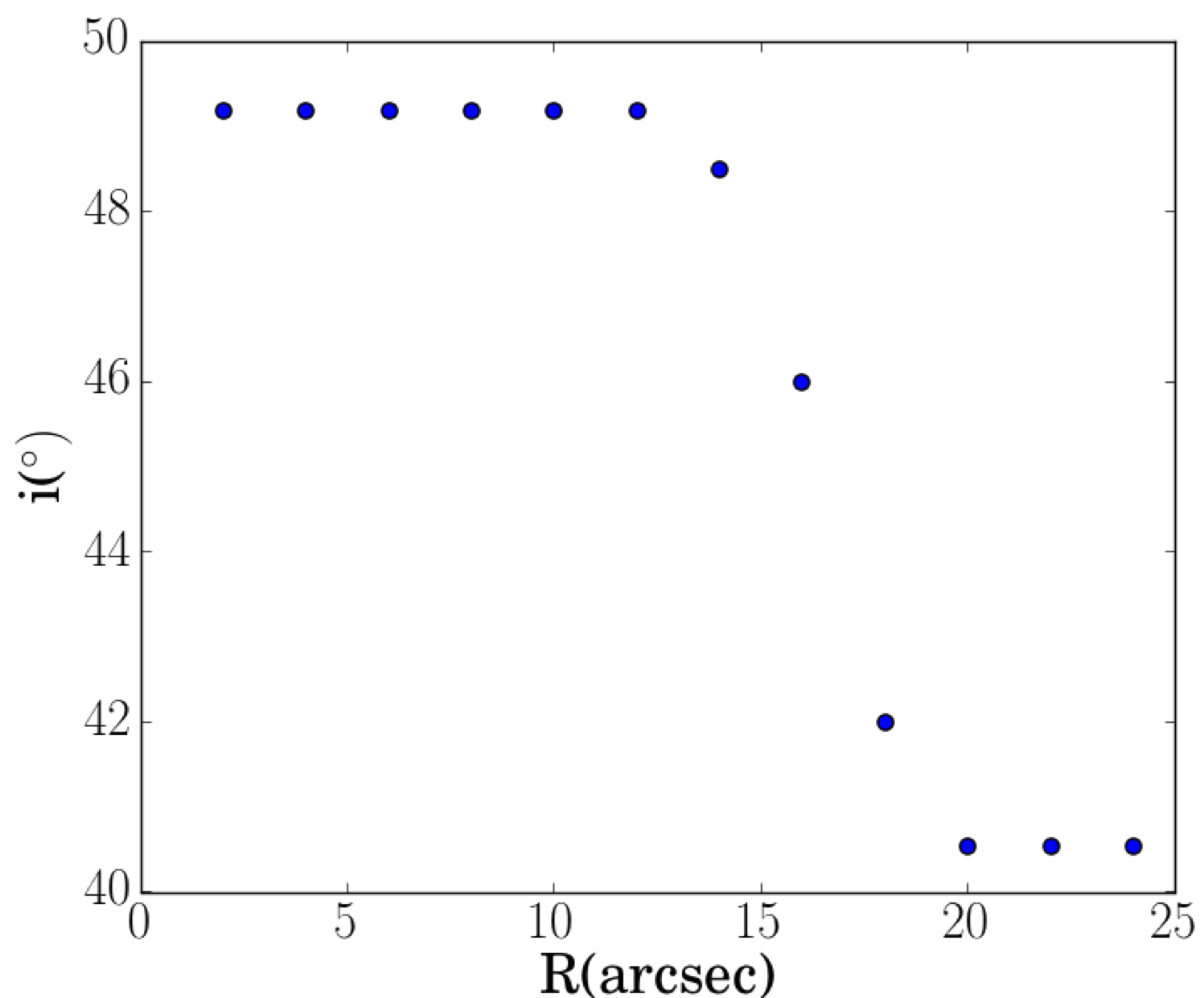}
	\caption{Values of the inclination angle at different radii used in the tilted ring fitting process for the derivation of the rotation velocity curve of NGC\,0774.}
	\label{inc0774}
\end{figure}

This is the only galaxy of our sample for which the 2 component decomposition was not sufficient. As explained in Section~\ref{ph:n0774}, its surface brightness is described using a Sérsic profile for the bulge and 2 exponential profiles for the disc, D1 and D2. The kinematic decomposition is performed in this galaxy considering as disc the sum of these 2 exponential profiles. For example in equation~(\ref{id}) the expression for $I_{\mathrm{d}}(\xi,\eta)$ becomes in this case: $I_{\mathrm{D,1}}(\xi,\eta)+I_{\mathrm{D,2}}(\xi,\eta)$, where the dependence on $(\xi,\eta)$ allowed us to take into account for different geometrical parameters of the 2 discs ($q_{\mathrm{d}}=0.65$ and 0.76 for D1 and D2 respectively, corresponding to inclination angle $i\sim$ of $49^{\circ}$ and $40^{\circ}$). At the same time in the derivation of the rotation velocity profile we consider different inclination angles for different rings: $i=49^{\circ}$ out to radius at which $I_{\mathrm{D,1}}(\xi,\eta)\lesssim\,I_{\mathrm{D,2}}(\xi,\eta)$, $i=40^{\circ}$ from radius at which $I_{\mathrm{D,2}}(\xi,\eta)\gtrsim2*I_{\mathrm{D,1}}(\xi,\eta)$ and intermediate smoothed values in the transition regions between the others 2 (see Fig.~\ref{inc0774}). 

\section{Stellar masses}\label{app:mass}
\begin{figure}
	\includegraphics[width=\columnwidth]{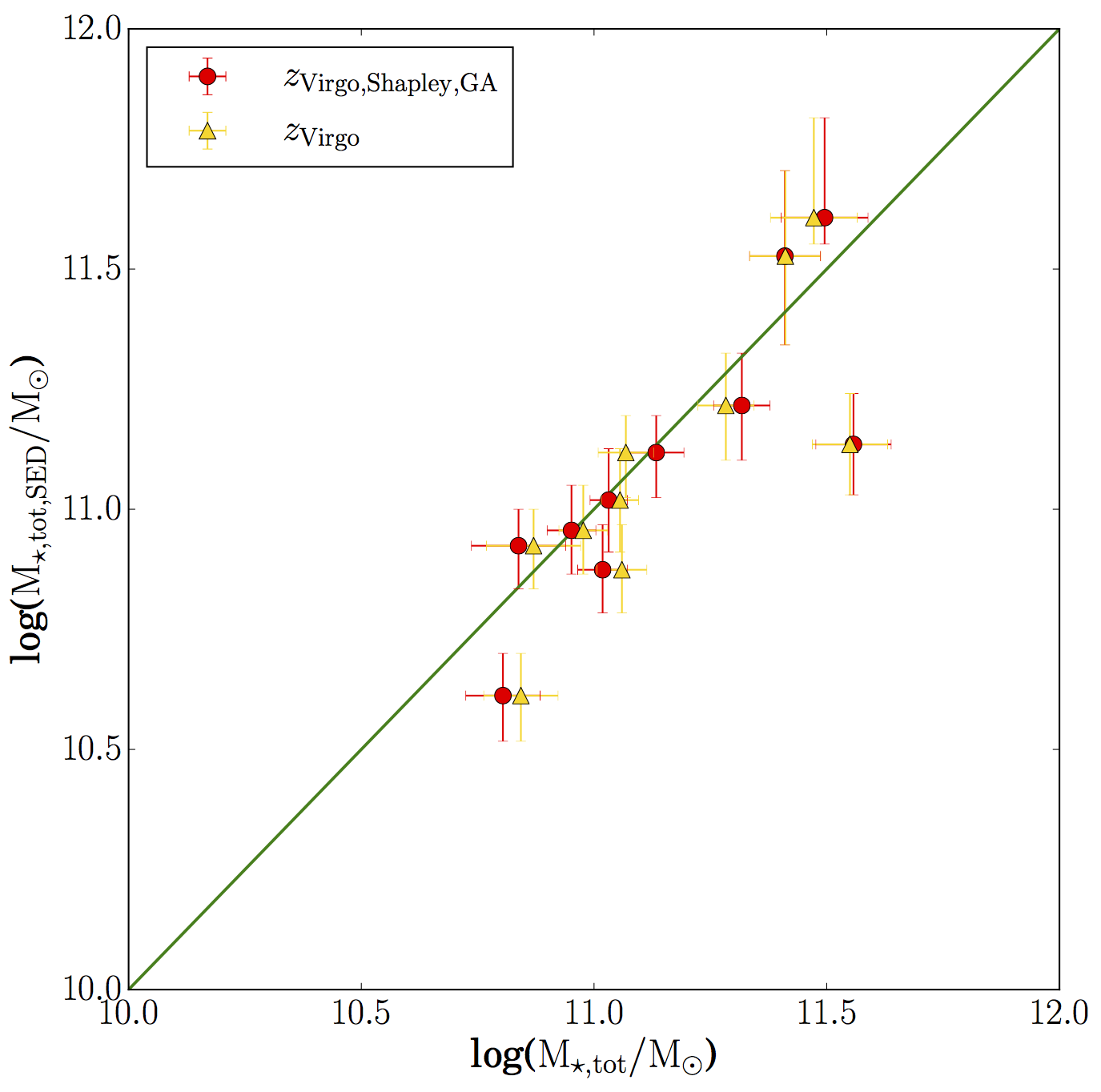}
	\caption{Total stellar masses derived from SED fitting (from CALIFA stellar masses catalogue, \citet{walcher2014}, vs stellar masses obtained from the colour-$M_{\star}/L$ relation \citep{into}. This last quantity is derived summing the bulge and disc masses, obtained through relation between $g-i$ colour and $L_{r}$ (equation~(\ref{color})). The yellow traingles are masses derived from luminosities in which the redshifts were corrected for infall into Virgo cluster, which are those used in this work, while the red circles are those derived assuming the same distances used to obtain $M_{\mathrm{\star\,tot,SED}}$, which are corrected for Virgo, Great Attractor and Shapley. The green line represents a 1:1 relation between quantities on x and y axis.}
	\label{confronto}
\end{figure}

In Section~\ref{sec:mind} we calculate the stellar masses using a colour-$M_{\star}/L$ relation. In Fig.~\ref{confronto} we show the values of $M_{\mathrm{\star\,tot}}$ obtained as explained in Section~\ref{sec:mind} and the total masses of our S0s obtained through SED fitting  (see Section~\ref{sec:mref}), $M_{\mathrm{\star\,tot,SED}}$. For consistency, for each galaxy we show both the total stellar masses obtained using distances derived from redshifts corrected for the Virgo infall (Table~\ref{nome}) (yellow triangles) and those obtained using distances derived from redshifts corrected for infall into Virgo, Great Attractor and Shapley (red circles), that are those appropriate for a right comparison with $M_{\mathrm{\star\,tot,SED}}$. The uncertainty estimations are obtained taking into account for standard deviations on ($g-i$) colours, uncertainties on the structural parameters of bulges and discs (Table~\ref{tab:photo}) used for calculation of their respective $\mathrm{L}_{r}$ and errors on distances (Table~\ref{nome}). The visual inspection of Fig.~\ref{confronto} reveals that the two ways to estimate the stellar masses give consistent results with the exception of two galaxies: UGC\,08234 and IC\,1652. For this second galaxy $M_{\mathrm{\star\,tot}}$ and $M_{\mathrm{\star\,tot,SED}}$ differ by a factor $\sim1.5$, but the two values are compatible within 2$\sigma$. For UGC\,08234 there is a factor of $\sim2.5$ between $M_{\mathrm{\star\,tot}}$ and $M_{\mathrm{\star\,tot,SED}}$.\\

\bsp	% typesetting comment
\label{lastpage}
\end{document}